\documentclass[12pt]{article}

\usepackage[left=0.8in,right=0.8in,top=1.0in,bottom=1.0in]{geometry}
\usepackage{color}
\usepackage{amsmath,amssymb}
\usepackage{bm}
\usepackage{graphicx, color}
\usepackage{wrapfig}
\usepackage{here}
\usepackage{cite}
\usepackage{ulem}

\numberwithin{equation}{section} 

\definecolor{green}{rgb}{0, 0.5, 0} 

\usepackage[
colorlinks=true,linkcolor=black,citecolor=blue,urlcolor=blue]{hyperref}

\begin{document}
\title{
		\begin{flushright}
		\ \\*[-80pt] 
		\begin{minipage}{0.22\linewidth}
			\normalsize
			EPHOU-21-014\\
			CTPU-PTC-21-39\\
			KEK-TH-2369\\
			HUPD-2113\\*[30pt]
		\end{minipage}
	\end{flushright}
	{\Large \bf 
	 Modular symmetry in  the SMEFT
		\\*[20pt]}}

\author{ 
	\centerline{
		Tatsuo Kobayashi $^{1}$,   Hajime Otsuka $^{2,3}$,
		Morimitsu Tanimoto $^{4}$,  Kei Yamamoto  $^{5}$
	} 
	\\*[20pt]
	\centerline{
		\begin{minipage}{\linewidth}
			\begin{center}
				$^1${\it \normalsize
					Department of Physics, Hokkaido University, Sapporo 060-0810, Japan} \\*[5pt]
				$^2${\it \normalsize
				    KEK Theory Center, Institute of Particle and Nuclear Studies,
				    KEK, 1-1 Oho, Tsukuba, Ibaraki 305-0801, Japan}\\*[5pt]
				$^3${\it \normalsize
				    Center for Theoretical Physics of the Universe, Institute for Basic Science, Dajeon 34126, Korea}\\*[5pt]
				$^4${\it \normalsize
Department of Physics, Niigata University, Niigata 950-2181, Japan} \\*[5pt]
				$^5${\it \normalsize
					Core of Research for the Energetic Universe, Hiroshima University,
					Higashi-Hiroshima 739-8526, Japan} \\*[5pt]
			\end{center}
	\end{minipage}}
	\\*[80pt]}
\date{
	\centerline{\small \bf Abstract}
	\begin{minipage}{0.9\linewidth}
		\medskip 
		\medskip 
		\small
		\mbox{}
We study the modular symmetric standard-model 
effective field theory.
We employ the stringy Ansatz on coupling structure that 
4-point couplings $y^{(4)}$ of matter fields are written by 
a product of 3-point couplings $y^{(3)}$ of matter fields, 
i.e., $y^{(4)} = y^{(3)}y^{(3)}$. 
In this framework, we discuss the flavor structure of bilinear fermion operators
 and 4-fermion operators,
where the holomorphic and anti-holomorphic modular forms appear.
From the Ansatz,  
the $A_4$ modular-invariant semileptonic four-fermion  operator 
$[\,\bar E_R \Gamma  E_R][\,\bar D_R  \Gamma  D_R\,]$ 
does not lead to the flavor changing (FC) processes
since this operator would be 
constructed in terms of gauge couplings $g$ as $y^{(3)} \sim g$.
	The chirality flipped bilinear operator $[\,\bar D_R \Gamma  D_L\,]$ 
	also does not lead  FC if the mediated mode   corresponds to the Higgs	boson $H_d$.
	In this case, the flavor structure of this operator is  the exactly same as the mass matrix.
	On the other hand, if the flavor structure of the operator is not  the exactly same as the mass matrix, the situation would change drastically.
	Then,
	we obtain the non-trivial relations of  FC transitions
	at nearby fixed points $\tau=i,\,\omega\,, i\infty$, which are testable in the future.
As an application, we discuss the relations of the lepton flavor violation processes
$  \mu\to e\gamma,\, \tau\to \mu\gamma$ and 
$\tau\to e\gamma$ at nearby $\tau_e=i$,
where the successful lepton mass matrix was obtained.
We also study the flavor changing 4-quark operators 
in the $A_4$ modular symmetry of quarks. 
As a result, the minimal flavor violation could be realized by taking relevant specific parameter sets of order one.
	\end{minipage}
}

\begin{titlepage}
	\maketitle
	\thispagestyle{empty}
\end{titlepage}


{\color{black}
\tableofcontents
}

\newpage
\section{Introduction}
 The Standard Model (SM) has been regarded as the low-energy limit of an  full theory at the high energy scale.  
  After the first years of running of the LHC, no new physics (NP) has been
  discovered. That is,  there is a mass gap between the SM spectrum and these hypothetical additional degrees of freedom such as new particles.
   
Describing possible physics beyond the SM in general terms gets increasingly important, 
and the systematic study goes under the name of the SMEFT: the SM Effective Field Theory (EFT) based on the 
  $SU(3)_c \otimes SU(2)_L \otimes U(1)_Y$ gauge symmetry, and the SM field content
  as dynamical degrees of freedom below a cut-off scale $\mu  >  G^{-1/2}_F$.
Several years after the pioneering analysis in~Ref.\cite{Buchmuller:1985jz}, 
the first complete list of non-redundant SMEFT Lagrangian terms up to dimension-six
 has been presented in~Ref.\cite{Grzadkowski:2010es}.

When all the possible flavor structures are taken into account in absence of any flavor symmetry, 
a large proliferation in the number of independent coefficients in the SMEFT occurs:
there are 1350 CP-even and 1149 CP-odd independent coefficients for the dimension-six operators~\cite{Alonso:2013hga}.  
The flavor symmetry is a challenging hypothesis to reduce the number of independent parameters of the flavor sector.  
Above all, the flavor symmetries $U(3)^5$ and $U(2)^5$ have been successfully 
  applied to the SMEFT~\cite{Faroughy:2020ina}.
The $U(3)^5$ flavor symmetry is the maximal flavor symmetry allowed by the SM gauge sector, 
 while $U(2)^5$ is the corresponding  subgroup acting only on the first two (light) families. 
 The $U(3)^5$ allows us to apply the Minimal Flavor Violation (MFV) hypothesis~\cite{Chivukula:1987py,DAmbrosio:2002vsn}, 
 which is the most restrictive hypothesis consistent with the SMEFT,
 and suppress non-standard contributions to flavor-violating observables~\cite{DAmbrosio:2002vsn}. 
 In the $U(2)^5$ symmetry~\cite{Barbieri:2011ci,Barbieri:2012uh,Blankenburg:2012nx}, 
  it retains most of the MFV virtues and allows us to have a much richer structure 
  as far as the dynamics of third family is concerned.  
  
In the $U(3)^5$ and $U(2)^5$ flavor symmetric scenario,
the Yukawa couplings are understood as spurions (symmetry-breaking terms) 
which have non-trivial representations under the symmetry. 
Assuming that flavor- and CP-violating effects in NP is also controlled by the Yukawa,
the flavor structure of higher-dimensional operators are expressed by the spurions. 
Then, by counting different order in the breaking terms of these symmetries,
we can classify the number of independent dimension-six operators,  
as studied in Ref.~\cite{Faroughy:2020ina}. 
 These flavor symmetries are not the only options to efficiently suppress flavor-violating observables in the SMEFT,
 and the non-Abelian discrete symmetry would be one of the alternative choice. 
   
  The non-Abelian discrete groups have been discussed to 
  challenge the flavor problem  of quarks and leptons
  \cite{Altarelli:2010gt,Ishimori:2010au,Ishimori:2012zz,Hernandez:2012ra,King:2013eh,King:2014nza,Tanimoto:2015nfa,King:2017guk,Petcov:2017ggy,Feruglio:2019ktm}.
  Indeed,
  supersymmetric (SUSY) modular invariant theories give us an attractive framework to address the flavor symmetry of quarks and leptons
  with non-Abelian discrete groups \cite{Feruglio:2017spp}.
  In this approach, the quark and lepton mass matrices are written in terms of modular forms which are holomorphic functions of the modulus $\tau$.
  The arbitrary symmetry breaking sector of the conventional models based on  flavor  symmetries is replaced by the moduli space, and then Yukawa couplings are given by modular forms.  
  
  The well-known finite groups $S_3$, $A_4$, $S_4$ and $A_5$
  are isomorphic to the finite modular groups 
  $\Gamma_N$ for $N=2,3,4,5$, respectively\cite{deAdelhartToorop:2011re}.
  The lepton mass matrices have been given successfully  in terms of $A_4$ modular forms \cite{Feruglio:2017spp}.
  Modular invariant flavor models have been also proposed on the $\Gamma_2\simeq  S_3$ \cite{Kobayashi:2018vbk},
  $\Gamma_4 \simeq  S_4$ \cite{Penedo:2018nmg} and  
  $\Gamma_5 \simeq  A_5$ \cite{Novichkov:2018nkm}.
  Based on these modular forms, the flavor mixing of quarks and leptons has been discussed intensively in these years.
  Phenomenological studies of the lepton flavors have been done
  based on  $A_4$ \cite{Criado:2018thu,Kobayashi:2018scp,Ding:2019zxk}, 
  $S_4$ \cite{Novichkov:2018ovf,Kobayashi:2019mna,Wang:2019ovr} and 
  $ A_5$ \cite{Ding:2019xna}.
  A clear prediction of the neutrino mixing angles and the Dirac CP phase was given in  the  simple lepton mass matrices with
  the $A_4$ modular symmetry \cite{Kobayashi:2018scp}.
  The Double Covering groups  $\rm T'$~\cite{Liu:2019khw,Chen:2020udk}
  and $S_4'$ \cite{Novichkov:2020eep,Liu:2020akv} were also
  realized in the modular symmetry.
  Furthermore, phenomenological studies have been developed  in many works
   \cite{deMedeirosVarzielas:2019cyj,
  	Asaka:2019vev,Ding:2020msi,Asaka:2020tmo,Behera:2020sfe,Mishra:2020gxg,deAnda:2018ecu,Kobayashi:2019rzp,Novichkov:2018yse,Kobayashi:2018wkl,Okada:2018yrn,Okada:2019uoy,Nomura:2019jxj, Okada:2019xqk,
  	Kariyazono:2019ehj,Nomura:2019yft,Okada:2019lzv,Nomura:2019lnr,Criado:2019tzk,
  	King:2019vhv,Gui-JunDing:2019wap,deMedeirosVarzielas:2020kji,Zhang:2019ngf,Nomura:2019xsb,Kobayashi:2019gtp,Lu:2019vgm,Wang:2019xbo,King:2020qaj,Abbas:2020qzc,Okada:2020oxh,Okada:2020dmb,Ding:2020yen,Nomura:2020opk,Nomura:2020cog,Okada:2020rjb,Okada:2020ukr,Nagao:2020azf,Nagao:2020snm,Yao:2020zml,Wang:2020lxk,Abbas:2020vuy,
  	Okada:2020brs,Yao:2020qyy,Feruglio:2021dte,King:2021fhl,Chen:2021zty,Novichkov:2021evw,Du:2020ylx,Kobayashi:2021jqu,Ding:2021zbg,Kuranaga:2021ujd,Li:2021buv,Tanimoto:2021ehw,Okada:2021aoi,Kobayashi:2021ajl,Dasgupta:2021ggp,Nomura:2021ewm,Nagao:2021rio,Nomura:2021yjb,Nomura:2021aep,Okada:2021qdf,Ding:2021eva,Qu:2021jdy,Zhang:2021olk,Wang:2021mkw,Wang:2020dbp
  }.
  

Superstring theory is a promising candidate for the unified theory including gravity.
Various string compactifications lead to four-dimensional low energy field {theories} with the specific structure, 
where 4-point couplings $y^{(4)}_{ijk\ell}$ {of matter fields} can be written by a product of 3-point couplings $y^{(3)}_{ijm}$ {of matter fields}, 
\begin{equation}
\label{eq:Ansatz}
y^{(4)}_{ijk\ell}=\sum_m y^{(3)}_{ijm}y^{(3)}_{mk\ell}\,,
\end{equation}
up to an overall factor, where the modes corresponding to $m$ may be light or heavy modes.
Furthermore, $n$-point couplings $y^{(n)}$ can also be written by products of 3-point couplings $y^{(3)}$, 
i.e., $y^{(n)} = (y^{(3)})^{n-2}$.
Thus, the symmetries in 3-point couplings are still symmetries even for higher-dimensional operators, and 
the flavor structures of higher-dimensional operators are controlled by 3-point couplings.
This structure in the string-derived low-energy effective field theory meets the MFV hypothesis \cite{Kobayashi:2021uam}. Note that the string EFTs satisfy the relation (\ref{eq:Ansatz}) at the compactification scale or the string scale, but it holds at the low-energy scale. This is because new operators appearing through the vacuum expectation value (VEV) of scalar fields and integrating out heavy states keep the relation (\ref{eq:Ansatz}).

In addition, these couplings in {the} string-derived effective field theory depend on moduli, which represent 
geometrical characters of string compact spaces such as size and shape.
When we ignore {the} dynamic of moduli fields, these moduli-dependent couplings behave as 
spurions.
Then, the geometrical symmetry, under which moduli transform non-trivially, would be important 
from the viewpoint of the MFV, although Yukawa spurions transform 
$({\bf 3},{\bf \bar 3},1,1,1)$, $({\bf 3},1,{\bf \bar 3},1,1)$, and $(1,1,1,{\bf 3},{\bf \bar 3})$ in 
the $U(3)^5 = U(3)_Q \otimes U(3)_U \otimes U(3)_D \otimes U(3)_L \otimes U(3)_E$  flavor symmetric MFV scenario,
where $\{ Q,U,D,L,E \}$ denote the five independent types of the SM fermions.
The $U(2)^5$ flavor symmetric scenario can also be realized in string models due to the fact that matter Yukawa couplings have rank 1 at the leading order\cite{Ibanez:2012zz}.
The modular symmetry is the geometrical symmetry, which changes the basis of cycles of 
the torus $T^2$ as well as the orbifold $T^2/\mathbb{Z}_2$.\footnote{{The possible discrete modular symmetries on higher-dimensional toroidal orbifolds were classified in the context of Type IIB string theory\cite{Kobayashi:2020hoc}.}}.
Moreover, zero-modes transform non-trivially under the modular symmetry 
(see, e.g., for heterotic string theory on orbifolds Ref.~\cite{Ferrara:1989qb} and for magnetized brane models  
Ref.~\cite{Kobayashi:2018rad}).
That is, the modular symmetry corresponds to the flavor symmetry in the low-energy effective field theory.
Yukawa couplings as well as other couplings transform non-trivially under the modular symmetry, 
because these couplings depend on the modulus.
Calabi-Yau {threefolds} have many moduli, and their geometrical symmetries are 
symplectic modular symmetries \cite{Strominger:1990pd,Candelas:1990pi}.
Their phenomenological implications were studied in Refs. \cite{Ishiguro:2020nuf,Ishiguro:2021ccl}. 
Hence, these observations strongly support that flavor structures of higher-dimensional operators as well as Yukawa couplings in the string EFTs are determined by the modular flavor symmetry.

  
A drawback of the MFV hypothesis is that it does not allow us to define  a clear power-counting in the SMEFT. This is because one of the breaking term, namely Yukawa coupling $Y_t$, is large. It is therefore not obvious why one should not consider more powers of $Y_t$ in the counting of independent operators. 
On the other hand, it defines a clear power counting in the modular symmetry due to the modular weights.   

In this paper, we study the SMEFT with the $\Gamma_N$ modular flavor symmetry. 
The modular flavor symmetry is regarded as a remnant of the geometrical symmetry of 
the extra-dimensional space. 
Constraints on higher-dimensional operators only by $\Gamma_N$ would be weak, in particular for $\Gamma_N$ singlets 
and many parameters would be allowed.
Hence, we employ the relation (\ref{eq:Ansatz}) as Ansatz in the SMEFT.
Through this Ansatz, higher-dimensional operators are related with 3-point couplings, 
although the $m$ mode in Eq. (\ref{eq:Ansatz})  may be known or unknown.

We  take the level 3 finite modular groups, 
  $\Gamma_3$ for the flavor symmetry since 
  the property of $A_4$ flavor symmetry has been well known
  \cite{Ma:2001dn,Babu:2002dz,Altarelli:2005yp,Altarelli:2005yx,
  	Shimizu:2011xg,Petcov:2018snn,Kang:2018txu}.
Based on the tensor decomposition of $A_4$ modular group, we discuss the bilinear and 4-fermion operators with flavor changing (FC) process
at nearby fixed points.
As an application, we discuss the lepton flavor violation (LFV).

  
The paper is organized as follows.
In section \ref{sec:mod}, we review on the  $A_4$ modular flavor symmetry for flavors.
In section \ref{sec:SMEFT}, we discuss the   4-fermion operators of the SMEFT in
 $A_4$ modular  symmetry.
In sections \ref{sec:RL} and \ref{sec:LL}, we study the flavor structure of
the bilinear fermion operators with chirality flip and chirality conserve,
respectively.
In section \ref{sec:qqqq}, we discuss the  4-quark operator with  $\Delta F=2$.
Section \ref{sec:summary} is devoted to the summary.
In Appendix \ref{App:SMEFT}, we summarizes the SMEFT operators.
In Appendix \ref{massmatrixmodel}, we give a brief review on an explicit $A_4$ modular flavor model. 
The  $S$ and $ST$ transformations of its mass matrix are given in Appendix \ref{appe-base}.
In addition, the  mass matrix at nearby $\tau = i$ is given in 
Appendix \ref{massmatrix-i}.
In Appendix \ref{decomposition}, we present tensor products of $A_4$ including modular forms.
Appendix \ref{U2} summarizes briefly the  $U(2)$ flavor symmetry of the quark sector.

\section{$A_4$ modular symmetry and flavor of quarks and leptons}
\label{sec:mod}

In this section, we briefly review the models with $A_4$ modular symmetry.

  \subsection{Modular flavor symmetry}
 
The modular group $\bar\Gamma$ is the group of linear fractional transformations
$\gamma$ acting on the modulus  $\tau$, 
belonging to the upper-half complex plane as:
\begin{equation}\label{eq:tau-SL2Z}
\tau \longrightarrow \gamma\tau= \frac{a\tau + b}{c \tau + d}\ ,~~
{\rm where}~~ a,b,c,d \in \mathbb{Z}~~ {\rm and }~~ ad-bc=1, 
~~ {\rm Im} [\tau]>0 ~ ,
\end{equation}
which is isomorphic to  $PSL(2,\mathbb{Z})=SL(2,\mathbb{Z})/\{\rm I,-I\}$ transformation.
This modular transformation is generated by $S$ and $T$, 
\begin{eqnarray}
S:\tau \longrightarrow -\frac{1}{\tau}\ , \qquad\qquad
T:\tau \longrightarrow \tau + 1\ ,
\label{symmetry}
\end{eqnarray}
which satisfy the following algebraic relations, 
\begin{equation}
S^2 ={\rm I}\ , \qquad (ST)^3 ={\rm I}\ .
\end{equation}

We introduce the series of groups $\Gamma(N)$, called principal congruence subgroups, where  $N$ is the level $1,2,3,\dots$.
These groups are defined by
\begin{align}
\begin{aligned}
\Gamma(N)= \left \{ 
\begin{pmatrix}
a & b  \\
c & d  
\end{pmatrix} \in SL(2,\mathbb{Z})~ ,
~~
\begin{pmatrix}
a & b  \\
c & d  
\end{pmatrix} =
\begin{pmatrix}
1 & 0  \\
0 & 1  
\end{pmatrix} ~~({\rm mod} N) \right \}
\end{aligned} .
\end{align}
For $N=2$, we define $\bar\Gamma(2)\equiv \Gamma(2)/\{\rm I,-I\}$.
Since the element $\rm -I$ does not belong to $\Gamma(N)$
for $N>2$, we have $\bar\Gamma(N)= \Gamma(N)$.
The quotient groups defined as
$\Gamma_N\equiv \bar \Gamma/\bar \Gamma(N)$
are  finite modular groups.
In these finite groups $\Gamma_N$, $T^N={\rm I}$  is imposed.
The  groups $\Gamma_N$ with $N=2,3,4,5$ are isomorphic to
$S_3$, $A_4$, $S_4$ and $A_5$, respectively \cite{deAdelhartToorop:2011re}.

Modular forms $f_i(\tau)$ of weight $k$ are the holomorphic functions of $\tau$ and transform as
\begin{equation}
f_i(\tau) \longrightarrow (c\tau +d)^k \rho(\gamma)_{ij}f_j( \tau)\, ,
\quad \gamma\in \bar \Gamma\, ,
\label{modularforms}
\end{equation}
under the modular symmetry, where
$\rho(\gamma)_{ij}$ is a unitary matrix under $\Gamma_N$.

Under the modular transformation of Eq.\,(\ref{eq:tau-SL2Z}), chiral superfields $\psi_i$ ($i$ denotes flavors) with weight $-k$
transform as \cite{Ferrara:1989bc}
\begin{equation}
\psi_i\longrightarrow (c\tau +d)^{-k}\rho(\gamma)_{ij}\psi_j\, .
\label{chiralfields}
\end{equation}

We study global SUSY models.
The superpotential which is built from matter fields and modular forms
is assumed to be modular invariant, i.e., to have 
a vanishing modular weight. For given modular forms, 
this can be achieved by assigning appropriate
weights to the matter superfields.

The kinetic terms  are  derived from a K\"ahler potential.
The K\"ahler potential of chiral matter fields $\psi_i$ with the modular weight $-k$ is given simply  by 
\begin{equation}\label{eq:Kahler}
\frac{1}{[i(\bar\tau - \tau)]^{k}} \sum_i|\psi_i|^2,
\end{equation}
where the superfield and its scalar component are denoted by the same letter, and $\bar\tau =\tau^*$ after taking VEV of $\tau$.
The canonical form of the kinetic terms  is obtained by 
changing the normalization of parameters \cite{Kobayashi:2018scp}.
The general K\"ahler potential consistent with the modular symmetry possibly contains additional terms \cite{Chen:2019ewa}. However, we consider only the simplest form of
the K\"ahler potential.

For $\Gamma_3\simeq \rm A_4$, the dimension of the linear space 
${\cal M}_k(\Gamma{(3)})$ 
of modular forms of weight $k$ is $k+1$ \cite{Gunning:1962,Schoeneberg:1974,Koblitz:1984}, i.e., there are three linearly 
independent modular forms of the lowest non-trivial weight $2$,
which form a triplet of the $A_4$ group.
These modular forms have been explicitly given \cite{Feruglio:2017spp} in the symmetric base of the $A_4$ generators $S$ and $T$ for the triplet representation
as shown in the next subsection.  
 

\subsection{Modular forms}
The holomorphic and anti-holomorphic modular forms with weight 2 
compose the $A_4$ triplet as: 
\begin{align}
&\begin{aligned}
{ Y^{\rm (2)}_{\bf 3}}(\tau)=
\begin{pmatrix}
Y_1(\tau)  \\
Y_2(\tau) \\
Y_3(\tau)
\end{pmatrix}\,,\qquad 
\overline {Y^{\rm (2)}_{\bf 3}(\tau)}\equiv Y^{\rm (2)*}_{\bf 3}(\tau)=
\begin{pmatrix}
Y_1^*(\tau)  \\
Y_3^*(\tau) \\
Y_2^*(\tau)
\end{pmatrix}\,.
\end{aligned}
\end{align}
In the representation of 
the generators $S$ and $T$  for $A_4$ triplet:
\begin{align}
\begin{aligned}
S=\frac{1}{3}
\begin{pmatrix}
-1 & 2 & 2 \\
2 &-1 & 2 \\
2 & 2 &-1
\end{pmatrix},
\end{aligned}
\qquad\quad 
\begin{aligned}
T=
\begin{pmatrix}
1 & 0& 0 \\
0 &\omega& 0 \\
0 & 0 & \omega^2
\end{pmatrix}, 
\end{aligned}
\label{ST}
\end{align}
where $\omega=e^{i\frac{2}{3}\pi}$,
modular forms  are given explicitly
in terms of  the Dedekind eta function 
$\eta(\tau)$ and its derivative \cite{Feruglio:2017spp}:
\begin{eqnarray} 
\label{eq:Y-A4}
Y_1(\tau) &=& \frac{i}{2\pi}\left( \frac{\eta'(\tau/3)}{\eta(\tau/3)}  +\frac{\eta'((\tau +1)/3)}{\eta((\tau+1)/3)}  
+\frac{\eta'((\tau +2)/3)}{\eta((\tau+2)/3)} - \frac{27\eta'(3\tau)}{\eta(3\tau)}  \right), \nonumber \\
Y_2(\tau) &=& \frac{-i}{\pi}\left( \frac{\eta'(\tau/3)}{\eta(\tau/3)}  +\omega^2\frac{\eta'((\tau +1)/3)}{\eta((\tau+1)/3)}  
+\omega \frac{\eta'((\tau +2)/3)}{\eta((\tau+2)/3)}  \right) , \label{Yi} \\ 
Y_3(\tau) &=& \frac{-i}{\pi}\left( \frac{\eta'(\tau/3)}{\eta(\tau/3)}  +\omega\frac{\eta'((\tau +1)/3)}{\eta((\tau+1)/3)}  
+\omega^2 \frac{\eta'((\tau +2)/3)}{\eta((\tau+2)/3)}  \right)\,.
\nonumber
\end{eqnarray}
%
Those are also  expressed in the $q$ expansions, $q=\exp (2i\pi\tau)$:
\begin{align}
\begin{pmatrix}Y_1(\tau)\\Y_2(\tau)\\Y_3(\tau)\end{pmatrix}=
\begin{pmatrix}
1+12q+36q^2+12q^3+\dots \\
-6q^{1/3}(1+7q+8q^2+\dots) \\
-18q^{2/3}(1+2q+5q^2+\dots)
\end{pmatrix}\,.
\label{Y(2)}
\end{align}
\subsection{Representation of down-type quarks and charged leptons}

Assign the left-handed down-type quarks to $A_4$ triplets $\bf  3$ 
and the three right-handed ones to $A_4$ three different singlets.
Then, those are expressed as follows:
\begin{align}
\begin{aligned}  D_L=
\begin{pmatrix}
d_L  \\
s_L \\
b_L
\end{pmatrix}\, ,
\quad 
\bar D_L=
\begin{pmatrix}
\bar d_L  \\
\bar b_L \\
\bar s_L
\end{pmatrix}\, , \quad 
(d^c_R,\,s^c_R,\,b^c_R)=(1,\, 1'',\, 1')\,, \quad
( d_R,\, s_R,\, b_R)=(1,\, 1',\, 1'')\,.
\end{aligned}
\end{align}
It is noticed that quarks of second and third families 
 are exchanged each other in $\bar D_L$.
 The weight of $D_L$ and  $\bar D_L$,  $k$ are assigned to  $2$ and 
  $-2$, respectively.
  On the other hand, $k=0$ for singlets  $d^c_R$, $d_R$, etc..

The  charged leptons are like down-type quarks as:
\begin{align}
\begin{aligned}  E_L=
\begin{pmatrix}
e_L  \\
\mu_L \\
\tau_L
\end{pmatrix}\, ,
\quad 
\bar E_L=
\begin{pmatrix}
\bar e_L  \\
\bar \tau_L \\
\bar \mu_L
\end{pmatrix}\, , \quad 
(e^c_R,\,\mu^c_R,\,\tau^c_R)=(1,\, 1'',\, 1')\,, \quad
( e_R,\, \mu_R,\,\tau_R)=(1,\, 1',\, 1'')\,.
\end{aligned}
\end{align}
The weight of $E_L$ and  $\bar E_L$,  $k$, are also  $2$ and 
and $-2$, respectively.
On the other hand, $k=0$ for  $e^c_R$, $e_R$, etc..

Most of modular flavor models, which have been studied, are 
supersymmetric models.
In the following sections, we study models below the  supersymmetry breaking scale.
We assume that the light modes are exactly the same as the SM with two doublet Higgs models.

\section{SMEFT  4-fermion operators in $A_4$ modular symmetry}
\label{sec:SMEFT}

We write down 4-fermion operators as well as dipole operators 
in terms of modular forms $Y(\tau)$.
We also follow the Ansatz (\ref{eq:Ansatz}) that 
those higher-dimensional operators are related with 3-point couplings, e.g., Yukawa couplings 
with Higgs fields. 
Here, the higher-dimensional operators are supposed to be generated by integrating out heavy superpartners, massive gauge bosons and stringy modes. 
We have already many modular flavor symmetric models, which lead to realistic quark and lepton mass 
matrices separately.
However, when we use the common value of the modulus $\tau$ for both quark and lepton sectors, 
the models are severely constrained and very difficult to realize all the experimental values of 
quark and lepton masses and their mixing angles at the same time.
In order to cover many modular flavor models, we assume that the $A_4$ modular flavor symmetry in the lepton sector 
is independent of the $A_4$ symmetry in the quark sector, i.e., $A^E_{4}\otimes A^Q_{4}$ symmetry.
They have two independent moduli, $\tau_q$ and $\tau_e$ for the quark sector and the lepton sector, 
respectively.
Such a setup can be realized through the compactification, that the compact space includes $T^2 \times T^2$, and 
the flavor structure in the quark sector originates from one $T^2$, while the lepton flavor structure originates 
from the other $T^2$.
Indeed, a similar setup was studied e.g., in Ref. \cite{Kobayashi:2018wkl}.
Using this setup and Ansatz, we study their implications on flavor changing processes.


As  examples, 
consider the semileptonic flavor changing neutral processes,
\begin{align}
	b\rightarrow s \, \bar\mu \mu\ (s\,\bar e e) \,,
	\qquad b\rightarrow d \, \bar\mu \mu\ (d\,\bar e e) \,,
	\qquad s\rightarrow d \, \bar\mu \mu\ (d\,\bar e e) \,,
\end{align}
which are caused by the flavor changing  $\Delta F=1$ operator.
Impose the modular  $A_4$ symmetry on quarks and leptons, respectively, that is  $A^E_{4}\otimes A^Q_{4}$.
The triplet modular forms with weight $2$ are denoted as 
$Y(\tau_q)$ and $Y(\tau_e)$, which are different for quarks and charged leptons because $\tau_q$ and $\tau_e$ are  different.
In order to discuss relevant operators, 
 we take a $A_4$ modular model, which leads to the successful 
  fermion mass matrices.
Suppose that 
three left-handed quark and lepton doublets are of a triplet of the $A_4$ group.
The three right-handed quarks and charged leptons are different singlets of $A_4$.
On the other hand, the Higgs doublets are supposed to be singlets of $A_4$.
The generic assignments of representations and modular weights to the fields 
are presented in Table \ref{quark-model0},
where right-handed up-type quarks are omitted 
since those are not necessary in following discussions.
\begin{table}[h]
	\centering
	\begin{tabular}{|c||c|c|c|c|c|cc|} \hline
		&$Q_L$&$(d^c_R,\,s^c_R,\,b^c_R)$& $L_L$ & $(e^c_R,\,\mu^c_R,\,\tau^c_R)$&$H_d$&
		$Y(\tau_q)$, & $Y(\tau_e)$ \\  \hline\hline 
		\rule[14pt]{0pt}{0pt}
		$SU(2)$&$\bf 2$&$\bf 1$& $\bf 2$& $\bf 1$&$\bf 2$
		&\multicolumn{2}{c|}{$\bf 1$} \\
		$A_4$&$\bf 3$& \bf (1,\ 1$''$,\ 1$'$)&$\bf 3$&\bf (1,\ 1$''$,\ 1$'$)&$\bf 1$
		& \multicolumn{2}{c|}{$\bf 3$} \\
		$k$ & $2$ &$(0,\ 0,\ 0)$& $2$ & $(0,\, 0,\, 0)$&0& \multicolumn{2}{c|}{$2$} \\
		\hline
	\end{tabular}
	\caption{The assignment of $A_4$ representations and weights
		$k$ for down-type quarks, charged leptons, down-type Higgs doublet and the modular forms.
	}
	\label{quark-model0}
\end{table}

	We discuss the semileptonic 4-fermion operators, which are categorized as
	\begin{align}
[\,\bar E_L \Gamma E_L\,][\,\bar D_L  \Gamma  D_L\,] \notag 
	&: Q_{\ell q}^{(1)},Q_{\ell q}^{(3)}\,, \\
	 [\,\bar E_R \Gamma  E_R\,] [\,\bar D_R  \Gamma  D_R\,] \notag 
	&: Q_{ed}\,,\\
  [\,\bar E_L \Gamma  E_L][\,\bar D_R  \Gamma  D_R\,]
	&: Q_{\ell d}\,, \\ 
  [\,\bar E_R \Gamma  E_R\,][\,\bar D_L  \Gamma  D_L\,] \notag 
	&: Q_{qe}\,, \\
 [\,\bar E_L   \Gamma E_R\,] [\,\bar D_R  \Gamma   D_L\,] \notag 
	& : Q_{\ell edq} \,,
	\end{align}
	where $L$ and $R$ denote the left-handed and the right-handed fields,
	and $\Gamma$ denotes a generic combination of Dirac matrices, color and 
	$SU(2)_L$ generators, which play
	no role as far as the flavor structure is concerned.
	Corresponding SMEFT operators $Q$ of which explicit expression are shown in Appendix \ref{App:SMEFT}, 
	are also listed. 

First of all, let us construct the modular $A_4$-invariant 4-fermion operators $[\,\bar E_L  \Gamma  E_R\,] [\,\bar D_R   \Gamma  D_L\,] $ 
  by using the  holomorphic and anti-holomorphic modular forms:
\begin{align}
  [\,\bar E_L   \Gamma E_R\,] [\,\bar D_R  \Gamma   D_L\,]
\Rightarrow   [\,\bar E_L Y^*(\tau_e)\Gamma E_R\,]_1[\,\bar d_R \Gamma Y(\tau_q) d_L\,]_1
  \,,
\end{align}
where
the subscript $1$ denotes the $A_4$ trivial singlet.
After the decomposition of $A_4$ tensor product, 
we obtain coefficients for each operator, which are written explicitly
in the following sections.
From the viewpoint of the Ansatz, $y^{(4)}=y^{(3)}y^{(3)}$,  
these $[\,\bar E_L  \Gamma  E_R\,] [\,\bar D_R   \Gamma  D_L\,]$ operators would be constructed in terms of 
Yukawa couplings with the Higgs fields.
Thus, we assume that these operators include the same coefficients as 
the Yukawa couplings, i.e. mass matrices, as written explicitly later.

Next, the $A_4$ modular-invariant semileptonic 4-fermion  operators 
$[\,\bar E_R \Gamma  E_R][\,\bar D_R  \Gamma  D_R\,]$ 
can be written by 
\begin{align}
 [\,\bar E_R \Gamma  E_R\,] [\,\bar D_R  \Gamma   D_R\,] 
\Rightarrow\left [\,\sum_{i=1}^3 r_{ei}\ \bar e_{iR}  \Gamma   e_{iR}\,\right ]
\left [\, \sum_{i=1}^3 r_{qi}\ \bar d_{iR}  \Gamma   d_{iR} \,\right ] 
\,,
\label{RRRR} 
\end{align}
where $r_{ei}$ and  $r_{qi}$ are arbitrary real constants.
From the viewpoint of the Ansatz {(\ref{eq:Ansatz})}, $y^{(4)}=y^{(3)}y^{(3)}$,  
these operators would be 
constructed in terms of gauge couplings $g$ as $y^{(3)} \sim g$.
Therefore, one expects $r_{e(q)1}=r_{e(q)2}=r_{e(q)3}$,
	which  do not lead to the FC processes. 
Similarly, for  the other operators, 
one of possible modular $A_4$-invariant operators in our Ansatz  (\ref{eq:Ansatz}) would be the type like 4-fermion operators 
mediated by gauge bosons.
\footnote{Note that our kinetic terms (\ref{eq:Kahler}) for left-handed fermions are not canonical. 
They couple with gauge bosons as $\frac{1}{(\tau_e-\bar \tau_e)^2}A^\mu [\,\bar E_L  \Gamma  E_L\,]_1$ and $\frac{1} {(\tau_q-\bar \tau_q)^2}A^\mu  [\,\bar D_L   \Gamma  D_L\,]_1$.}
However, they do not lead to the FC processes 
similar to the $ [\,\bar E_R \Gamma  E_R\,] [\,\bar D_R  \Gamma   D_R\,] $ operator.
Operators including holomorphic and anti-holomorphic modular forms would lead to 
the FC processes.
Hence, another possibility for  
 $A_4$ modular-invariant semileptonic 4-fermion operators are constructed  by using the  holomorphic and anti-holomorphic modular forms:
\begin{align}
&[\,\bar E_L  \Gamma  E_L\,][\,\bar D_L   \Gamma   D_L\,] \Rightarrow 
[\,\bar E_L  Y^*  (\bar{\tau_e}) \Gamma Y(\tau_e) E_L\,]_1[\,\bar D_L     Y^*(\bar{\tau_q}) \Gamma Y(\tau_q) D_L\,]_1 
 \,, \\
&  [\,\bar E_L \Gamma   E_L][\,\bar D_R  \Gamma   D_R\,]
\Rightarrow [\,\bar E_L    Y^*(\bar{\tau_e}) \Gamma Y(\tau_e) E_L\,]_1 
\left [\, \sum_{i=1}^3 r_{qi}\, \bar d_{iR} \Gamma   d_{iR}\right ]\,,
 \label{RRLL}\\
&
 [\,\bar E_R  \Gamma  E_R\,][\,\bar D_L  \Gamma   D_L\,] 
\Rightarrow \left [\,\sum_{e_i=e,\,\mu,\,\tau} r_{e_i} \bar e_{iR}  \Gamma  e_{iR}\,\right ]\left [\,\bar D_L   Y^*(\bar{\tau_q}) \Gamma Y(\tau_q) D_L\,\right ]_1
 \,.
\end{align}
After decomposition of $A_4$ tensor products, we will
give coefficients explicitly in the following sections.
These  operators could be consistent with the Ansatz, 
but the mode $m$ in Eq.~(\ref{eq:Ansatz}) is unknown.
Then, many free parameters, which are not related with couplings 
in the renormalizable SM Lagrangian, appear in these operators.
At any rate, these operators are $A_4$ modular invariant.
The above $[\,\bar E_L \Gamma   E_L][\,\bar D_R  \Gamma   D_R\,]$ and $[\,\bar E_R \Gamma   E_R][\,\bar D_L  \Gamma   D_L\,]$ operators do not lead to the FC processes in the quark sector and lepton sector, respectively.
In the following sections, we concentrate on these $[\,\bar E_L \Gamma   E_R][\,\bar D_R  \Gamma   D_L\,]$,
$[\,\bar E_R \Gamma   E_L][\,\bar D_L  \Gamma   D_R\,]$ and 
$[\,\bar E_L \Gamma   E_L][\,\bar D_L  \Gamma   D_L\,]$ operators leading to the FC processes as well as dipole operators.
Moreover, 4-quark operators are also constructed in a similar way.


\section{Bilinear fermion operators $[\,\bar D_R  \Gamma  D_L\,]$
	and  $ [\,\bar D_L \Gamma  D_R\,]$}	
\label{sec:RL}	

The 4-fermion scalar operator
$[\,\bar E_L    E_R\,][\,\bar D_R    D_L\,] $
does not appear at the tree level in the standard model (SM).
Indeed,  it is not allowed by the exact $U(3)$ flavor symmetry.
On the other hand, it appears in  three-point couplings of two fermions and  modular forms in the modular flavor symmetry. Moreover,
due to the modular weights, they define a clear power counting of modular forms.

Since the 4-fermion operators are given by the products of bilinear fermion operators,  interesting features from  Ansatz (\ref{eq:Ansatz})
appear in the bilinear fermion operators.
For example, the 3-point couplings are realized in terms of the modular forms typically  in $[\,\bar D_R  \Gamma  D_L\,] $ operators. 

In  this section, we discuss the bilinear operators
of quarks, 
 $[\,\bar D_R  \Gamma  D_L\,]$ and  $ [\,\bar D_L \Gamma  D_R\,]$,
 and corresponding ones of charged leptons.
 Those correspond the SMEFT operators $Q$ of which explicit expression are shown in Appendix \ref{App:SMEFT} as follows: 
	\begin{align}
 [\,\bar D_L  \Gamma D_R\,] \notag 
	&: Q_{dH},Q_{G},Q_{dW}, Q_{dB}\,,\\
	 [\,\bar D_R \Gamma D_L\,] \notag 
	&: Q_{dH}^\dag,Q_{G}^\dag,Q_{dW}^\dag, Q_{dB}^\dag\,,\\
 [\,\bar E_L  \Gamma E_R\,] \notag 
	&: Q_{eH},Q_{eW}, Q_{eB}\,,\\
 [\,\bar E_R  \Gamma E_L\,] \notag 
	&: Q_{eH}^\dag,Q_{eW}^\dag, Q_{eB}^\dag\,.
	\end{align}

 \subsection{$[\,\bar D_R \Gamma  D_L\,]$ and $[\,\bar D_L \Gamma  D_R\,]$ bilinears in the flavor space}
\label{bilinear-flavor}	
 
At first,
let us begin with discussing the holomorphic operator
$[\,\bar D_R \Gamma  D_L\,]$ and
anti- holomorphic  operator $[\,\bar D_L \Gamma  D_R\,]$
in the flavor space.
The magnitudes of  $LR$ couplings are proportional to modular forms.
Taking account of $\bar D_R=(d^c,s^c,b^c)$, we can decompose  the operator
in the base of Eq.\,\eqref{ST} for $S$ and $T$ as:
	\begin{align}
	 [\,\bar D_R\Gamma  D_L\,]
	&\Rightarrow  [\,\bar D_R \Gamma Y({\tau_q}) D_L\,]_{\bf 1}
	= \nonumber\\
	& [\,{\alpha_{d}\,  \bar d_R }\Gamma
	(Y_{1}({\tau_q}) d_L+Y_{2}({\tau_q}) b_L
	+Y_{3}({\tau_q})s_L)_1+{\beta_{d}\, \bar s_R}\Gamma
	(Y_{3}({\tau_q}) b_L+Y_{1}({\tau_q}) s_L +Y_{2}({\tau_q}) d_L)_{\bf 1'}
	\nonumber \\
	&+ {\gamma_{d}\,  \bar b_R} \Gamma  (Y_{2}({\tau_q}) s_L +Y_{1}({\tau_q}) b_L +Y_{3}({\tau_q}) d_L)_{\bf 1''}]
	\,,\nonumber\\
 [\,\bar D_L  \Gamma D_R\,]
	&\Rightarrow [\,\bar D_L   Y^*({\tau_q})\Gamma D_R\,]_{\bf 1}
	= \nonumber\\
	&[\,{ \alpha_{d}^*\,   d_R } \Gamma (Y_{1}^*({\tau_q})\bar d_L+Y_{2}^*({\tau_q})\bar b_L
	+Y_{3}^*({\tau_q}) \bar s_L)_{\bf 1}+ 
	{ \beta_{d}^*\, s_R } \Gamma
	(Y_{3}^*({\tau_q}) \bar b_L+Y_{1}^*({\tau_q}) \bar{s}_L +Y_{2}^*({\tau_q}) \bar d_L)_{\bf 1''}   \nonumber\\
	&+{ \gamma_{d}^*\,  b_R} \Gamma 
	(Y_{2}^*({\tau_q}) \bar{s}_L+Y_{1}^*({\tau_q}) \bar{b}_L +Y_{3}^*({\tau_q})\bar d_L)_{\bf 1'}]
	\,,
	\label{DLR}
	\end{align}	
	where 
	the subscript $1$, $1'$, $1''$ denote the $A_4$ singlets, respectively.
The parameters	$\alpha_{d}$,
	$\beta_{d}$ and	$\gamma_{d}$ are constants.   
These expressions are written in the matrix representation as 
	\begin{align} 
	[\,\bar D_R \Gamma Y({\tau_q}) D_L\,]_{\bf 1} 
	&= ( \bar d_R,  \bar s_R,  \bar b_R)\Gamma
	\begin{pmatrix}
	 \alpha_d & 0 & 0 \\
	0 & \beta_d & 0\\
	0 & 0 & \gamma_d
	\end{pmatrix}
	\begin{pmatrix}
	Y_1(\tau_q) & Y_3(\tau_q)& Y_2(\tau_q)\\
	Y_2(\tau_q) & Y_1(\tau_q) &  Y_3(\tau_q) \\
	Y_3(\tau_q) &  Y_2(\tau_q)&  Y_1(\tau_q)
	\end{pmatrix}
	\begin{pmatrix}
	d_L\\
	s_L\\
	b_L
	\end{pmatrix}
	\,, \\ %
	[\,\bar D_L   Y^*({\tau_q})\Gamma D_R\,]_{\bf 1} 
		&= (\bar d_L, \bar s_L, \bar b_L)\Gamma
	\begin{pmatrix}
	Y^*_1(\tau_q) & Y^*_2(\tau_q)& Y^*_3(\tau_q)\\
	Y^*_3(\tau_q) & Y^*_1(\tau_q) &  Y^*_2(\tau_q) \\
	Y^*_2(\tau_q) &  Y^*_3(\tau_q)&  Y^*_1(\tau_q)
	\end{pmatrix}
	\begin{pmatrix}
	\alpha_d^* & 0 & 0 \\
	0 & \beta_d^* & 0\\
	0 & 0 & \gamma_d^*
	\end{pmatrix}
	\begin{pmatrix}
	d_R\\
	s_R\\
     b_R
	\end{pmatrix}
	\,.
	\end{align}  
It is useful to compare them with  the down-type quark mass matrix $M_d$
 in the assignment of Table \ref{quark-model0}.
 The mass matrix is given in terms of weight 2 modular forms as:
 \begin{align} 
 M_d&= v_d
 \begin{pmatrix}
 \alpha_{d(m)} & 0 & 0 \\
 0 & \beta_{d(m)} & 0\\
 0 & 0 &\gamma_{d(m)}
 \end{pmatrix}
 \begin{pmatrix}
 Y_1(\tau_q) & Y_3(\tau_q)& Y_2(\tau_q)\\
 Y_2(\tau_q) & Y_1(\tau_q) &  Y_3(\tau_q) \\
 Y_3(\tau_q) &  Y_2(\tau_q)&  Y_1(\tau_q)
 \end{pmatrix}_{RL}\,, 
 \label{quarkmassmatrix} 
 \end{align} 
 where 
 the VEV of the Higgs field $H_d$ is denoted by $ v_d$.
Parameters $\alpha_{d(m)}$, $\beta_{d(m)}$,  $\gamma_{d(m)}$ can be taken to be  real constants.
Since the bilinear operators appear in four-field operators, 
it is reasonable to assume 
\begin{align}
\label{eq:alpha-relation}
\alpha_d=c\alpha_{d(m)}, \qquad \beta_d=c\beta_{d(m)}, \qquad \gamma_d=c\gamma_{d(m)},
\end{align}
from the viewpoint of the Ansatz Eq.\,(\ref{eq:Ansatz}),  
where the mode $m$ may correspond to $H_d$.
Here, $c$ is a common constant.
Hereafter , we set $c=1$ for simplicity. 
In this case, the matrix structure of bilinear operators $[\,\bar D_R \Gamma  D_L\,]$ appearing four-field operators is 
exactly the same as the mass matrix.
Obviously, the bilinear operator matrix is diagonal in the basis for mass eigenstates.
The FC processes such $b \to s$, $b \to d$, $s \to d$ never happen. 
Hence, we obtain the very clear results in the modular symmetric SMEFT with the Ansatz Eq.\,(\ref{eq:Ansatz}).

If the relation is violated, the situation would change drastically.
When such violations are small such as
\begin{align}
\label{eq:alpha-relation2}
\alpha_d-\alpha_{d(m)} \ll \alpha_d, \qquad \beta_d-\beta_{d(m)} \ll \beta_d, \qquad \gamma_d-\gamma_{d(m)} \ll \gamma_d,
\end{align}
FC processes are still suppressed.

In what follows, we study larger violations such that $\alpha_d$, $\beta_d$, and $\gamma_d$ are 
of ${\cal O}(\alpha_{d(m)})$, ${\cal O}(\beta_{d(m)})$, and ${\cal O}(\gamma_{d(m)})$, respectively, 
but they are different by factors from $\alpha_{d(m)}$, $\beta_{d(m)}$, and $\gamma_{d(m)}$, i.e.
\begin{align}
\label{eq:alpha-relation3}
\alpha_d-\alpha_{d(m)} \sim \alpha_d, \qquad \beta_d-\beta_{d(m)} \sim \beta_d, \qquad \gamma_d-\gamma_{d(m)} \sim \gamma_d.
\end{align}
Unknown modes $m$ in Eq.\,(\ref{eq:Ansatz}) may contribute to such violations.


For the charged lepton operators 
 $[\,\bar E_R \Gamma  E_L\,]$ and $[\,\bar E_L \Gamma  E_R\,]$,
 we obtain the decompositions by
  replacing $\tau_q$, $\alpha_{d}$, $\beta_{d}$ and  $\gamma_{d}$  
  with   $\tau_e$, 
$\alpha_{e}$, $\beta_{e}$ and  $\gamma_{e}$  in Eq.\,\eqref{DLR}.
As in the down-sector quarks, the FC processes such as $\mu \to e$, $\tau \to e$, and $\tau \to \mu $ never occur 
when we assume $\alpha_e=\alpha_{e(m)}$, $\beta_e=\beta_{e(m)}$, and $\gamma_e=\gamma_{e(m)}$, 
where $\alpha_{e(m)}$, $\beta_{e(m)}$, and $\gamma_{e(m)}$ are parameters in the charged lepton mass matrix 
as shown in Appendix \ref{massmatrixmodel}.
This is the clear result in the modular symmetric SMEFT with the Ansatz Eq.\,(\ref{eq:Ansatz}).

On the other hand, violations of the above parameter relation may lead to the FC processes.
In what follows, we study such violations such as 
\begin{align}
\label{eq:alpha-relation3e}
\alpha_e-\alpha_{e(m)} \sim \alpha_e, \qquad \beta_e-\beta_{e(m)} \sim \beta_e, \qquad \gamma_e-\gamma_{e(m)} \sim \gamma_e.
\end{align}

The $A_4$ flavor  coefficients 
are given  in Table \ref{LR-Table} for  relevant  bilinear  operators
of down-type quarks and charged leptons, where 
the overall strength of the NP effect is not included.
\footnote{The overall strength of the NP effect
is omitted in coefficients of other Tables.}
Hereafter, without specifying them, 
we denote $\alpha_{d,e(m)}$, $\beta_{d,e(m)}$, and $\gamma_{d,e(m)}$ by 
 $\alpha_{d,e}$, $\beta_{d,e}$, and $\gamma_{d,e}$, too,  
because they are the same orders.
\begin{table}[h]
\small
		\centering
		\begin{tabular}{|c|c|c|c||c|c|c|c|c|} \hline 
			\rule[14pt]{0pt}{2pt}   		
			$\begin{matrix}
		{\bar R L}\\ {\bar LR}
		\end{matrix} $ 
			&		$\begin{matrix}
			\bar s_R \Gamma  b_L^{}\\\bar s_L \Gamma  b_R^{}
			\end{matrix} $ 
			& 	$\begin{matrix}\bar d_R \Gamma  b_L^{}\\\bar d_L \Gamma  b_R^{}
			\end{matrix} $ 
			&	$\begin{matrix}\bar d_R \Gamma  s_L^{}\\\bar d_L \Gamma  s_R^{}
			\end{matrix} $ 
			&
				$\begin{matrix}
			\bar \mu_R \Gamma  \tau_L^{}\\\bar \mu_L \Gamma  \tau_R^{}
			\end{matrix} $ 
			& 	$\begin{matrix}\bar e_R \Gamma  \tau_L^{}\\\bar e_L \Gamma  \tau_R^{}
			\end{matrix} $ 
			&	$\begin{matrix}\bar e_R \Gamma  \mu_L^{}\\\bar e_L \Gamma \mu_R^{}
			\end{matrix} $ 
				& 	$\begin{matrix}\bar e_R \Gamma  e_L^{}\\\bar e_L \Gamma  e_R^{}
			\end{matrix} $ 
			&	$\begin{matrix}\bar \mu_R \Gamma  \mu_L^{}\\\bar \mu_L \Gamma \mu_R^{}
			\end{matrix} $ 
			\\  \hline
			\rule[14pt]{0pt}{1pt} 			
				$\begin{matrix}
		{\rm Coeff.}
			\end{matrix} $ 
			&	$\begin{matrix}	{\beta_{d}\,  Y_{3}({\tau_q}) }\\
				{\gamma_{d}\,  Y_{2}^*({\tau_q}) }
			\end{matrix} $ 
			& 	$\begin{matrix}	{ \alpha_{d}\,  Y_{2}({\tau_q}) }\\
			{\gamma_{d}\,  Y_{3}^*({\tau_q}) }
			\end{matrix} $  &
			$\begin{matrix}{ \alpha_{d}\,  Y_{3}({\tau_q}) }\\
			{\tilde\beta_{d}\,  Y_{2}^*({\tau_q}) }
			\end{matrix} $ 
				&
					$\begin{matrix}	{\beta_{e}\,  Y_{3}({\tau_e}) }\\
				{\gamma_{e}\,  Y_{2}^*({\tau_e}) }
				\end{matrix} $ 
				& 	$\begin{matrix}	{ \alpha_{e}\,  Y_{2}({\tau_e}) }\\
				{\gamma_{e}\,  Y_{3}^*({\tau_e}) }
				\end{matrix} $  &
				$\begin{matrix}{ \alpha_{e}\,  Y_{3}({\tau_e}) }\\
				{\beta_{e}\,  Y_{2}^*({\tau_e}) }
				\end{matrix} $ 
				& 	$\begin{matrix}	{ \alpha_{e}\,  Y_{1}({\tau_e}) }\\
				{\alpha_{e}\,  Y_{1}^*({\tau_e}) }
				\end{matrix} $  &
				$\begin{matrix}{ \beta_{e}\,  Y_{1}({\tau_e}) }\\
				{\beta_{e}\,  Y_{1}^*({\tau_e}) }
				\end{matrix} $ \\ \hline
		\end{tabular}
		\caption{$A_4 $ flavor coefficients of the  bilinear operators of down-type quarks and charged leptons.}
	\label{LR-Table}
\normalsize
\end{table}



	Therefore, the flavor structure of 
	 the these operators is predicted if the modulus $\tau_{q,e}$ is fixed.
	It is noticed that above operators are given in the flavor base.
 In order to move the mass eigenstate of the left-handed quarks and leptons,
 we must fix the modulus $\tau_{q,e}$.
 The value of $\tau_{q,e}$ depends on  models, for example, 
  Eqs.\,(\ref{quark-mass-matrix}) and  (\ref{ME222}).
  The interesting value of $\tau_{q,e}$ is fixed points 
  of the modulus in the fundamental domain
  of $SL(2,\mathbb{Z})$ since 
   the moduli stabilization is realized in a controlled way at nearby fixed points\cite{Abe:2020vmv,Kobayashi:2020uaj}. Furthermore, the fixed points are 
   statistically favored in the string landscape \cite{Ishiguro:2020tmo}.
   We discuss the phenomenology at nearby fixed points in the next subsection.


\subsection{Diagonal matrix $M_E^\dagger M_E$ and $M_q^\dagger M_q$ 
	at fixed points}

Residual symmetries arise whenever the VEV of the modulus $\tau$ breaks
the modular group $\overline{\Gamma}$ only partially.
Here and in what follows, 
 we denote $\tau =\tau_{q,e}$ unless we specify it.
Fixed points of modulus are the case.
There are only 2 inequivalent finite points in the fundamental domain
of $\overline{\Gamma}$,
namely,  $\tau = i$ and 
$ \tau =\omega=-1/2+ i \sqrt{3}/2$.
The first point is  invariant under the $S$ transformation
$\tau=-1/\tau$.  In the case of $A_4$ symmetry, the subgroup $\mathbb{Z}_2^{S}=\{ I, S \}$ is preserved at $ \tau = i$.
The second point is the left cusp in the fundamental domain of the modular group,
which is invariant under the  $ST$ transformation $\tau=-1/(\tau+1)$.
Indeed, $\mathbb{Z}_3^{ST}=\{  I, ST,(ST)^2 \}$ is  one of  subgroups of 
$A_4$ group. 
The right cusp at  
$ \tau =-\omega^2=1/2+ i \sqrt{3}/2$
is related to $\tau=\omega$ by the $T$ transformation.
There is also infinite point $ \tau = i \infty$,
in which  the subgroup  $\mathbb{Z}^T_3=\{ I,T,T^2 \}$ of $A_4$
is preserved. We summarize at three cases of the transformation:
\begin{align}
S\ {\rm  invariant} \ :\ \tau=i\,,\qquad     ST\ {\rm invariant}\,   : \ \tau=\omega\,,  \quad 
 T\ {\rm invariant}\ :\  \tau=i\infty\,.
\label{fixed-points}
\end{align}

If  a residual symmetry of $S$ and $T$ in $A_4$ 
	is preserved in  mass matrices of leptons and quarks,
	we have commutation relations 
	between the mass matrices and the generator $ G \equiv S, T, ST$ as:
	\begin{align}
	[M_{RL}^\dagger M_{RL},\, { G}]=0 \, , 
	\label{commutator}
	\end{align}
where $M_{RL}$ denotes the  mass matrix of charged leptons and quarks, $M_E$ and  $M_q\,(q=u,d)$.


Then the mass matrices
$M_E^\dagger M_E$ and  $M_q^\dagger M_q$
could be diagonal in  the diagonal basis of G at the fixed points.
Therefore, the hierarchical structures of flavor mixing are easily realized
near those fixed points.

\subsubsection{Mass matrix and operators $\bar D_L  \Gamma   D_R$,
	 $\bar D_R  \Gamma   D_L$ at the fixed point $\tau=i$}

At $\tau=i$,
holomorphic and  anti-holomorphic modular forms of weight $2$ are given  as:
\begin{align}
&\begin{aligned}
{ Y}(\tau_q=i)=Y_1(i)
\begin{pmatrix}
1  \\ 1-\sqrt{3} \\ -2+\sqrt{3}
\end{pmatrix}\,,\qquad 
\ Y^{*}(\tau_q=i)=Y_1(i)
\begin{pmatrix}
1  \\ -2+\sqrt{3} \\ 1-\sqrt{3}
\end{pmatrix}\,,
\end{aligned}\nonumber\\
&\begin{aligned}
{ Y}(\tau_e=i)=Y_1(i)
\begin{pmatrix}
1  \\ 1-\sqrt{3} \\ -2+\sqrt{3}
\end{pmatrix}\,,\qquad 
\ Y(\tau_e=i)=Y_1(i)
\begin{pmatrix}
1  \\ -2+\sqrt{3} \\ 1-\sqrt{3}
\end{pmatrix}\,,
\end{aligned}
\label{modularS}
\end{align}
in the base of Eq.\,\eqref{ST}.
At this fixed point, we transform the left-handed quarks and charged lepton fields  as:
\begin{align}
&D_L \rightarrow  D_L^S\equiv  U_{S}\, D_L\,, \ \ \qquad 
\bar D_L \rightarrow \bar D_L^S \equiv \bar D_L U_{S}^\dagger  \,,\nonumber\\
&E_L \rightarrow  E_L^S\equiv  U_{S}\, E_L\,, \quad\qquad
\bar E_L \rightarrow \bar E_L^S \equiv 
{\bar E}_L\,U_{S}^\dagger \,,
\label{dL-S}
\end{align}
where the unitary matrix $U_{S}$  is
\begin{align}
U_S =
\frac{1}{2\sqrt{3}}
\begin{pmatrix}
2  & 2 & 2\\
\sqrt{3}+1  &  -2 & \sqrt{3}-1\\
\sqrt{3}-1 & -2  & \sqrt{3}+1
\end{pmatrix}\,,
\label{U-S}
\end{align}
(see  Eq.\,\eqref{U-S0} of Appendix \ref{massmatrix-i}) .
On the other hand,
the right-handed quarks and charged lepton fields are unchanged
since they are $A_4$ singlets.

At $\tau_q=i$,
 under the transformation of Eq.\,\eqref{dL-S},
 the down-type quark mass matrix  is given as  (see in Appendix \ref{massmatrix-i})
\begin{align} 
&M_{d}\,=\frac12 v_d
\begin{pmatrix}
0& 3(\sqrt{3}-1) \tilde\alpha_{d(m)} &-(3-\sqrt{3}) \tilde\alpha_{d(m)}\\
0 & -3(\sqrt{3}-1) \tilde\beta_{d(m)} & -(3-\sqrt{3}) \tilde\beta_{d(m)}\\
0 &0& 2(3-\sqrt{3})\tilde\gamma_{d(m)}
\end{pmatrix}_{RL}\,, \nonumber\\
&M_{d}^\dagger M_{d}=\frac12 v_d^2
\begin{pmatrix}
0  & 0 & 0\\
0  &  9(2-\sqrt{3}) (\tilde\alpha_{d(m)}^2+\tilde\beta_{d(m)}^2) &
3(3-2\sqrt{3}) (\tilde\alpha_{d(m)}^2-\tilde\beta_{d(m)}^2)\\
0 &  3(3-2\sqrt{3}) (\tilde\alpha_{d(m)}^2-\tilde\beta_{d(m)}^2) & 
3(2-\sqrt{3}) (\tilde\alpha_{d(m)}^2+\tilde\beta_{d(m)}^2+4\tilde\gamma_{d(m)}^2)
\end{pmatrix}_{LL}\,,
\label{matrix-S}
\end{align}
where $\tilde \alpha_{d(m)}=(6-3\sqrt{3})  Y_1(i)^2\alpha_{d(m)}$,
$\tilde \beta_{d(m)}= (6-3\sqrt{3})  Y_1(i)^2 \beta_{d(m)}$ and 
$\tilde \gamma_{d(m)} =(6-3\sqrt{3})Y_1(i)^2  \gamma_{d(m)}$
and 
 $\tilde \gamma_{d(m)}$ is supposed to be  much larger than  $\tilde \alpha_{d(m)}$
 and  $\tilde \beta_{d(m)}$.  

Since two eigenvalues of $S$ are degenerate
such as $(1,-1,-1)$, there is still a freedom of the  $2$--$3$ family rotation.
Therefore, $M_{d}^\dagger M_{d}$ could be diagonal
 after the small $2$--$3$ family rotation of
  ${\cal O}(\tilde \alpha_{d(m)}^2\tilde/ \gamma_{d(m)}^2,\,\tilde \beta_{d(m)}^2/\tilde \gamma_{d(m)}^2)$.
The charged lepton mass matrix is the same one 
in Eq.\,(\ref{matrix-S}) by replacing
the subscript $d$ with $e$.

 Let us consider the  bilinear operators of the subsection \ref{bilinear-flavor}
in the diagonal base of the generator $S$.
The $A_4$ triplet left-handed fields are transformed as in 
Eq.\,(\ref{dL-S}). 
Putting the modular forms of Eq.\,(\ref{modularS}) into 
the coefficients of Table \ref{LR-Table},
we can predict the flavor structure of the FC bilinear operators in the new base of $S$.
Those coefficients are listed  in Table \ref{Co-S} at $\tau=i$.
The   left-handed fields are not yet  the mass eigenstate,
  but close to it. We should move the left-handed fields to the mass eigenstate
   by the small rotation in the flavor space.


\subsubsection{Mass eigenstate at nearby $\tau=i$}
In order to get the observed fermion masses and CKM elements,
 the modulus $\tau$ is deviated from the fixed points $\tau=i$.
 Indeed, the successful quark mass matrices
 have been obtained  at nearby $\tau=i$ \cite{Okada:2019uoy}.
  By using a small dimensionless parameter $\epsilon$,
   we put the modulus value as  $\tau=i+\epsilon$.
Then, approximate behaviors of the ratios of modular forms
are \cite{Okada:2020ukr}: 
\begin{align}
\begin{aligned}
\frac{Y_2(\tau)}{Y_1(\tau)}\simeq (1+\epsilon_1)\, (1-\sqrt{3}) \, , \quad 
\frac{Y_3(\tau)}{Y_1(\tau)}\simeq (1+\epsilon_2)\, (-2+\sqrt{3}) \, ,
\quad \epsilon_1=\frac{1}{2} \epsilon_2\simeq 2.05\,i\,\epsilon\,.
\end{aligned}
\label{epS12}
\end{align}
These approximate  forms are  agreement with exact numerical values within  $0.1\,\%$
for $|\epsilon|\leq 0.05$.
Since the modulus $\tau$ is different  ones for the quark and lepton sectors each other,
we use the notation $\epsilon_1^q$ for the quark sector
and  $\epsilon_1^\ell$ for the lepton sector hereafter.

The quark mass matrix is diagonalized 
 by the transformation which is  shown in Appendix \ref{massmatrix-i}:	
\begin{align}
D_L \rightarrow  D_L^m\equiv U_{md}^T U_{12}^T(90^\circ) U_{S}\, D_L\,, \quad
\bar D_L \rightarrow \bar D_L^m \equiv 
{\bar D}_L\,U_{S}^\dagger U_{12}(90^\circ) U_{md} \,,
\nonumber\\
\label{massbase-S}
\end{align}	
where
\begin{align}
U_{md}\simeq
\begin{pmatrix}
1 & s_{12}^{d}e^{i\eta_d} & 0\\
-s_{12}^{d}e^{-i\eta_d} & 1 & s_{23}^{d} \\
s_{12}^d s_{23}^d & -s_{23}^{d} & 1
\end{pmatrix}
\simeq 
\begin{pmatrix}
1 & {\cal O}(\epsilon_1^q) & 0\\
{\cal O}(\epsilon_1^q) & 1 &  {\cal O}(\epsilon_1^q)\\
 {\cal O}(\epsilon_1^{q\,2}) &{\cal O}(\epsilon_1^q) & 1
\end{pmatrix}
\label{mixing-Sd}\,.
\end{align}

In these transformations, $U_{12}^T(90^\circ)$  denotes an extra rotation of $90^\circ$ between the first and second families. It is required
to realize the hierarchy of three CKM mixing angles simply.
Owing to   $U_{12}^T(90^\circ)$, the mixing angle $s_{13}^{d}$ 
between the first and third families is negligibly small.
Details are presented  
 in Eq.\eqref{downmassmatrix-m} of Appendix \ref{massmatrix-i}.
In the quark sector, $s_{12}^d$ may be expressed in terms of CKM elements.
Since $V_{CKM}=U_{mu}^\dagger U_{md}$, where $U_{mu}$ is the mixing matrix of the up-type quark mass matrix,
$s_{12}^d$ is approximately  the ratio of CKM elements 
 $|{V_{td}}/V_{ts}|$  if the mixing angle $s_{13}^u$
 is also  negligibly small as well as $s_{13}^d$.
  Then, we have 
\begin{align}
 s_{12}^d e^{i\eta_d}\simeq -\frac{V_{td}^*}{V_{ts}^* }\,.
 \label{s12d}
 \end{align}
 
In the mass eigenstate,
$A_4$ flavor coefficients of quark bilinear operators in Eq.\,\eqref{DLR}
and Table \ref{LR-Table}
are given in terms of mixing angles $s_{12}^d$, $s_{23}^d$
and $\epsilon_1^q$ at $\tau_q=i+\epsilon_1^q$
as well as the case  at  $\tau_q=i$ as seen in Table \ref{Co-S}.
Due to the extra rotation of the left-handed quarks $U_{12}(90^\circ)$ of Eq. \eqref{massbase-S},
the magnitudes of coefficients between  $\tau_q=i$ and  $\tau_q=i+\epsilon$
are exchanged
with respect to $d_L$ and $s_L$ ($\bar d_L$ and $\bar s_L$)
in Table \ref{Co-S}.

\begin{table}[h]
\hskip 1.5 cm
	\begin{tabular}{|c|c|c|c|} \hline 
		\rule[14pt]{0pt}{2pt}   		
	$\tau_q$
	&		$\begin{matrix}\bar s_R \Gamma  b_L^{}\\\bar s_L \Gamma  b_R^{}
	\end{matrix} $ 
	& 	$\begin{matrix}\bar d_R  \Gamma b_L^{}\\\bar d_L \Gamma  b_R^{}
	\end{matrix} $ 
	&	$\begin{matrix}\bar d_R \Gamma  s_L^{}\\\bar d_L \Gamma  s_R^{}
	\end{matrix} $ 
		\\  \hline
		\rule[14pt]{0pt}{1pt} 			
		$i$	
		&	$\begin{matrix}-\frac12  \,(3-\sqrt{3})\tilde \beta_d\\	0
		\end{matrix} $ 
		& 	$\begin{matrix}	-\frac12  \,(3-\sqrt{3})\tilde \alpha_d\\
			\rule[14pt]{0pt}{1pt} 	
		0
		\end{matrix} $  &
		$\begin{matrix}		\frac32  \,(\sqrt{3}-1)\tilde \alpha_d
		\\	0
		\end{matrix} $ 
		\\ \hline
		\rule[14pt]{0pt}{1pt} 			
		$i+\epsilon$	
		&$\begin{matrix}-\frac12  \,(3-\sqrt{3})\tilde \beta_d\\
		  (1-\sqrt{3})(\sqrt{3}s_{23}^d+\epsilon_1^*)\tilde \gamma_d
		\end{matrix} $  
		& $\begin{matrix}- \frac12  \,[3-\sqrt{3}]
		\,\tilde \alpha_d\\
			\rule[14pt]{0pt}{1pt} 	
		(\sqrt{3}-1)s_{12}^d \epsilon_1^*\tilde \gamma_d
		\end{matrix} $ 
		&$\begin{matrix}\frac12  \,(\sqrt{3}-1)(3s_{12}^d-2\epsilon_1^q)\,
		\, \tilde \alpha_d\\
		\frac32 (1-\sqrt{3})\tilde \beta_d
		\end{matrix} $  \\
		\hline 
	\end{tabular}
	\caption{$A_4$ flavor coefficients of the FC quark bilinear operators in $Y_1(i)$ unit
		at $\tau_q=i$ and  $\tau_q=i+\epsilon$.}
	\label{Co-S}
\end{table}



The mixing angles are
 $s_{12}^d={\cal O}(0.1)$, $s_{23}^d={\cal O}(\epsilon_1^q)$ and 
 $s_{23}^d={\cal O}(\epsilon_1^q)$ as seen in Appendix \ref{massmatrix-i}.
 t is remarked that   ratios among the   $b\to s $ and $b\to d$  transitions are
\begin{align} -\frac12\,\frac{\tilde \alpha_d}{\tilde \beta_d}
\frac{[\,\bar d_R \Gamma  b_L^{m}\,]}{[\,\bar s_R \Gamma  b_L^{m}\,]}\,,
\qquad   -\frac{s_{12}^d \epsilon_1^* } {\sqrt{3}s_{23}^d+\epsilon_1^*}
\frac{[\,\bar d_L \Gamma  b_R^{m}\,]}{[\,\bar s_L\Gamma   b_R^{m}\,]}\,,
\qquad 
-\frac{\sqrt{3} }{2}\frac{1} {\sqrt{3}s_{23}^d+\epsilon_1^*}
\frac{\tilde \beta_d}{\tilde \gamma_d}
\frac{[\,\bar s_R \Gamma  b_L^{m}\,]}{[\,\bar s_L\Gamma   b_R^{m}\,]}\,,
\label{LR-ratio}
\end{align}
where  $\tilde \alpha_{d}=(6-3\sqrt{3})  Y_1(i)\alpha_{d}$,
$\tilde \beta_{d}= (6-3\sqrt{3})  Y_1(i) \beta_{d}$ and 
$\tilde \gamma_{d} =(6-3\sqrt{3}) Y_1(i)  \gamma_{d}$.
The superscript $m$ of the left-handed quarks denotes
the mass eigenstate of the transformation in Eq.\,(\ref{massbase-S}). 
The magnitude of this ratio depends on the detail of
	the model, especially, the up-type quark sector. 
For example,
	we obtained the best fit parameters: 
	\begin{align}
	\frac{\tilde \beta_{d(m)}}{\tilde \gamma_{d(m)}}=4.26\times 10^{-3}\,,\qquad
	\qquad 
	\frac{\tilde \alpha_{d(m)}}{\tilde \beta_{d(m)}}=3.40\,,
	\label{alphabetagamma}
	\end{align}
in the model of Appendix \ref{massmatrixmodel} \cite{Okada:2020ukr}.
	Since $ \tilde\alpha_{d}\sim \tilde\alpha_{d(m)}$,
	$ \tilde\beta_{d}\sim \tilde\beta_{d(m)}$ and 
	$ \tilde\gamma_{d}\sim \tilde\gamma_{d(m)}$,
	it is found that 
the $\bar s_L   b_R^{m}$ transition is the dominant among others
	 as seen in Table \ref{Co-S}. 
	The  $\bar d_L   b_R^{m}$ transition is smaller of one order  than
	 the $\bar s_L   b_R^{m}$ transition.
	 Both $\bar d_R   s_L^{m}$ and
	  $\bar d_L   s_R^{m}$ transitions are significantly suppressed
	  compared with $\bar s_L   b_R^{m}$.

 For charged leptons $e$, $\mu$ and $\tau$, 
 the transformation is somewhat different from the down-type quark sector
 in Eq.\,\eqref{massbase-S}.
The charged lepton mass matrix is diagonalized 
by the transformation which is also  shown in Appendix \ref{massmatrix-i}:	
\begin{align}
E_L \rightarrow  E_L^m\equiv U_{me}^T  U_{S}\, E_L\,,
\quad \  \,
\bar E_L \rightarrow \bar E_L^m \equiv 
{\bar E}_L\,U_{S}^\dagger  U_{me} \,,
\label{massbase-Se}
\end{align}	
where

\begin{align}
U_{me}\simeq
\begin{pmatrix}
1 & s_{12}^{e}  &s_{13}^{e}\\
-s_{12}^{e} & 1 & 0\\
-s_{13}^d  & 0 & 1
\end{pmatrix}
\simeq 
\begin{pmatrix}
1 & {\cal O}(0.1) & {\cal O}(|\epsilon_1^e|) \\
{\cal O}(0.1) & 1 & 0 \\
{\cal O}(|\epsilon_1^e|) &0 & 1
\end{pmatrix}
\label{mixing-Se}\,.
\end{align}
Indeed, the numerical fit was succeeded 
as shown  in Appendix \ref{massmatrix-i}\cite{Okada:2020brs}.
In these transformations, an extra rotation of $U_{12}^T(90^\circ)$ 
 is not required in the transformation of  Eq.\,\eqref{massbase-Se}
 because the observed lepton mixing angles are quite large.
The large mixing angles are realized in the neutrino mass matrix
  of Appendix \ref{massmatrixmodel}.

In the mass eigenstate,
the $A_4$  coefficients of charged lepton bilinear operators 
are given in terms of mixing angles $s_{12}^e$, $s_{13}^e$
and $\epsilon_1^e$ at $\tau_e=i+\epsilon$ in Table \ref{Co-Se}.
These coefficients are different from the quark ones in Table \ref{Co-S}.

\begin{table}[h]
	\hskip 1 cm	\begin{tabular}{|c|c|c|c|} \hline 
			\rule[14pt]{0pt}{2pt}   		
			$\tau_e$
			&		$\begin{matrix}\bar \mu_R \Gamma  \tau_L^{}\\
			\bar \mu_L \Gamma  \tau_R^{}
			\end{matrix} $ 
			& 	$\begin{matrix}\bar e_R  \Gamma \tau_L^{}\\
			\rule[14pt]{0pt}{1pt} 	
			\bar e_L \Gamma  \tau_R^{}
			\end{matrix} $ 
			&	$\begin{matrix}\bar e_R \Gamma  \mu_L^{}\\
			\bar e_L \Gamma  \mu_R^{}
			\end{matrix} $ 
			\\  \hline
			\rule[14pt]{0pt}{1pt} 			
			$i$	
			&	$\begin{matrix}-\frac12  \,(3-\sqrt{3})\tilde \beta_e\\	0
		\end{matrix} $ 
		& 	$\begin{matrix}	-\frac12  \,(3-\sqrt{3})\tilde \alpha_e\\
		\rule[14pt]{0pt}{1pt} 	
		0
		\end{matrix} $  &
		$\begin{matrix}		\frac32  \,(\sqrt{3}-1)\tilde \alpha_e
		\\	0
		\end{matrix} $ 
			\\ \hline
			\rule[14pt]{0pt}{1pt} 			
			$i+\epsilon$	
			&	$\begin{matrix}-\frac12  \,(3-\sqrt{3})\tilde \beta_e\\
				(1-\sqrt{3})s_{12}^e\epsilon_1\tilde \gamma_e
		\end{matrix} $ 
		& 	$\begin{matrix}	-\frac12  \,(3-\sqrt{3})\tilde \alpha_e\\
		\rule[14pt]{0pt}{1pt} 	
		(1-\sqrt{3})(\sqrt{3}s_{13}^e+\epsilon_1)\tilde \gamma_e
		\end{matrix} $  &
		$\begin{matrix}		\frac32  \,(\sqrt{3}-1)\tilde \alpha_e
		\\		\frac12  \,(\sqrt{3}-1)(3s_{12}^e+\sqrt{3}s_{13}^e-2\epsilon_1)
		\tilde \beta_e
		\end{matrix} $ \\
			\hline 
	\end{tabular}
	\caption{$A_4$ flavor coefficients of the FC lepton bilinear operators in $Y_1(i)$ unit
		at $\tau_e=i$ and  $\tau_e=i+\epsilon$.}
	\label{Co-Se}
\end{table}

\subsection{Diagonal bases of $ST$ for quarks}
\subsubsection{At $\tau=\omega$ }

At $\tau=\omega $, as presented in Eq. (\ref{fixed-points}),
holomorphic and  anti-holomorphic  modular forms are given  as:
\begin{align}
&\begin{aligned}
{ Y^{\rm (2)}}(\tau_q=\omega)=Y_{1}(\omega)
\begin{pmatrix}
1  \\\omega\\ -\frac{1}{2}\, \omega^2
\end{pmatrix}\,,\qquad 
\ Y^{\rm (2)*}(\tau_q=\omega)=Y_{1}(\omega)
\begin{pmatrix}
1  \\ -\frac{1}{2}\, \omega \\\omega^2
\end{pmatrix}\,.
\end{aligned}
\label{modular-ST}
\end{align}

The left-handed quark fields are transformed as:
\begin{align}
&D_L \rightarrow  D_L^{ST}\equiv U_{STi}\, D_L\,, \qquad 
\bar D_L \rightarrow \bar D_L^{ST} \equiv {\bar D}_L\,U_{STi}^\dagger \,,
\label{dL-ST}
\end{align}
where $U_{STi}$ is presented in Eq. (\ref{STdiagonal}) in Appendix \ref{appe-ST}.
It is noticed there are independent $6$ unitary transformation $U_{STi}$ 
to diagonalize $M_d^\dagger M_d$.
At this stage, we cannot fix $U_{STi}$ among six ones.
  Once the up-type quark mass matrix is given to reproduce the CKM matrix,
 $U_{STi}$ is fixed.

Putting modular forms of Eq.\,\eqref{modular-ST} into
coefficients of  bilinear operators in 
 Table \ref{LR-Table}, we obtain coefficients of FC bilinear operators
 in the diagonal base of $ST$.
  We summarize them for each $U_{STi}$.
  They correspond to the mass matrix  as discussed below Eq.\,\eqref{quarkmassmatrix}.

\begin{table}[h]
\footnotesize
	\centering
	\begin{tabular}{|c|c|c|c|c|} \hline 
		 \rule[14pt]{0pt}{2pt}
			& $M_d$ \qquad\qquad\qquad\quad   $M_d^\dagger M_d$
			 & $\begin{matrix}\bar s_R \Gamma b_L^{ST}\\\bar s_L\Gamma  b_R^{ST}
			 \end{matrix}$ 
			 & $\begin{matrix}\bar d_R \Gamma b_L^{ST}\\\bar d_L \Gamma b_R^{ST}
			 \end{matrix}$ 
			 &
			 $\begin{matrix}\bar d_R \Gamma s_L^{ST}\\\bar d_L\Gamma s_R^{ST}
			\end{matrix}$ 
			 \\  \hline			
	  $U_{ST1}$ &
		{$ \frac32 v_d\begin{pmatrix}
			0& 0& \omega \tilde\alpha_{d(m)}\\
			-\omega^2 \tilde\beta_{d(m)} & 0 & 0\\
			0 & \tilde\gamma_{d(m)} &0\end{pmatrix} $}, \quad
		{	$ \frac94v_d^2 \begin{pmatrix}
		\tilde\beta_{d(m)}^2  & 0 & 0\\
		0  & \tilde\gamma_{d(m)}^2 & 0\\
		0 & 0 & \tilde\alpha_{d(m)}^2\end{pmatrix}$}
	 & $\begin{matrix}  0\\ \frac32\tilde{\gamma_d}\end{matrix}$
	 & $\begin{matrix}\frac32\omega\tilde{\alpha_d}\\0\end{matrix}$
	  & $\begin{matrix}0\\
	  -\frac32\omega\tilde{\beta_d}\end{matrix}$\\
		\hline 
		 \rule[14pt]{0pt}{1pt} 
		 $U_{ST2}$ &
		{$ \frac32 v_d \begin{pmatrix}
		0& 0& \omega \tilde\alpha_{d(m)}\\
		0&-\omega^2 \tilde\beta_{d(m)} & 0\\
		\tilde\gamma_{d(m)}&0&0\end{pmatrix} $}, \quad
		{ 	$ \frac94 v_d^2\begin{pmatrix}
		\tilde\gamma_{d(m)}^2  & 0 & 0\\
		0  & \tilde\beta_{d(m)}^2 & 0\\
		0 & 0 & \tilde\alpha_{d(m)}^2\end{pmatrix}$}
		& $\begin{matrix}  0\\ 0\end{matrix}$
		& 	$\begin{matrix} \frac32\omega\tilde{\alpha_d} \\\frac32\tilde\gamma_d\end{matrix}$ 
		&  $\begin{matrix}  0\\ 0\end{matrix}$\\
		\hline 
		 \rule[14pt]{0pt}{1pt} 
		$U_{ST3}$ &
		{$ \frac32 v_d\begin{pmatrix}
		0& \omega \tilde\alpha_{d(m)}&0\\
		-\omega^2 \tilde\beta_{d(m)} &0& 0\\
		0&0&\tilde\gamma_{d(m)}\end{pmatrix} $}, \quad
		{ 	$ \frac94 v_d^2\begin{pmatrix}
			\tilde\beta_{d(m)}^2  & 0 & 0\\
			0  & \tilde\alpha_{d(m)}^2 & 0\\
			0 & 0 & \tilde\gamma_{d(m)}^2\end{pmatrix}$}
		& $\begin{matrix}  0 \\0\end{matrix}$ 
		&
		 $\begin{matrix}  0\\0\end{matrix}$ 
		 & $\begin{matrix}\frac32\omega\tilde{\alpha_{d}}\\
		 -\frac32\omega\tilde{\beta_{d}}\end{matrix}$ \\
		\hline  
		 \rule[14pt]{0pt}{1pt} 
		$U_{ST4}$ &
		{$ \frac32 v_d\begin{pmatrix}
		\omega \tilde\alpha_{d(m)}&0&0\\
		0&0&-\omega^2 \tilde\beta_{d(m)}\\
		0&\tilde\gamma_{d(m)}&0\end{pmatrix} $}, \quad
		{ 	$ \frac94 v_d^2\begin{pmatrix}
		\tilde\alpha_{d(m)}^2  & 0 & 0\\
		0  & \tilde\gamma_{d(m)}^2 & 0\\
		0 & 0 & \tilde\beta_{d(m)}^2\end{pmatrix}$}
		& $\begin{matrix}-\frac32\omega^2\tilde{\beta_d}\\
		\frac32\tilde{\gamma_d}\end{matrix}$
		 & $\begin{matrix}0\\0\end{matrix}$
		  & $\begin{matrix}0\\0\end{matrix}$\\
		\hline  
		 \rule[14pt]{0pt}{1pt} 
		$U_{ST5}$ &
		{$ \frac32 v_d\begin{pmatrix}
		\omega \tilde\alpha_{d(m)}&0&0\\
		0&-\omega^2 \tilde\beta_{d(m)}&0\\
		0&0&\tilde\gamma_{d(m)}\end{pmatrix} $}, \quad
		{ 	$ \frac94 v_d^2 \begin{pmatrix}
		\tilde\alpha_{d(m)}^2  & 0 & 0\\
		0  & \tilde\beta_{d(m)}^2 & 0\\
		0 & 0 & \tilde\gamma_{d(m)}^2\end{pmatrix}$}
		& $\begin{matrix}0\\0
		\end{matrix}$
		 & $\begin{matrix}0\\0\end{matrix}$ 
		 & 	$\begin{matrix}0\\0\end{matrix}$\\
		\hline   
			 \rule[14pt]{0pt}{1pt} 
		$U_{ST6}$ &
		{$ \frac32 v_d \begin{pmatrix}
			0&\omega \tilde\alpha_{d(m)}&0\\
			0&0&-\omega^2 \tilde\beta_{d(m)}\\
			\tilde\gamma_{d(m)}&0&0\end{pmatrix} $}, \quad
		{ 	$ \frac94 v_d^2 \begin{pmatrix}
		\tilde\gamma_{d(m)}^2  & 0 & 0\\
		0  & \tilde\alpha_{d(m)}^2 & 0\\
		0 & 0 & \tilde\beta_{d(m)}^2\end{pmatrix}$}
		&$\begin{matrix}-\frac32\omega^2\tilde{\beta_d}\\0
		\end{matrix}$
		 & $\begin{matrix}0\\\frac32\tilde{\gamma_d}
		 \end{matrix}$
		  &$\begin{matrix}\frac32\omega\tilde{\alpha_d}\\0
		  \end{matrix}$ \\
		\hline         
	\end{tabular}
	\caption{Down-type quark mass matrices and 
	 $A_4$ flavor coefficients of the FC quark  bilinear operators 
	 in $Y_1(\omega)$ unit for each  	$U_{STi}$ at $\tau_q=\omega$.
Here,  $\tilde\alpha_{d}= Y_1(\omega)\alpha_{d}$,
 $\tilde\beta_{d}= Y_1(\omega)\beta_{d}$ and $\tilde\gamma_{d}= Y_1(\omega)\gamma_{d}$.
}
	\label{tb:ST-omega}
\normalsize
\end{table}

\subsubsection{Mass eigenstate at nearby $\tau=\omega$}

In order to get the observed fermion masses and CKM elements,
the modulus $\tau$ is deviated from the fixed points $\tau=\omega$.
By using a small dimensionless parameter $\epsilon$,
we put the modulus value as  $\tau=i+\epsilon$.
Then, approximate behaviors of the ratios of modular forms
are given in Ref. \cite{Okada:2020ukr}. 
Small deviation of $\tau$ like $\tau=\omega+\epsilon$
leads to the  approximate behavior of the ratios of modular forms:
\begin{align}
\begin{aligned}
\frac{Y_2(\tau)}{Y_1(\tau)}\simeq \omega\,(1+\epsilon_1) \, , \quad 
\frac{Y_3(\tau)}{Y_1(\tau)}\simeq -\frac{1}{2}\omega^2\,(1+\epsilon_2) \, ,
\quad \epsilon_1\simeq\frac{1}{2} \epsilon_2\simeq 2.1\,i\,\epsilon\,.
\end{aligned}
\label{epST12}
\end{align}
These approximate  forms are  agreement with exact numerical values within  a few $\%$ for $|\epsilon|\leq 0.05$.

As a representative,
we show the $M_d^\dagger M_d$ including corrections up to ${\cal O}(\epsilon)$
for the case of $U_{ST4}$ in Table \ref{tb:ST-omega}:
\begin{align}
M_d^\dagger M_d\simeq  \frac{9}{4} v_d^2\,
\begin{aligned} P
	\begin{pmatrix}
		\tilde\alpha_{d(m)}^2& -\frac{2}{3}\tilde\gamma_{d(m)}^2|\epsilon_1^q| & 0 \\
		-\frac{2}{3}\tilde\gamma_{d(m)}^2|\epsilon_1^q|&\tilde\gamma_{d(m)}^2  &  \frac{2}{3}\tilde\beta_d^2|\epsilon_1^q| \\
		0 & \frac{2}{3}\tilde\beta_{d(m)}^2|\epsilon_1^q| & \tilde\beta_{d(m)}^2 \\
	\end{pmatrix} P^*
\end{aligned}\,, \label{downquark-ST}
\end{align}
where $\tilde\alpha_{d(m)}= Y_1(\omega)\alpha_{d(m)}$,
 $\tilde\beta_{d(m)}= Y_1(\omega)\beta_{d(m)}$ and $\tilde\gamma_{d(m)}= Y_1(\omega)\gamma_{d(m)}$,
and ${\cal O}(\tilde\beta_{d(m)}^2\,\epsilon_1^q,\tilde\alpha_{d(m)}^2\,\epsilon_1^q)$,
 terms are neglected.
The phase matrix $P$ is
\begin{align}
P=
\begin{pmatrix}
e^{ 2i\eta }& 0 & 0 \\
0&e^{ i\eta } & 0\\
0 & 0 & 1 \\
\end{pmatrix} \,, \qquad \eta=\arg \epsilon_1^* \,.
\end{align}
The mass matrix $M_d^\dagger M_d$ is diagonalized as:
\begin{align}
U_{m}^T P^* M_d^\dagger M_d P U_{m}= {\rm diag }\,(m_d^2,\, m_s^2,\, m_b^2)\,,
\end{align}
where  we have up to ${\cal O}(|\epsilon_1^q|)$:
\begin{align}
U_{md}\simeq
\begin{pmatrix}
1 & s_{12}^{d} & s_{13}^d\\
-s_{12}^{d} & 1 & s_{23}^{d} \\
s_{12}^d s_{23}^d-s_{13}^d & -s_{23}^{d} & 1
\end{pmatrix}\simeq
\begin{pmatrix}
1& -2/3|\epsilon_1^q| & {\cal O}(|\epsilon_1^q|^2) \\
2/3|\epsilon_1^q|&1 & 2/3|\epsilon_1^q|\\
{\cal O}(|\epsilon_1^q|^2) &-2/3|\epsilon_1^q| & 1 \\
\end{pmatrix} \,.
\label{mixingdown-ST}
\end{align}
Here, the 1-3 mixing angle  $s_{13}^d$ negligibly small.
The magnitude of $s_{12}^d$ is also approximately  the ratio of CKM elements 
$|{V_{td}}/V_{ts}|$  if the mixing angle $s_{13}^u$
is also  negligibly small as well as $s_{13}^d$.
For other cases of $U_{STi}$ in Table \ref{tb:ST-omega},
the mixing matrix $U_{md}$ is the same one as in Eq. \eqref{mixingdown-ST}.

The mass eigenstate is realized by the transformation:	
\begin{align}
&d_L \rightarrow  d_L^m\equiv U_{md}^T U_{STi}\, d_L\,,\qquad\qquad \ \
\bar d_L \rightarrow \bar d_L^m \equiv 
{\bar d}_L\,U_{STi}^\dagger  U_{md} \,.
\label{massbase-ST}
\end{align}


In Table \ref{tb:STmass-omega},
 $A_4$ flavor coefficients of the FC operators are summarized. 
\begin{table}[h]
	\centering
	\begin{tabular}{|c|c|c|c|} \hline 
		\rule[14pt]{0pt}{2pt}
		& 
		$\begin{matrix}\bar s_R  \Gamma b_L^{m}\\\bar s_L \Gamma  b_R^{m}
		\end{matrix} $ 
		& $\begin{matrix}\bar d_R \Gamma  b_L^{m}\\\bar d_L \Gamma  b_R^{m}
		\end{matrix} $
		&
		$\begin{matrix}\bar d_R \Gamma s_L^{m}\\\bar d_L \Gamma  s_R^{m}
		\end{matrix} $
		\\  \hline
		\rule[14pt]{0pt}{1pt} 			
		$U_{ST1}\,(\tilde{\alpha_d}\gg \tilde{\gamma_d}\gg \tilde{\beta_d})$ 	
		& $\begin{matrix}  \omega^2s^d_{23}\epsilon_1\tilde{\gamma_d}\\ \frac32\tilde{\gamma_d}\end{matrix}$
		& $\begin{matrix}\frac32\omega\tilde{\alpha_d}\\
		-\frac32s^d_{12}\tilde{\gamma_d}\end{matrix}$
		& $\begin{matrix} -\frac32\omega s^d_{23}\tilde{\alpha_d}\\
		-\frac32\omega\tilde{\beta_d}\end{matrix}$\\
		\hline 
		\rule[14pt]{0pt}{1pt} 
		$U_{ST2}\,(\tilde{\alpha_d}\gg \tilde{\beta_d}\gg \tilde{\gamma_d})$
		& $\begin{matrix}  -\frac32\omega^2s^d_{23}\tilde{\beta_d}\\ \frac32 s^d_{12}\tilde{\gamma_d}\end{matrix}$
		& 	$\begin{matrix} \frac32\omega\tilde{\alpha_d} \\\frac32\tilde\gamma_d\end{matrix}$ 
		&  $\begin{matrix} 
		 -\frac12\omega (3s^d_{23}-2\epsilon_1)\tilde{\alpha_d}\\
		 \frac12\omega (3s^d_{23}-2\epsilon_1^*)\tilde{\beta_d}\end{matrix}$\\
		\hline 
		\rule[14pt]{0pt}{1pt} 
		$U_{ST3}\,(\tilde{\gamma_d}\gg \tilde{\alpha_d}\gg \tilde{\beta_d})$
		& $\begin{matrix} -\omega^2\epsilon_1\tilde{\beta_d} \\ -\frac12 (3s^d_{23}+2\epsilon_1^*)\tilde{\gamma_d}\end{matrix}$ 
		&
		$\begin{matrix}  \frac32\omega s^d_{23}\tilde{\alpha_d}\\
		 \frac12 s^d_{12}(3s^d_{23}+2\epsilon_1^*)\tilde{\gamma_d}\end{matrix}$ 
		& $\begin{matrix}\frac32\omega\tilde{\alpha_d}\\
		-\frac32\omega\tilde{\beta_d}\end{matrix}$ \\
		\hline  
		\rule[14pt]{0pt}{1pt} 
		$U_{ST4}\,(\tilde{\beta_d}\gg \tilde{\gamma_d}\gg \tilde{\alpha_d})$
		& $\begin{matrix}-\frac32\omega^2\tilde{\beta_d}\\
		\frac32\tilde{\gamma_d}\end{matrix}$
		& $\begin{matrix}
		 \omega\epsilon_1\tilde{\alpha_d}\\
		 -\frac12 (3s^d_{12}+2\epsilon_1^*)\tilde{\gamma_d}\end{matrix}$
		& $\begin{matrix} \frac32\omega s^d_{12}\tilde{\alpha_d}\\
		-\frac12\omega s^d_{12}(3s^d_{23}-2\epsilon_1^*)\tilde{\beta_d}\end{matrix}$\\
		\hline  
		\rule[14pt]{0pt}{1pt} 
		$U_{ST5}\,(\tilde{\gamma_d}\gg \tilde{\beta_d}\gg \tilde{\alpha_d})$ 
		& $\begin{matrix}-\frac12\omega^2 (3s^d_{23}+2\epsilon_1)\tilde{\beta_d}\\
		-\frac32 s^d_{23}\tilde{\gamma_d}
		\end{matrix}$
		& $\begin{matrix}\omega s^d_{23}\epsilon_1\tilde{\alpha_d}\\
		-\epsilon_1^*\tilde{\gamma_d}\end{matrix}$ 
		& 	$\begin{matrix}\frac12\omega (3s^d_{12}+2\epsilon_1)\tilde{\alpha_d}\\\frac32 \omega s^d_{12}\tilde{\beta_d}\end{matrix}$\\
		\hline   
		\rule[14pt]{0pt}{1pt} 
		$U_{ST6}\,(\tilde{\beta_d}\gg \tilde{\alpha_d}\gg \tilde{\gamma_d})$ 
	&$\begin{matrix}-\frac32\omega^2\tilde{\beta_d}\\
	\frac12 (3s^d_{12}-2\epsilon_1^*)\tilde{\gamma_d}
	\end{matrix}$
	& $\begin{matrix}
	\frac12\omega (3s^d_{23}+2\epsilon_1)\tilde{\alpha_d}\\\frac32\tilde{\gamma_d}
	\end{matrix}$
	&$\begin{matrix}\frac32\omega\tilde{\alpha_d}\\
	-\omega \epsilon_1^*\tilde{\beta_d}
	\end{matrix}$ \\
		\hline         
	\end{tabular}
	\caption{$A_4$ flavor coefficients of the FC quark bilinear operators 
	 in $Y_1(\omega)$ unit for each  $U_{STi}$  at nearby $\tau_q=\omega$.}
	\label{tb:STmass-omega}
\end{table}


\subsection{Diagonal bases of $T$ for quarks}
\subsubsection{At $\tau=i\infty$ }

At $\tau=i\infty$ as presented in Eq. (\ref{fixed-points}),
holomorphic and anti-holomorphic modular forms are given as:
\begin{align}
&\begin{aligned}
 Y^{\rm (2)}(\tau_q=i\infty)=
\begin{pmatrix}
1  \\0\\ 0
\end{pmatrix}\,,\qquad 
\ Y^{\rm (2)*}(\tau_q=i\infty)=
\begin{pmatrix}
1  \\0 \\0
\end{pmatrix}\,.
\end{aligned}
\label{modular-T}
\end{align}


Since the generator $T$ is already diagonal in Eq. \eqref{ST},
the unitary matrix to transform $D_L$ is the unit matrix.
Then,
\begin{align}
&D_L \rightarrow   D_L\,, \qquad \qquad\qquad\qquad\ \
\bar D_L \rightarrow   {\bar D}_L\,.
\label{T-transform}
\end{align}

The down-type quark mass matrix is simply given as:
\begin{align} 
M_{d}\,=v_d
\begin{pmatrix}
\alpha_{d(m)} &0 & 0\\
0 & \beta_{d(m)} & 0\\
0 &0& \gamma_{d(m)}
\end{pmatrix}_{RL}\,, \qquad 
M_{d}^\dagger M_{d}=v_d^2
\begin{pmatrix}
\alpha_{d(m)}^2 &0 & 0\\
0 & \beta_{d(m)}^2 & 0\\
0 &0& \gamma_{d(m)}^2
\end{pmatrix}\,,
\label{matrix-T}
\end{align}
where  $Y_1(i\infty)=1$ is taken.

At $\tau_q=i\infty$,
$A_4$ flavor coefficients of the relevant operators are summarized in Table \ref{tb:T-infty}.

\begin{table}[h]
	\centering
	\begin{tabular}{|c|c|c|c|} \hline 
		\rule[14pt]{0pt}{2pt}
			&$\begin{matrix}\bar s_R \Gamma  b_L^{}\\\bar s_L \Gamma  b_R^{}
  	\end{matrix} $ 
  	 & $\begin{matrix}\bar d_R \Gamma  b_L^{}\\\bar d_L  \Gamma  b_R^{}
  	 \end{matrix} $ 
  	 &$\begin{matrix}\bar d_R \Gamma  s_L^{}\\\bar d_L \Gamma  s_R^{}
  	 \end{matrix} $ 
		\\  \hline
		\rule[14pt]{0pt}{1pt} 			
 $\tau_q=i\infty$	& $\begin{matrix}0\\0
 \end{matrix} $ 
 & $\begin{matrix}0\\0
 \end{matrix} $ 
 &$\begin{matrix}0\\0
 \end{matrix} $ \\\hline
 	\rule[14pt]{0pt}{1pt} 			
 $\tau_q\to i\infty$
 	& $\begin{matrix}s^d_{23}{\beta_d}\\
 		-(s^d_{23}+\delta^*){\gamma_d}\end{matrix} $ 
 	&  $\begin{matrix}-\delta{\alpha_d}\\
 	(s^d_{12}s^d_{23}+s^d_{12}\delta^*-\frac12\delta^{*2}){\gamma_d}
 	\end{matrix} $ 
 	&$\begin{matrix}s^d_{12}{\alpha_d}\\
 -(s^d_{12}+\delta^*){\beta_d}\end{matrix} $\\
		\hline 
		\end{tabular}
	\caption{$A_4$ flavor coefficients of the FC quark bilinear operators
		at $\tau_q=i\infty$ and toward  $\tau_q=i\infty$.}
	\label{tb:T-infty}
\end{table}

\subsubsection{Mass eigenstate towards $\tau_q=i\infty$}
Taking leading terms of  Eq.\,(\ref{Y(2)}), we can express modular forms 
approximately  as:
\begin{align}
Y_1(\tau)\simeq 1+ 12 p\,\epsilon\,,\quad  
Y_2(\tau)\simeq -6 p^{\frac{1}{3}}\,\epsilon^{\frac{1}{3}}\,,
\quad  Y_3(\tau)\simeq -18 p^{\frac{2}{3}}\,\epsilon^{\frac{2}{3}}\,, \quad
p=e^{2\pi i\, {\rm Re}\, \tau}\,,\quad 
\epsilon=e^{-2\pi\, {\rm Im}\, \tau}\,,
\label{epsilonT}
\end{align}
where $ {\rm Im}\, \tau \gg 1$.
Then, the down-type quark mass matrix 

\begin{align}
\begin{aligned}
M_d^\dagger M_d\simeq v_d^2 
\begin{pmatrix}
\alpha_{d(m)}^2 &  \beta_{d(m)}^2\, \delta^*&
\alpha_{d(m)}^2\, \delta\\ 
\beta_{d(m)}^2\, \delta&  \beta_{d(m)}^2 &  
\gamma_{d(m)}^2\,  \delta^*\\ 
\alpha_{d(m)}^2\, \delta^*
& \gamma_{d(m)}^2\,  \delta
&\gamma_{d(m)}^2
\end{pmatrix} \, ,
\end{aligned}
\label{M2Q1T}
\end{align}
where $\delta=-6 p^{\frac{1}{3}}\,\epsilon^{\frac{1}{3}}$.

Taking account of the quark mass hierarchy,
 that is  $\gamma_q^2\gg \beta_q^2\gg \alpha_q^2$,
the mass matrix
 $M_d^\dagger M_d$  is rewritten as:
\begin{align}
M_d^\dagger M_d\simeq  v_d^2
\begin{aligned} P
\begin{pmatrix}
\alpha_{d(m)}^2& \beta_{d(m)}^2|\delta| & 0 \\
\beta_{d(m)}^2|\delta|&\beta_{d(m)}^2  &  \gamma_{d(m)}^2|\delta| \\
0 & \gamma_{d(m)}^2|\delta| & \gamma_{d(m)}^2 \\
\end{pmatrix} P^*
\end{aligned}\,,
\label{}
\end{align}
where ${\cal O}(\alpha_d^2 \delta)$ is neglected and
\begin{align}
P=
\begin{pmatrix}
e^{ 2i\eta }& 0 & 0 \\
0&e^{ i\eta } & 0\\
0 & 0 & 1 \\
\end{pmatrix} \,, \qquad \eta=\arg \delta^* \,.
\end{align}
Then, the mass matrix is diagonalized as:
\begin{align}
U_{md}^T P^* M_d^\dagger M_d P U_{md}= {\rm diag }\,(m_d^2,\, m_s^2,\, m_b^2)\,,
\end{align}
where  we have up to ${\cal O}(|\delta|)$:
\begin{align}
U_{md}\simeq
\begin{pmatrix}
1 & s_{12}^{d} & s_{13}^d\\
-s_{12}^{d} & 1 & s_{23}^{d} \\
s_{12}^d s_{23}^d-s_{13}^d & -s_{23}^{d} & 1
\end{pmatrix}\simeq
\begin{pmatrix}
1& |\delta| & {\cal O}(\delta^2)\\
-|\delta|&1 & |\delta|\\
 {\cal O}(\delta^2) &-|\delta| & 1 \\
\end{pmatrix} \,.
\label{mixingdown-T}
\end{align}
Here, the 1-3 mixing angle  $s_{13}^d$  is negligibly small.
The mixing angle $s_{12}^d$ is also  approximately  the ratio of CKM elements  $|{V_{td}}/V_{ts}|$  if the mixing angle $s_{13}^u$
is also  negligibly small as well as $s_{13}^d$.

The mass eigenstate of the down-type quarks is obtained by the transformation:	
\begin{align}
&d_L \rightarrow  d_L^m\equiv U_{md}^T \, d_L\,,\qquad\qquad \ \
\bar d_L \rightarrow \bar d_L^m \equiv 
{\bar d}_L\, U_{md} \,.
\label{massbase-T}
\end{align}	
Then,
towards  $\tau_q=i\infty$,
$A_4$ flavor coefficients of the FC operators are obtained.
We summarize them in Table \ref{tb:T-infty}.
Since $ \gamma_{d(m)} \gg \beta_{d(m)} \gg \alpha_{d(m)}$, 
 the	$[\bar s_L \Gamma  b_R]$ transition
 is much larger than others.

  \subsection{Bilinear FC operators in  $U(2)$ flavor symmetry}
  As well known, the $U(2)$ flavor symmetry  is a powerful hypothesis to reduce the number of independent parameters of the quark sector.  Indeed,
  the flavor symmetries $U(2)^5$ have been successfully applied to the SMEFT \cite{Faroughy:2020ina}.
   Let us compare our  predictions with ones of  $U(2)^5$ flavor symmetry.
  The down-type quark transitions with chirality flip  allowed by different $U(2)$ breaking terms 
  are summarized in Table \ref{tab:U2RLbs},  
  where we follow the result of Table 4 in Ref.\cite{Faroughy:2020ina}. 
  Details of parameters are given as presented in Appendix \ref{U2}.
  The magnitudes of parameters are
  \begin{eqnarray}
  s_d   = \mathcal O(10^{-1})\,,  \qquad
  \epsilon_q   = \mathcal O(10^{-1})\,,  \qquad
  \delta_d  = \mathcal O(10^{-2})\,,\qquad  
  \delta'_d  =   \mathcal O(10^{-3})\,,
  \end{eqnarray}
  and others are of  $\mathcal O(1)$.

\begin{table}[h]
	\centering
	\begin{tabular}{|c|c|c|c|}
		\hline
			\rule[14pt]{0pt}{2pt}
		& $s \to d$ & $b \to s$ & $b \to d$ \\ \hline
		\rule[14pt]{0pt}{2pt}	
			$\bar R L$ 
		& $(\rho_1 s_d \delta_d')^*[ \bar d_R s_L ]$ 
		& $(\sigma_1 \epsilon_q \delta_d)^* [ \bar s_R b_L ] $
		& $(\sigma_1 \epsilon_q s_d \delta_d')^* [ \bar d_R b_L ] $ \\
			\rule[14pt]{0pt}{2pt}	
		$\bar L R$ 
		& $- \rho_1 s_d \delta_d [ \bar d_L s_R ]$ 
		& $\beta_1 \epsilon_{q} [ \bar s_L b_R ]$ 
		& - \\ 
	 \hline			
	\end{tabular}
	\caption{Left-right fermion bilinears allowed by different $U(2)$ breaking terms. 
	The $\epsilon_q, \delta_d , \delta_d'$ stand for an order of spurion, 
	$s_d$ is a mixing angle and $\rho_1,\sigma_1,\beta_1$ are independent coefficients.  
	The detail definition of the parameters are presented in Appendix \ref{U2}.}
	\label{tab:U2RLbs}
\end{table}

Those results are compared with  our ones
at  $\tau=i+\epsilon$ of Table \ref{Co-S},
at  $\tau=\omega+\epsilon$ of Table \ref{tb:STmass-omega}
and towards    $\tau=i\infty$ of Table \ref{tb:T-infty}.
In $U(2)$ flavor symmetry,  the $[ \bar s_L b_R ]$ transition
is much larger than other ones as seen in Table \ref{tab:U2RLbs}.
The $[ \bar d_R s_L ]$ transition is suppressed of  $10^{-5}$.
These results are contrast to our ones.
 For example,  the $\bar d_L   b_R^{m}$ transition is the dominant one in our case of  Table \ref{Co-S}.
 The $\bar s_L   b_R^{m}$ transition is one order suppressed.
 Taking numerical values in
Eq,\,\eqref{alphabetagamma} and $\epsilon_1^q={\cal O}(0.1)$, it is found that  $\bar d_R   s_L^{m}$ and 
$\bar d_L   s_R^{m}$ transitions are suppressed of ${\cal O}(10^{-3})$
 and ${\cal O}(10^{-5})$, respectively as seen in Table \ref{Co-S}.

 Thus,  the predictions of $A_4$ modular symmetry are considerably different from ones of $U(2)$.
 This difference comes from the flavor structure of the right-handed fermions. In our framework, the right-handed fermions are assigned
 to be $A_4$ singlets contrast to the left-handed ones of the $A_4$ triplet typically, that is,
 the flavor structure is drastically different from the left-handed one.
 On the other hand, both left-handed and right-handed ones are 
 singlets (3rd family) and doublets (1st and 2nd families)
  in the framework of $U(2)^5$ symmetry.
 Then, operators $\bar RL$ and  $\bar LR$ have 
 the typical flavor structures like CKM, respectively.


\subsection{Application to the lepton flavor violation}

	As well known, the LFV such as 
	the $\mu \rightarrow e  \gamma$ decay is suppressed enough in the SM, and is  a good probe to discover the NP.
	If the flavor structure of NP is ruled with the $A_4$ modular symmetry, we can expect the correlation
	among decays of  $\mu \rightarrow e  \gamma$, $\tau\rightarrow \mu  \gamma$ and $\tau \rightarrow e  \gamma$.
	

	For $\mu \rightarrow e  \gamma$, 
	the relevant effective Lagrangian which is valid below some scale $\Lambda$ with $m_W  \ge \Lambda \gg m_b$, 
	is given as \cite{Crivellin:2017rmk,Davidson:2020hkf}
	\begin{eqnarray}
	{\cal L_{\rm eff}}=\frac{1}{\Lambda^2} (C^{di}_{RL} O^{di}_{RL}+C^{di}_{LR} O^{di}_{LR})+h.c.\,,
	\label{Lagrangian}
	\end{eqnarray}
	with the dipole operators $O^{di}_{RL}=e m_\mu(\bar e_R \sigma^{\mu\nu}\mu_L) F_{\mu \nu}$
	and  $O^{di}_{LR}=e m_\mu(\bar e_L \sigma^{\mu\nu}\mu_R) F_{\mu \nu}$,
	while  $C^{di}_{RL}$ and $C^{di}_{LR}$ are the dimensionless Wilson coefficients. 
	Here, we consider tree-level matching between the SMEFT and this effective Lagrangian. 
Then, $O^{di}_{LR}$  matches to the SMEFT operators $Q_{eW}$ and $Q_{eB}$
	(see Table \ref{tab:SMEFTOpe} in Appendix \ref{App:SMEFT}). 
	The matching relation for $C^{di}_{LR}$ is given as the linear combination of
	  $C_{eW}$ and $C_{eB}$, which are Wilson coefficients of the SMEFT operators.				
Since both $C_{eW}$ and $C_{eB}$ are proportional to $A_4$ flavor coefficients  of the bilinear operators
 $\bar E_R\Gamma E_L$ and $\bar E_L\Gamma E_R$, respectively,
 we can simply take  $C^{di}_{LR}$ to be equal to the coefficients in Table \ref{LFV} apart from the absolute values.
In Table \ref{LFV}, $A_4$ flavor coefficients  are listed for the decay processes  $\mu \rightarrow e  \gamma$, $\tau\rightarrow \mu  \gamma$ and $\tau \rightarrow e  \gamma$ at nearby $\tau_e=i$.
	\begin{table}[h]
		\centering
		{\footnotesize
			\begin{tabular}{|c|c|c|} \hline 
				\rule[14pt]{0pt}{2pt}				
				 $\mu \to e\gamma$	 &
				$\tau\to \mu \gamma$ &
				$\tau\to e\gamma$
				\\  \hline
				\rule[14pt]{0pt}{1pt} 	
				 $\frac32  \,(\sqrt{3}-1)\tilde \alpha_e	[\bar e_R \sigma^{\mu\nu}  \mu_L]$& 
				$-\frac12  \,(3-\sqrt{3})\tilde \beta_e
			 [\bar \mu_R \sigma^{\mu\nu}  \tau_L]$&
				$	\frac12  \,(\sqrt{3}-1)(3s_{12}^e+\sqrt{3}s_{13}^e-2\epsilon_1)
				\tilde \beta_e
			[\bar e_R \sigma^{\mu\nu}  \tau_L]$\\
				\rule[14pt]{0pt}{1pt} 					
		 $(1-\sqrt{3}) \,(\sqrt{3}s_{13}^e+\epsilon_1^e)\,
				\, \tilde \beta_e
				[\bar e_L \sigma^{\mu\nu}  \mu_R]$& 
				$		(1-\sqrt{3})s_{12}^e\epsilon_1\tilde \gamma_e[\bar \mu_L \sigma^{\mu\nu}  \tau_R]$&
				$	(1-\sqrt{3})(\sqrt{3}s_{13}^e+\epsilon_1)\tilde \gamma_e[\bar e_L \sigma^{\mu\nu}  \tau_R]$\\
				\hline 
			\end{tabular}
	}
			\caption{$A_4$ flavor coefficients of the charged lepton FC bilinear operators 
				at nearby  $\tau_e=i$, where
				$\tilde{\gamma_e}\gg\tilde{\alpha_e}\gg\tilde{\beta_e}$. }
			\label{LFV}
	\end{table}
		
			At the tree level, the Lagrangian in Eq.\,\eqref{Lagrangian}
	results in the branching ratio
	\begin{eqnarray}
	{\rm BR}(\mu\to e\gamma) =
	\frac{\alpha m_\mu^5}{\Lambda^4\Gamma_\mu}(|C^{di}_{RL}|^2+|C^{di}_{LR}|^2)\,,
	\end{eqnarray}
	where $\Gamma_\mu$ is the width of the muon
	and $\alpha$ is the electromagnetic fine-structure constant. 
	The flavor structure of the dipole operators
	$[\,\bar E_R  \sigma^{\mu\nu}  E_L F_{\mu\nu}\,]$
	and  $[\,\bar E_L  \sigma^{\mu\nu}  E_R F_{\mu\nu}\,]$
	 in the flavor space 
	are presented in Table \ref{LFV}.
	Therefore, we can calculate Wilson coefficients
	at nearby fixed points of the modulus $\tau_e=i$,
	where  the successful lepton mass matrix
	has been reproduced \cite{Okada:2020brs}.

	At nearby $\tau_e=i$,
	 the best fit point is given by the parameters \cite{Okada:2020brs}:

	\begin{eqnarray}
	\tau_e=-0.080+ 1.007\, i\,,\qquad 
	\frac{\tilde \alpha_{e(m)}}{\tilde \gamma_{e(m)}}=6.82\times 10^{-2}\,,\qquad
	\frac{\tilde \beta_{e(m)}}{\tilde \gamma_{e(m)}}=1.02\times 10^{-3}\, .
	\label{Lepton-model}
	\end{eqnarray}
	That is  $\tilde \gamma_{e(m)}\gg \tilde \alpha_{e(m)}\gg \tilde \beta_{e(m)}$.
	Since the  Wilson coefficients are proportional to the coefficients 
	in Table \ref{LFV},
	the $O^{di}_{RL}$ operator dominates the decay amplitude 
	for  $\mu \rightarrow e  \gamma$. On the other hand,
	the $O^{di}_{LR}$ operator dominates the decay amplitude 
	for $\tau\rightarrow \mu  \gamma$ and $\tau \rightarrow e  \gamma$.
	Then, we predict
	\begin{eqnarray}
   {\rm BR}(\mu\to e\gamma)\geqq {\rm BR}(\tau\to e\gamma)\gg
	{\rm BR}(\tau\to \mu\gamma)\,.
	\label{BR-relation}
	\end{eqnarray}
	Those ratios are:
	\begin{align}
	&\frac{{\rm BR}(\tau\to \mu\gamma)}{{\rm BR}(\tau\to e\gamma) } 
	\simeq 
	\left |\frac{ s_{12}^e \epsilon_1}{\sqrt{3} s_{13}^e+\epsilon_1}
	\right |^2
	 \simeq  1.3\times 10^{-3}\, ,
	\nonumber\\
	&\frac{{\rm BR}(\mu\to e\gamma)}{{\rm BR}(\tau\to e\gamma) }
	\simeq \left(\frac{m_\mu}{m_\tau}\right )^5 
	\left(\frac{\Gamma_\tau}{\Gamma_\mu}\right )
	\frac34 
	\left |\frac{1}{\sqrt{3} s_{13}^e+\epsilon_1}\right |^2
\left (\frac{\tilde \alpha_e}{\tilde \gamma_e}\right )^2
	\simeq 4\,, 
	\end{align}
	where $s_{12}^e=s_{13}^e=\epsilon_1=0.1$  are put 
	for a bench mark as seen in Appendix \ref{massmatrix-i},
	and $\tilde \alpha_e/\tilde \gamma_e\simeq
	\tilde \alpha_{e(m)}/\tilde \gamma_{e(m)}$  is put.
	Inputting the current upperbound of the branching ratio $\mu\to e\gamma$,  
	${\rm BR}(\mu\to e\gamma)<4.2\times 10^{-13}$ \cite{TheMEG:2016wtm}, we obtain 
	${\rm BR}(\tau\to \mu\gamma)<1.4\times 10^{-16}$
	and  ${\rm BR}(\tau\to e\gamma)<1.1\times 10^{-13}$.
	These current experimental upper bounds are  $3.3\times 10^{-8}$ and $4.4\times 10^{-8}$,
	respectively \cite{Zyla:2020zbs},
	 which are far away  from our predictions.

Let us compare above  predictions with ones of  $U(2)^5$ flavor symmetry.
The chirality-flipped $\ell_i \to \ell_j \gamma$ transitions
are summarized in Table \ref{tab:leptonLRatU2},  
where we follow the result of Table 4 in Ref.\cite{Faroughy:2020ina}. 
Magnitudes of parameters are given as presented 
in Appendix \ref{U2},
\begin{eqnarray}
 s_e=\mathcal O(10^{-1})\,,  \qquad\epsilon_\ell   = \mathcal O(10^{-1})\,,  \qquad
\delta_e  = \mathcal O(10^{-2})\,,\qquad  
 \delta'_e  =   \mathcal O(10^{-3})\,,
\end{eqnarray}
and others are of  $\mathcal O(1)$.
As seen  in Table \ref{tab:leptonLRatU2},  
$O^{di}_{LR}$ operator dominates the decay amplitude 
for $\mu\rightarrow e \gamma$ and $\tau \rightarrow \mu  \gamma$
while $O^{di}_{RL}$ operator dominates the decay amplitude 
for  $\tau \rightarrow e  \gamma$.
Then, we predict
\begin{eqnarray}
{\rm BR}(\tau\to \mu\gamma)\gg {\rm BR}(\mu\to e\gamma)
\gg {\rm BR}(\tau\to e\gamma) \,.
\end{eqnarray}
This relation is completely  different from the prediction  
at nearby $\tau=i$.

\begin{table}[h]
\centering
\begin{tabular}{|c|c|c|c|}
	\hline 
	\rule[14pt]{0pt}{2pt}
	& $\mu \to e \gamma$ & $\tau \to \mu \gamma$ & $\tau \to e \gamma$ \\
	 \hline
	 \rule[14pt]{0pt}{2pt}
		$\bar R L$ 
	& $(\rho_1 s_e \delta_e')^*[ \bar e_R \sigma^{\mu\nu} \mu_L ]$ 
	& $(\sigma_1 \epsilon_{\ell} \delta_e)^* [ \bar \mu_R\sigma^{\mu\nu}  \tau_L ] $
	& $(\sigma_1 \epsilon_{\ell} s_e \delta_e')^* [ \bar e_R\sigma^{\mu\nu} \tau_L ] $
	\\\rule[14pt]{0pt}{2pt}
	$\bar L R$ 
	& $- \rho_1 s_e \delta_e [ \bar e_L\sigma^{\mu\nu}  \mu_R ]$ 
	& $\beta_1 \epsilon_{\ell} [ \bar \mu_L \sigma^{\mu\nu} \tau_R ]$ 
	& -  \\ 
	\hline			
\end{tabular}
\caption{Left-right fermion bilinears allowed by different $U(2)$ breaking terms.
	The $\epsilon_\ell, \delta_e , \delta_e'$ stand for an order of spurion, 
	$s_e$ is a mixing angle and $\rho_1,\sigma_1,\beta_1$ are independent coefficients.  
	The detail definition of the parameters are presented in Appendix \ref{U2}.}
\label{tab:leptonLRatU2}
\end{table}

\section{Bilinear fermion operators $[\,\bar D_L  \Gamma  D_L\,]$
	and  $ [\,\bar E_L \Gamma  E_L\,]$ }
\label{sec:LL}

In order to study the semileptonic operators 
 $[\,\bar E_L  \Gamma  E_L\,] [\,\bar D_L \Gamma  D_L\,]$,
 which dominates   $b\to s \bar\mu\mu$, etc. transitions,
 we discuss the bilinear  operators $[\,\bar D_L  \Gamma  D_L\,]$
 and  $ [\,\bar E_L \Gamma  E_L\,]$. 
Corresponding SMEFT operators $Q$ including fermion bilinear are 
	\begin{align}
[\,\bar D_L  \Gamma  D_L\,] \notag 
	&: Q_{Hq}^{(1)},Q_{Hq}^{(3)}\,, \\
 [\,\bar E_L \Gamma  E_L\,] 
	&: Q_{H \ell}^{(1)},Q_{H \ell}^{(3)}\,.
	\end{align}
	
\subsection{$[\,\bar D_L  \Gamma  D_L\,]$
	and  $ [\,\bar E_L \Gamma  E_L\,]$ in flavor space}
\label{bilinear-flavorL}

At first, we construct the $A_4$ invariant bilinear quark operator
$\bar D_L  \Gamma D_L$.
As presented in Appendix \ref{decomposition},
the $A_4$ invariant bilinear quark operator is obtained
by the two type of combinations.
\begin{align}
(Y^* \bar D_L)_{\bf 1,1',1''} \otimes  (Y D_L)_{\bf 1,1',1''}\,,
\qquad 
[(Y^* \bar D_L)_{\bf 3s,3a} \otimes  (Y D_L)_{\bf 3s, 3a}]_{\bf 1,1',1''}\,.
\label{}
\end{align}
The singlets combination gives three $A_4$ invariant bilinear quark operators as
\begin{align}
( Y^* \bar D_L)_{\bf 1}\gamma^\mu (Y D_L)_{\bf 1}\,, \qquad
( Y^* \bar D_L)_{\bf 1'}\gamma^\mu (Y D_L)_{\bf 1''}\,, \qquad
( Y^* \bar D_L)_{\bf 1''}\gamma^\mu (Y D_L)_{\bf 1'}\,.
\end{align}
On the other hand, the triplets combination gives
four $A_4$ invariant bilinear quark operators as
\begin{align}
&( Y^*\bar D_L \Gamma Y D_L)_{\bf 1(s,s)}=
(Y^* \bar D_L)_{3s1} \Gamma (Y D_L)_{3s1} + 
(Y^* \bar D_L)_{3s2} \Gamma (Y D_L)_{3s3} 
+ (Y^* \bar D_L)_{3s3} \Gamma (Y D_L)_{3s2}\,, \nonumber \\
&(Y^*\bar D_L \Gamma Y D_L)_{\bf 1(a,a)}=
(Y^* \bar D_L)_{3a1} \Gamma (Y D_L)_{3a1} + 
(Y^* \bar D_L)_{3a2} \Gamma (Y D_L)_{3a3} 
+ (Y^* \bar D_L)_{3a3}\Gamma (Y D_L)_{3a2}\,, \nonumber \\
&(Y^*\bar D_L \Gamma Y D_L)_{\bf 1(s,a)}=
(Y^* \bar D_L)_{3s1} \Gamma (Y D_L)_{3a1} + 
(Y^* \bar D_L)_{3s2} \Gamma (Y D_L)_{3a3} 
+ (Y^* \bar D_L)_{3s3} \Gamma (Y D_L)_{3a2}\,, \nonumber \\
&(Y^*\bar D_L \Gamma Y D_L)_{\bf 1(a,s)}=
(Y^* \bar D_L)_{3a1} \Gamma (Y D_L)_{3s1} + 
{(Y^* \bar D_L)_{3a2} \Gamma (Y D_L)_{3s3} + 
(Y^* \bar D_L)_{3a3} \Gamma (Y D_L)_{3s2}\,.}
\label{}
\end{align}

Finally, we have the $A_4$ invariant bilinear quark operator, 
which is expressed as:
\begin{align}
&\bar D_L  \Gamma  D_L \Rightarrow 
s_{q1}\,(Y^* \bar D_L)_{\bf 1}\Gamma (Y D_L)_{\bf 1}+
s_{q2}\,(Y^* \bar D_L)_{\bf 1''}\Gamma (Y D_L)_{\bf 1'}+
s_{q3}\, (Y^* \bar D_L)_{\bf 1'}\Gamma(Y D_L)_{\bf 1''}
\nonumber\\
&
+t_{qss}\,(Y^*\bar D_L \Gamma Y D_L)_{\bf 1(s,s)}
+ t_{qaa}\,(Y^*\bar D_L \Gamma Y D_L)_{\bf 1(a,a)}+
t_{qsa}\,(Y^*\bar D_L \Gamma Y D_L)_{\bf 1(s,a)}+
t_{qas}\,(Y^*\bar D_L \Gamma Y D_L)_{\bf 1(a,s)}\,.
\label{quarkLL0}
\end{align}
Here, $s_{qi}(i=1,2,3)$ and $t_{qss}$, etc., are arbitrary constants.
For the lepton sector, we have similar forms as:
\begin{align}
&\bar E_L  \Gamma  E_L =
s_{e1}\,( Y^* \bar E_L)_{\bf 1}\Gamma (Y E_L)_{\bf 1}+
s_{e2}\,(Y^* \bar E_L)_{\bf 1''}\Gamma (Y E_L)_{\bf 1'}+
s_{e3}\, (Y^* \bar E_L)_{\bf 1'}\Gamma (Y E_L)_{\bf 1''}
\nonumber\\
&
+t_{ess}\,(Y^*\bar E_L \Gamma Y E_L)_{\bf 1(s,s)}
+ t_{eaa}\,(Y^*\bar E_L \Gamma YE_L)_{\bf 1(a,a)}+
t_{esa}\,(Y^*\bar E_L \Gamma Y E_L)_{\bf 1(s,a)}+
t_{eas}\,(Y^*\bar E_L\Gamma Y E_L)_{\bf 1(a,s)}\,.
\label{leptonLL0}
\end{align}
Since both $\bar D_L  \gamma_\mu  D_L$ and $\bar E_L  \gamma_\mu  E_L$ 
are Hermitian, we have parameter relations as:
\begin{align}
{\rm Im}\,s_{q_1}=0\,,\quad  s_{q3}=s_{q2}^*\,,\quad t_{qsa}=t_{qas}^*\,,\qquad 
{\rm Im}\,s_{e_1}=0\,, \quad  s_{e3}=s_{e2}^*\,,\quad t_{esa}=t_{eas}^*\,.
\label{parameter-relation}
\end{align}
By using Eqs.\,(\ref{quarkLL0}) and (\ref{leptonLL0}), we can calculate
the bilinear  operators 
$[\,\bar D_L \Gamma  D_L\,]$ and 
$[\,\bar E_L \Gamma  E_L\,]$.

The $A_4$ flavor coefficients of the relevant operators are summarized
in the base of Eq. \eqref{ST} for $S$ and $T$ in Table  \ref{tb:LL-default},
where  coefficients of $\bar\mu_L \mu_L$ and $\bar e_L e_L$
are presented for the lepton sector 
since we focus on $b\to s \bar\mu\mu$, etc. transitions.

\begin{table}[h]
	\scriptsize{
		\centering
		\begin{tabular}{|c|c|} 
			\hline 
			\rule[14pt]{0pt}{2pt}
			Operator	&	$A_4$ coefficient
			\\  \hline
			\rule[14pt]{0pt}{2pt}
			$[\bar s_L \Gamma b_L]$	&
			{\scriptsize 
				$\frac{1}{36}\left [
				4(9s_{q3}-2t_{qss}-3t_{qsa})Y^*_{2}Y_{1}
				+(36s_{q1}+4t_{qss}+9t_{qaa}-6(t_{qas}+t_{qsa}))Y^*_{3}Y_{2}
				+4(9s_{q2}-2t_{q1}-3t_{q3})Y^*_{1}Y_{3}\right ] $}
			\\  \hline
			\rule[14pt]{0pt}{2pt}
			$[\bar d_L \Gamma  b_L]$	&
			{\scriptsize 
				$\frac{1}{36}
				\left [4(9s_{q1}-2t_{qss}+3t_{qsa})Y^*_{1}Y_{2}
				+4(9s_{q2}-2t_{qss}+3t_{qas})Y^*_{2}Y_{3} 
				+(36s_{q3}+4t_{qss}+9t_{qaa}+6(t_{qas}+t_{qsa}))
				Y^*_{3}Y_{1}
				\right]$ }
			\\  \hline
			\rule[14pt]{0pt}{2pt}
			$[\bar d_L \Gamma  s_L]$	&
			{\scriptsize 
				$\frac{1}{36}
				\left[
				(36s_{q2}+4t_{qss}+9t_{qaa}-6(t_{qas}+t_{qsa}))Y^*_{2}
				Y_{1}
				+2(-9s_{q3}+2t_{qss}+3t_{qas})Y^*_{3}Y_{2}
				+(36s_{q1}-8t_{qss}-12t_{qsa})Y^*_{1}Y_{3}
				\,\right] $ }
			\\  \hline
			\rule[14pt]{0pt}{2pt}
			$ [\,\bar \mu_L \Gamma  \mu_L\,]$	&
			{\scriptsize 
				$\frac{1}{36}
				\left [\,(36 s_{e2}+4t_{ess} - 9 t_{eaa} +
				6( t_{eas} - 6 t_{esa}))Y^*_{1}Y_{1}
				+4(9s_{e1}+4 t_{ess})  Y^*_{2}Y_{2} 
				+(36s_{e1}+4t_{e1} - 9 t_{eaa} - (6 t_{eas} - 6 t_{esa})) Y^*_{3}Y_{3}
				\right]$ }
			\\  \hline
			\rule[14pt]{0pt}{2pt}
			$ [\,\bar e_L \Gamma  e_L\,]$	&
			{\scriptsize 
				$\frac{1}{36}
				\left [4(9s_{e1}+4t_{e1})Y^*_{1}Y_{1}
				+( 36s_{e3}+4t_{ess} - 9 t_{eaa} - 6( t_{eas} -  t_{esa})) Y^*_{2}Y_{2}
				+(36s_{e3}+4t_{ess} - 9 t_{eaa} + 6( t_{eas} -  t_{esa})) Y^*_{3}Y_{3}
				\right]$ }
			\\  \hline
		\end{tabular}
	}
	\caption{$A_4$ flavor coefficients of the FC bilinear operators
		for arbitrary $\tau$. }
	\label{tb:LL-default}
\end{table}

\subsection{Diagonal base of $S$ for both quarks and leptons}
\label{Diag-S}
\subsubsection{Semileptonic operator at $\tau=i$}

In order to move the diagonal base of $S$ for both quark
$D_L$ and lepton $E_L$, we transform the quark and lepton fields like Eqs. \eqref{dL-S} and \eqref{U-S}. Furthermore, we put $\tau=i$ for modular forms as in Eq. \eqref{modularS}. 
Then,  $A_4$ triplet left-handed fields are transformed as in 
Eq. (\ref{dL-S}):
\begin{align}
D_L \rightarrow  D_L^{S}\equiv U_{S}\, D_L\,, \quad 
\bar D_L \rightarrow \bar D_L^{S} \equiv {\bar D}_L\,U_{S}^\dagger \,,
\quad  E_L \rightarrow  E_L^{S}\equiv U_{S}\, E_L\,, \quad 
\bar E_L \rightarrow \bar E_L^{S} \equiv {\bar E}_L\,U_{S}^\dagger \,,
\end{align}
where $U_{S}$ is presented in Eq. (\ref{U-S}) and
\begin{align}
D_L^{S}=
\begin{pmatrix}
d_L^{S}\\
s_L^{S}\\
b_L^{S}
\end{pmatrix}\, , \qquad 
E_L^{S}=
\begin{pmatrix}
e_L^{S}\\
\mu_L^{S}\\
\tau_L^{S}
\end{pmatrix}\, .
\end{align}

Then, we have relevant semileptonic operators
\begin{align}
C_{\alpha L\beta L}^{S} [\,\bar \alpha^{S} \Gamma\beta_{L}^{S}\,]\,,
\qquad 
\label{operators-S}
\end{align}
where $\alpha,\,\beta$ denote $d,\,s,\, b\,, e,\mu,\tau$.
Coefficients of Table \ref{tb:LL-default}
 are transformed to simple ones as:

	\begin{align}
	&C_{bLbL}^S=
	\frac{1}{2} Y_1(i)^2
	\left [\, (2-\sqrt{3})(s_{q1}+s_{q2}+{4s_{q3}}-t_{qaa} \right ]\,,\nonumber\\
	&C_{sLsL}^S=
	\frac{1}{2} Y_1(i)^2
	\left [\, (2-\sqrt{3})(9s_{q1}+9s_{q2}+{4t_{qss}} \right ]\,,\nonumber\\
	&  C_{dLdL}^S=
	\frac{1}{2}(2-\sqrt{3}) Y_1(i)^2 ({4t_{qss}}-3t_{qaa})\,,
	\nonumber\\
		&{C_{sLbL}^S=
		\frac{1}{2} Y_1(i)^2
		\left [\, (3-2\sqrt{3})(3 s_{q1}-3s_{q2}+2t_{qsa})
		\,\right ]\,,}
	\nonumber\\
		&{C_{bLsL}^S=
		\frac{1}{2} Y_1(i)^2
		\left [\, (3-2\sqrt{3})(3 s_{q1}-3s_{q2}-2t_{qas})
		\,\right ]\,,}
	\nonumber\\
	&C_{dLbL}^S= C_{dLsL}^S=0\,,\nonumber\\
	&C_{\mu L \mu L}^S=
	\frac{1}{2}(2-\sqrt{3}) Y_1(i)^2
	 (9s_{e1}+9s_{e2}+{4t_{ess}})\,, \nonumber\\
	&  C_{eLeL}^S=\frac{1}{2}(2-\sqrt{3}) Y_1(i)^2 ({4t_{ess}}-3t_{eaa})\,.
	\label{C-S}
	\end{align}
	  
	In this base of quarks and leptons,
	the off diagonal elements of the mass matrix   $(M_d^\dagger M_d)_{LL}$  vanish or are tiny  under the condition  
	${\tilde\gamma_d}\gg {\tilde\alpha_d},{\tilde\beta_d}$ as seen in Eq. \eqref{matrix-S}.
  We also expect that $C_{sLbL}^S$ and $C_{bLsL}^S$
  		are suppressed.
	Therefore, we impose the constraints of parameters in Eq. \eqref{C-S} as:
		\begin{align}
		3 s_{q1}-3s_{q2}+2t_{qsa}=0\,, \qquad   {\rm Re}\, t_{qsa}=0\,,
		\label{condition-S1}
		\end{align} 
which lead to $C_{sLbL}^S=C_{bLsL}^S=0$.

Next, we shift $\tau_q=i$ to $\tau_q=i+\epsilon$ and transform
$D_L \rightarrow  D_L^m\equiv U_{md}^T U_{12}^T(90^\circ) U_{S}\, D_L$ as well as in Eq. \eqref{massbase-S}, we obtain
	\begin{align}
	C_{sLbL}^{Sm}=&
		Y_1(i)^2
		\left [\,\frac12 (\sqrt{3}-2)[3(s_{q1}+s_{q2}+4s_{q3})-4t_{qss}] s_{23}^d
		-3(\sqrt{3}-2)(s_{q1}-s_{q2})\epsilon_1 \right .\nonumber\\
	& \left .	+ 3(2\sqrt{3}-3)(s_{q1}+s_{q2}-2s_{q3})\epsilon_1^* 
		\,\right ]\,,
	\nonumber\\
	C_{dLbL}^{Sm}=&
		Y_1(i)^2 
		\left [ \left \{\,\frac12 (\sqrt{3}-2)[3(s_{q1}+s_{q2}+4s_{q3})-4t_{qss}] s_{23}^d
		+3(\sqrt{3}-2)((s_{q1}-s_{q2})(\epsilon_1+\epsilon_1^*)\,\right \}
		\right .	
	 (-s_{12}^d) 	\nonumber\\
	 & + \left .\left\{3(\sqrt{3}-2)(s_{q1}-s_{q2})\epsilon_1+
	 \frac16 (3-2\sqrt{3}) (3t_{qaa}+4t_{qss})\epsilon_1^*  \right \}s_{23}^d 
	 \right ]
		\,, \nonumber\\
	C_{dLsL}^{Sm}=&
		Y_1(i)^2 
		\left [\,-\frac32 (\sqrt{3}-2)[3(s_{q1}+s_{q2})+4t_{qaa}] s_{12}^d
		+ 3 (\sqrt{3}-2)(s_{q1}-s_{q2})\epsilon_1 \right .
		\nonumber\\+
		& \left .\frac16 (3-2\sqrt{3}) (3t_{qaa}+4t_{qss})\epsilon_1^*
	\right ]\,,
	\label{CLL-Sm}
	\end{align}
	where $s_{12}^d$ is redefined to be complex one including a phase as seen in Eq.\,\eqref{s12d}, and $s_{23}^d$ is real.

{ For the coefficients of charged leptons, we transform
$E_L \rightarrow  E_L^m\equiv U_{me}^T U_{S}\, E_L$
as well as in Eq. \eqref{massbase-S}, we obtain
\begin{align}
C_{\mu L \mu L}^{Sm}\simeq
\frac{1}{2}(2-\sqrt{3}) Y_1(i)^2
(9s_{e1}+9s_{e2}+{4t_{ess}})\,, \qquad 
&  C_{eLeL}^{Sm}\simeq\frac{1}{2}(2-\sqrt{3}) Y_1(i)^2 ({4t_{ess}}-3t_{eaa})\,.
\label{C-Se}
\end{align}
Thus,  $C_{eLeL}^{Sm}$ is comparable to  $C_{\mu L \mu L}^{Sm}$
unless specific relations are set. }

It is found that there is no relation among three coefficients of quarks because
 of many parameters.
 As well known, the  $U(2)$ flavor symmetry predicts \cite{Faroughy:2020ina}
 \begin{align}
 \frac{C_{dLbL}^{Sm}}{C_{sLbL}^{Sm}}\simeq -s_{12}^d=\frac{V_{td}^*}{V_{ts}^*} \,,
 \label{U2-relation}
 \end{align}
where  Eq.\,\eqref{s12d} is taken account.
This relation  could be reproduced if we impose following relations
  in addition to ones in Eq.\,\eqref{condition-S1}
 \begin{align}
 s_{q1}=s_{q2}\,, \qquad\qquad 3t_{qaa}+4t_{qss}= 0\,,
 \end{align}
 which lead to  $t_{as}=t_{sa}=0$ and $s_{q1}=s_{q2}=s_{q3}$ finally. 
 We arrive at
  \begin{align}
  &	C_{sLbL}^{Sm}=
  \frac12 (\sqrt{3}-2)(9s_{q1}-t_{qss}) s_{23}^dY_1(i)^2\,, \nonumber\\
 & C_{dLbL}^{Sm}=-s_{12}^d C_{sLbL}^{Tm}Y_1(i)^2\,, \nonumber\\
& C_{dLbL}^{Sm}=
\frac32 (2-\sqrt{3})(6s_{q1}+t_{qaa}) s_{12}^d Y_1(i)^2\,.
 \end{align}
In addition, imposing $s_{q1}=0$,
we  obtain 
\begin{align}
\frac{C_{dLsL}^{Sm}}{C_{sLsL}^{Sm}}=-s_{12}^d=\frac{V_{td}^*}{V_{ts}^*} \,,
\label{U2-relation2}
\end{align}
where $C_{sLsL}^{Sm}\simeq C_{sLsL}^{S}$ is given in Eq.\,\eqref{C-S}.
This relation is also predicted in the  $U(2)$ flavor symmetry.
Thus,  a simple setup of parameters gives well known relations
of the coefficients in the  $U(2)$ symmetry.  However, we have no principle to choose such a parameter set at this stage unless an additional symmetry is put.

\subsection{Diagonal bases of $ST$ for quarks}
\label{Diag-ST}
We consider the  case of the diagonal   $ST$ only for  quarks
because the successful lepton mass matrices are realized at nearby $\tau_e=i$.
The $A_4$ triplet left-handed fields are transformed as in Eq. (\ref{dL-ST}):
\begin{align}
D_L \rightarrow  D_L^{ST}\equiv U_{STi}\, D_L\,, \quad 
\bar D_L \rightarrow \bar D_L^{ST} \equiv {\bar D}_L\,U_{STi}^\dagger \,,
\end{align}
where $U_{STi}$ is presented in Eq. (\ref{STdiagonal}) and
\begin{align}
D_L^{ST}=
\begin{pmatrix}
d_L^{ST}\\
s_L^{ST}\\
b_L^{ST}
\end{pmatrix}\, .
\end{align}

As a representative, 
we discuss the case of $U_{ST4}$, which is shown explicitly in  Eq.\,(\ref{STdiagonal}) of Appendix \ref{appe-ST}.
 The down-type quark mass matrix is given  in Table \ref{tb:ST-omega}.
Then, we have relevant semileptonic operators
\begin{align}
C_{\alpha L\beta L}^{ST}[\,\bar \alpha^{ST} \Gamma \beta_{L}^{ST}\,]\,,
\label{operators-ST}
\end{align}
where $\alpha,\,\beta$ denote $d,\,s,\, b$.
Putting $\tau_q=\omega$, those coefficients are given as: 
\begin{align}
&C_{bLbL}^{ST}=Y_1(\omega)^2
\,\frac14 (9 s_{q2}+4t_{qss}) \,,
\nonumber\\
&C_{sLsL}^{ST}=Y_1(\omega)^2
\frac{1}{16}
\left (\,36 s_{q3}+4 t_{qss}-9t_{qaa}
+6t_{qas}-6t_{qsa} \,\right )\,,
\nonumber\\
&C_{dLdL}^{ST}=Y_1(\omega)^2
\frac{1}{16}\left (\,36 s_{q1}+4 t_{qss}-9t_{qaa}
-6t_{qas}+6t_{qsa} \,\right )\,,
\nonumber\\
&C_{sLbL}^{ST}=C_{dLbL}^{ST}=C_{dLsL}^{ST}=0\,.
\label{CLL-ST}
\end{align}
Thus, the flavor changing operators vanish.

Next step, we shift $\tau_q=\omega$ to $\tau_q=\omega+\epsilon$
and transform $d_L \rightarrow  d_L^m\equiv U_{md}^T  U_{ST4}\, d_L$ and obtain
\begin{align}
&C_{sLbL}^{STm}=-Y_1(\omega)^2\left[\frac{3}{16}\,  s_{23}^d\,
 (12 s_{q2}+4 t_{qss}+ 3 t_{qaa}-12 s_{q3}-2t_{qas}+2t_{qsa}) -
 \frac{1}{6} \epsilon_1^*(9s_{q2}-2t_{qss}+3t_{qas})\right ]
 \,,
\nonumber\\
&C_{dLsb}^{STm}=
Y_1(\omega)^2\,  s_{12}^d
\left [ \,\frac{3}{16}s_{23}^d\, (12 s_{q2}+4 t_{qss}+ 3 t_{qaa}-12 s_{q3}-2t_{qas}+2t_{qsa})-\frac{1}{6} \epsilon_1^*(9s_{q2}
-2t_{qss}+3t_{qas})\right  ]
\nonumber\\
& +\frac{1}{24}Y_1(\omega)^2(\epsilon_1+\frac23 |\epsilon_1|^2)
	 (36s_{q1}+4 t_{qss}+9 t_{qaa} -6 t_{qas}-6t_{qsa})-\frac{1}{6}Y^2_1(\omega)\epsilon_1^* s_{23}^d 
	( 9s_{q3}-2t_{qss}+3t_{qsa})\,,
\nonumber\\
&C_{dLsL}^{STm}=Y_1(\omega)^2\left \{\frac{1}{4}\,  s_{12}^d
\left[\,9s_{q1}-9s_{q3}+3t_{qsa}-3t_{qas}
\right ]- \frac{1}{6}\,  \epsilon_1^*
(9s_{q3}-2t_{qss}+3t_{qsa})\right \}
\,.
\label{CLL-STm}
\end{align}

In order to obtain the $U(2)$ symmetry like relation in Eq. \eqref{U2-relation},
\begin{align}
\frac{C_{dLbL}^{STm}}{C_{sLbL}^{STm}}=-s_{12}^d=\frac{V_{td}^*}{V_{ts}^*} \,,
\end{align}
we impose relations of parameters:
\begin{align}
9s_{q3}-2t_{qss}+3t_{qsa}=0\,,\qquad 
36s_{q1}+4 t_{qss}+9 t_{qaa} -6 t_{qas}-6t_{qsa}=0\,.
\end{align}
Putting them into $C_{dLsL}^{STm}$ in Eq. \eqref{CLL-STm},
we obtain 
\begin{align}
C_{dLsL}^{STm}=\frac14 Y_1(\omega)^2\,  s_{12}^d\,
(9s_{q1}-9s_{q3}+3t_{qsa}-3t_{qas})
\,.
\end{align}
In addition, imposing $t_{sa}=t_{aa}=0$,
we  obtain 
\begin{align}
\frac{C_{dLsL}^{STm}}{C_{sLsL}^{STm}}=-s_{12}^d=\frac{V_{td}^*}{V_{ts}^*} \,,
\label{}
\end{align}
where $C_{sLsL}^{STm}\simeq C_{sLsL}^{ST}$ is given in Eq.\,\eqref{CLL-ST}.
This relation is also  predicted in  the $U(2)$ flavor symmetry.

\subsection{Diagonal bases of $T$ for quarks}
\label{Diag-T}

Let us consider the  case of the diagonal   $T$  only for  quarks.
In the diagonal base of $T$ of Eq.\,(\ref{ST}),
the mass matrix $M_q^\dagger M_q$ is given at $\tau_q=i\infty$ as:
\begin{align}
{\cal M}_q^{2(0)}\equiv   M_q^\dagger M_q 
= 
\begin{aligned} v_d^2
\begin{pmatrix}
\alpha_q^2&0 & 0 \\
0&\beta_q^2 &   0  \\
0 & 0 & \gamma_q^2 \\
\end{pmatrix}
\end{aligned}\,,
\label{quarkmassnewT}
\end{align}
where $Y_1(i\infty)=1$ is taken.
Mixing angles appear through the finite effect of ${\rm Im }\,[\tau]$.

Since the base of Eq.\,(\ref{ST})
is already the diagonal base of  $T$, the transformation is trivial
 as seen in Eq.\,\eqref{T-transform}.
 Then, we have 
\begin{align}
D_L^{T}=D_L=
\begin{pmatrix}
d_L^{T}\\
s_L^{T}\\
b_L^{T}
\end{pmatrix}\, .
\end{align}
Therefore,
 the relevant bilinear operators are given in subsection \ref{bilinear-flavorL}
for the diagonal case of $T$.

Then, we have relevant semileptonic operators
\begin{align}
C_{\alpha L\beta L}^{T} [\,\bar \alpha^{T} \Gamma  \beta_{L}^{T}\,]\,,
\label{operators-T}
\end{align}
where $\alpha,\,\beta$ denote $d,\,s,\, b$.
Those coefficients are given  at $\tau=i\infty$ as: 
\begin{align}
&C_{bLbL}^{T}=
\frac{1}{36}\left (\,36s_{q3}+4 t_{qss}-9t_{qaa}
-6t_{qas}+6t_{qsa} \,\right )\,,
\nonumber\\
&C_{sLsL}^{T}=
\frac{1}{36}\left (\,36s_{q2}+4 t_{qss}-9t_{qaa}
+6t_{qas}-6t_{qsa} \,\right )\,,
\nonumber\\
&C_{dLdL}^{T}=\frac{1}{9}\left (\,9 s_{q1}+4t_{qss}\,\right )\,,
\nonumber\\
&C_{sLbL}^{T}=C_{dLbL}^{T}=C_{dLsL}^T=0\,.
\label{CLL-T}
\end{align}
It is noticed that the flavor changing operators vanish at $\tau=i\infty$.


Next step, we include the finite effect of the modular forms
and  transform
$d_L \rightarrow  d_L^m\equiv U_{md}^T  \, d_L$ as given in Eq.\,\eqref{massbase-T} and obtain
\begin{align}
C_{sLbL}^{Tm} =&  \frac{1}{3}\,s_{23}^d\,
(3 s_{q2}-3 s_{q3}+ t_{qas}-  t_{qsa}) +
\frac{1}{9} \delta^*(9 s_{q3}-3 t_{qsa}-2 t_{qss})
\,,
\nonumber\\
C_{dLsb}^{Tm} =&
-\frac{1}{3}\,s_{12}^d s_{23}^d\,
(3 s_{q2}-3 s_{q3}+ t_{qas}-  t_{qsa}) 
-\frac{1}{9}s_{12}^d\delta^*(9 s_{q3}-3 t_{qsa}-2 t_{qss}) \nonumber\\
&+\frac{1}{9}\delta (9 s_{q1}+ 3t_{qsa}-t_{qss}\})+
\frac{1}{36}s_{23}^d (36 s_{q2}+4 t_{qss}+9 t_{qaa}-6 t_{qas}-6 t_{qsa} )
\,,
\nonumber\\
C_{dLsL}^{Tm} =& {s_{12}^d\,
( s_{q1}- s_{q2}+ \frac13 t_{qss}+ \frac14 t_{qaa}+\frac16  t_{qsa}
-\frac16  t_{qas}) } \nonumber\\
&+\delta^*(s_{q2}+ \frac19 t_{qss}+ \frac14 t_{qaa}-\frac16  t_{qsa}
-\frac16  t_{qas})
\,.
\label{CLL-Tm}
\end{align}
In order to obtain the $U(2)$ symmetry like relation
\begin{align}
\frac{C_{dLbL}^{Tm}}{C_{sLbL}^{Tm}}=-s_{12}^d=\frac{V_{td}^*}{V_{ts}^*} \,,
\label{}
\end{align}
we impose relations of parameters 
\begin{align}
t_{qss}= 9 s_{q1} + 3 t_{qsa}\,, \qquad t_{qaa}= -4 ( s_{q1} + 3 s_{q2})\,,
\end{align}
with all real parameters.
Then, we have a reasonable result for $C_{dLsL}^{Tm}$
\begin{align}
C_{dLsL}^{Tm}=s_{12}^d\,
( 3s_{q1}- 2s_{q2}+   t_{qsa}) \,,
\end{align}
which is proportional to $s_{12}^d$.

In addition, imposing $t_{sa}=t_{ss}=0$, which give $s_{q1}=0$,
we  obtain 
\begin{align}
\frac{C_{dLsL}^{Tm}}{C_{sLsL}^{Tm}}=-s_{12}^d=\frac{V_{td}^*}{V_{ts}^*} \,,
\label{}
\end{align}
which is also predicted in the  $U(2)$ flavor symmetry.

\subsection{$\Delta F=1$  semileptonic operators in $A^E_{4}\otimes A^Q_{4}$ symmetry}  \label{default}

Let us discuss the semileptonic flavor changing neutral processes,
\begin{align}
b\rightarrow s \, \bar\mu \mu\ (s\,\bar e e) \,,
\qquad b\rightarrow d \, \bar\mu \mu\ (d\,\bar e e) \,,
\qquad s\rightarrow d \, \bar\mu \mu\ (d\,\bar e e) \,,
\label{semileptonic-process}
\end{align}
which are caused by the  flavor changing  $\Delta F=1$ operators
of
$[\bar E_L  \gamma_{\mu}   E_L][\bar D_L  \gamma_{\mu}   D_L]$.

Suppose the  $A_4$ modular symmetry on quarks and leptons, respectively, that is  $A^E_{4}\otimes A^Q_{4}$.
Then, those operators are given simply by the products of bilinear fermion operators; 
We can predict the correlations among processes in Eq.\,\eqref{semileptonic-process}
 by using the results of the previous subsections.
  
  The $b_L\to s_L$,  $b_L\to d_L$  and  $s_L\to d_L$ transitions
   have been discussed in 
   subsections \ref{Diag-S},  \ref{Diag-ST}  and  \ref{Diag-T}.
   The transition ratio of 
   $b\rightarrow s \, \bar e e$ to  $b\rightarrow s \, \bar\mu \mu$ 
   is given by the ratio of  $C_{eLeL}^{Sm}$ to $C_{\mu L \mu L}^{Sm}$,
   which are presented in Eq.\,\eqref{C-Se}.
    Since   $C_{eLeL}^{Sm}$ is  comparable to
     $C_{\mu L \mu L}^{Sm}$,
     the $b\rightarrow s \, \bar e e$ process is not suppressed
     compared with $b\rightarrow s \, \bar\mu \mu$. 
     Since the  $A_4$ modular symmetry controls the flavor structure of NP in our framework,
   the investigation  of the ratio of 
  $B\rightarrow K^{(*)} \, \bar\mu \mu$ to $B\rightarrow K^{(*)}\bar e e$ 
    provides an important test.
    Also $B\rightarrow \bar\mu \mu$ and $B\rightarrow  \bar e e$
    are interesting processes
    because those include $[\bar E_L E_R][\bar D_R   D_L]$
    operator.

   %

\section{4-quark operator with $\Delta F=2$ in $A_{4}$ symmetry} 
\label{sec:qqqq}

Now we discuss the left-handed 4-quark operator of which corresponding SMEFT operators $Q$ are given as  
	\begin{align}
[\,\bar D_L  \Gamma  D_L\,][\,\bar D_L  \Gamma  D_L\,] \notag 
	&: Q_{qq}^{(1)},Q_{qq}^{(3)}\,.
	\end{align}
In the modular symmetry, the $A_4$ flavor coefficient of the 4-quark operator
 $\bar LL\bar LL$  is not 
given  by the products of coefficients of the bilinear quark operators.
Therefore, we discuss
the 4-quark operators directly, which are obtained  by
 extracting  the $A_4$ singlet component from 
\begin{align}
[\,\bar D_L  \gamma_\mu (Y^*({\tau_q})Y(\tau_q)) D_L\,] [\,\bar D_L \gamma_\mu (Y^*({\tau_q})Y(\tau_q)) D_L\,]\,.
\end{align}
This is expressed by three parts:
	\begin{align}
	\left\{[\,\bar D_L  \gamma_\mu (Y^*({\tau_q})Y(\tau_q)) D_L\,] [\,\bar D_L \gamma_\mu (Y^*({\tau_q})Y(\tau_q)) D_L\,]\right \}_1
	=\Omega _1 + \Omega_2 +\Omega_3 \,,
	\end{align}
where $\Omega_1$, $\Omega_2$ and $\Omega_3$ are decomposed as:
\begin{align}
\Omega_1=
&\left [(\bar D_L  Y^*)_1  \gamma_\mu(Y D_L)_1
\oplus(\bar D_L   Y^*)_{1'}\gamma_\mu(Y D_L)_{1''}
\oplus(\bar D_L  Y^*)_{1''} \gamma_\mu(Y D_L)_{1'}\right]_1 
\nonumber\\
\times
&\left [(\bar D_L   Y^*)_1\gamma_\mu(Y D_L)_1
\oplus(\bar D_L  Y^*)_{1'} \gamma_\mu(Y D_L)_{1''}
\oplus(\bar D_L   Y^*)_{1''}\gamma_\mu(Y D_L)_{1'}\right]_1 
\nonumber\\
+
&\left [(\bar D_L   Y^*)_1\gamma_\mu(Y D_L)_{1'}
\oplus(\bar D_L   Y^*)_{1'}\gamma_\mu(Y D_L)_{1}
\oplus(\bar D_L   Y^*)_{1''}\gamma_\mu(Y D_L)_{1''}\right]_{1'} 
\nonumber\\
\times
&\left [(\bar D_L   Y^*)_1\gamma_\mu(Y D_L)_{1''}
\oplus(\bar D_L   Y^*)_{1''}\gamma_\mu(Y D_L)_{1}
\oplus(\bar D_L   Y^*)_{1'}\gamma_\mu(Y D_L)_{1'}\right]_{1''} \,, 
\nonumber\\
\Omega_2
=
&\left [\sum_{\oplus\, i,j=s,a} (\bar D_L   Y^*)_{3i}\gamma_\mu(Y D_L)_{3j}
\right]_{1} 
\left [\sum_{\oplus\, i',j'=s,a} (\bar D_L  Y^*)_{3i'} \gamma_\mu
(Y D_L)_{3j'}
\right]_{1} 
\nonumber\\
\nonumber\\
+
&\left [\sum_{\oplus\, i,j=s,a} (\bar D_L  Y^*)_{3i} \gamma_\mu(Y D_L)_{3j}
\right]_{1'} 
\left [\sum_{\oplus\, i',j'=s,a} (\bar D_L  Y^*)_{3i'} \gamma_\mu
(Y D_L)_{3j'}
\right]_{1''} \,,
\nonumber\\
\Omega_3=
&\left\{\sum_{\oplus\, I,J=s,a}\left[\sum_{\oplus\, i,j=s,a} (\bar D_L   Y^*)_{3i}\gamma_\mu(Y D_L)_{3j}
\right]_{3I} 
\left [\sum_{\oplus\, i',j'=s,a} (\bar D_L   Y^*)_{3i'}
\gamma_\mu(Y D_L)_{3j'}
\right]_{3J} \right\}_1
\,. 
\label{QQQQ-decomposition}
\end{align}
As seen in Eq. \eqref{QQQQ-decomposition},
 $\Omega_1$, $\Omega_2$ and $\Omega_3$  have 
$18$ terms,   $32$ terms and 
$64$ terms, respectively.
Finally, we have $114$ terms in this decompositions 
(see also Appendix \ref{decomposition}).

We find that the MFV  cannot be  reproduced if all terms
contribute to the FC processes.
Therefore, we present three cases,
 in which a specific parameter set dominates the 
   4-quark operator with $\Delta F=2$.

\subsection{4-quark operator with $\Delta F=2$ at nearby $\tau_q=i$} 

 As far as $\Omega_2$ and $\Omega_3$ dominate the FC processes,
 the  MFV cannot be reproduced without fine tuning of parameters.
 As a simple example, if the following terms dominate the FC processes;
\begin{align}
&s_{Q1x} [(\bar D_L Y^*)_{1'}\gamma_\mu (Y D_L )_{1} ]
[(\bar D_L Y^*)_{1'} \gamma_\mu(Y D_L )_{1'}]
\nonumber\\
+& s_{Q1y} [(\bar D_L Y^*)_{1'}\gamma_\mu (Y D_L )_{1''} ]
[(\bar D_L Y^*)_{1'}\gamma_\mu (Y D_L )_{1''}]\,,
\end{align}
where $s_{Q1x}$ and $s_{Q1y}$ are complex parameters,
we obtain simple results:
\begin{align}
&C_{sLbLsLbL}^{Sm}=
(4\sqrt{3}-7)Y_1(i)^2(9s_{23}^{d\,2}-6\sqrt{3}\epsilon_1^*s_{23}^d
+\epsilon_1^{*2})
(s_{Q1x}+4s_{Q1y})\,, \nonumber\\
&C_{dLbLdLbL}^{Sm}=
(4\sqrt{3}-7)Y_1(i)^2(s_{12}^{d})^2(9s_{23}^{d\,2}-6\sqrt{3}\epsilon_1^*s_{23}^{d\,2}
+\epsilon_1^{*2})
(s_{Q1x}+4s_{Q1y})\,, \nonumber\\
&C_{dLsLdLsL}^{Sm}=-4(4\sqrt{3}-7)Y_1(i)^2(s_{12}^{d})^2 |\epsilon_1|^4 
(s_{Q1x}+4s_{Q1y})\,,
\end{align}
where $C_{sLbLsLbL}^{Sm}$ is  the coefficient of 
the operator $[(\bar s_L \gamma_{\mu} b_L) (\bar s_L \gamma_{\mu} b_L) ]$,
and so on.
Then, we have a typical  relation of the MFV:
\begin{align}
\left |\frac{C_{dLbLdLbL}^{Sm}}{C_{sLbLsLbL}^{Sm}}
\right |^2=(s_{12}^d)^2=\left |-\frac{V^*_{td}}{V^*_{ts}} \right |^2\,,
\label{MFV-relation}
\end{align}
where Eq.\,\eqref{s12d} is put.

\subsection{4-quark operator with $\Delta F=2$ at nearby $\tau_q=\omega$}

As a simple example, if the following terms dominate the FC processes ;
 \begin{align}
s_{Q1z} [(\bar D_L Y_q^*(\tau))_{1''}\gamma_\mu (Y_q(\tau) D_L )_{1'} ]
[(\bar D_L Y_q^*(\tau))_{1''} \gamma_\mu(Y_q(\tau) D_L )_{1'}]\,,
\label{Q1z}
\end{align}
where $s_{Q1z}$ is a complex parameter,
we obtain simple results:
\begin{align}
&C_{sLbLsLbL}^{STm}=\frac{9}{16}Y_1(\omega)^2
 (3s_{23}^{d}-2\epsilon_1^*)^2 s_{Q1z}\,, \nonumber\\
&C_{dLbLdLbL}^{STm}=
(s_{12}^{d})^2\frac{9}{16}Y_1(\omega)^2
(3s_{23}^{d}-2\epsilon_1^*)^2 s_{Q1z}\,,  \nonumber\\
&C_{dLsLdLsL}^{STm}=(s_{12}^{d})^2\frac{1}{16}Y_1(\omega)^2
|3s_{23}^{d}-2\epsilon_1^*|^4 s_{Q1z}\,.
\end{align}
These give the MFV like relation in Eq.\,\eqref{MFV-relation}.

\subsection{4-quark operator with $\Delta F=2$  towards  $\tau_q=i\infty$}

As a simple example, if the following terms dominate the FC processes;
\begin{align}
s_{Q1z}^* [(\bar D_L Y_q^*(\tau))_{1'}\gamma_\mu (Y_q(\tau) D_L )_{1''} ]
[(\bar D_L Y_q^*(\tau))_{1'} \gamma_\mu(Y_q(\tau) D_L )_{1''}]\,,
\end{align}
where $s_{Q1z}$ is the same one in Eq.\,\eqref{Q1z},
we obtain simple results:
\begin{align}
&C_{sLbLsLbL}^{Tm}=
(s_{23}^{d}+\delta^*)^2 s_{Q1z}^*\,, \nonumber\\
&C_{dLbLdLbL}^{Tm}=(s_{12}^{d})^2
(s_{23}^{d}+\delta^*)^2 s_{Q1z}^*\,,   \nonumber\\
&C_{dLsLdLsL}^{Tm}=(s_{12}^{d})^2
|s_{23}^{d}+\delta^*|^4 s_{Q1z}^*\,.
\end{align}
These also  give MFV like relation in Eq.\,\eqref{MFV-relation}.
\section{Summary}
\label{sec:summary}

We have studied the modular symmetric standard-model effective field theory. 
We have employed the stringy Ansatz on the coupling structure that 
4-point couplings $y^{(4)}$ of matter fields are written by a product of 3-point couplings $y^{(3)}$ of matter fields, i.e., $y^{(4)} = y^{(3)}y^{(3)}$.
In this framework,
 we have discussed the flavor structure of bilinear fermion operators
 and 4-fermion operators.

 In order to cover many modular flavor models, we take a setup that the $A_4$ modular flavor symmetry in the lepton sector 
 is independent of the $A_4$ symmetry in the quark sector, i.e., $A^E_{4}\otimes A^Q_{4}$ symmetry.
 They have two independent moduli, $\tau_q$ and $\tau_e$ for the quark sector and the lepton sector, respectively.
 Moreover,
  we take the  holomorphic and anti-holomorphic modular forms 
  with weight 2, which  couples to quarks and leptons.

   From the viewpoint of the Ansatz  $y^{(4)}=y^{(3)}y^{(3)}$,  
  the $A_4$ modular-invariant semileptonic 4-fermion  operator 
  $[\,\bar E_R \Gamma  E_R][\,\bar D_R  \Gamma  D_R\,]$ 
  does not lead to the FC processes
   since this operator would be 
  constructed in terms of gauge couplings $g$ as $y^{(3)} \sim g$.
  
  The bilinear operator $[\,\bar D_R \Gamma  D_L\,]$ 
  also does not lead FC if the mediated mode   corresponds to the Higgs
   boson $H_d$ in the viewpoint of the Ansatz.
  In this case, the flavor structure of this operator is  the exactly same as the mass matrix.
 Then, the bilinear operator matrix is diagonal in the basis for mass eigenstates.
  The FC processes such $b \to s$, $b \to d$, $s \to d$ never happen. 
On the other hand, if the flavor structure of the operator is not  the exactly same as the mass matrix, the situation would change drastically.
 Then,
 we have obtained the non-trivial relations of the FC transitions
  at nearby fixed points $\tau=i,\,\omega\,, i\infty$, which are testable in the future.
  
   As an application, we have  discussed the LFV processes,
   $\mu\to e \gamma$,  $\tau\to e \gamma$ and  $\tau\to \mu \gamma$.
   We have estimated
   the branching ratios by using $A_4$ flavor coefficients 
   at nearby $\tau_e=i$
   since  the successful lepton mass matrix
   has been obtained there.
   We have predicted 
   $  {\rm BR}(\mu\to e\gamma)\geqq {\rm BR}(\tau\to e\gamma)\gg
   {\rm BR}(\tau\to \mu\gamma)$ at nearby $\tau_e=i$.
   This prediction is different from the one of the $U(2)$
   flavor symmetry.

We also have studied the flavor changing 4-quark operators 
in the $A_4$ modular symmetry of quarks. Then, the MFV could be realized  by taking relevant  specific parameter sets
 of order one.
 
 The flavor symmetry is a useful way for a systematic NP analysis of the SMEFT. 
The finite modular group is one of the attractive choice for the flavor symmetry, 
and the characteristic predictions are derived.
This application on the SMEFT would be a first step of a systematic analysis for the modular symmetric model, 
connecting quark and lepton flavor observables.


\subsubsection*{Acknowledgments}

We thank Hiroshi Okada for useful discussions.
This work was supported by 
JSPS KAKENHI Grant Numbers JP19J00664 (HO), JP20K14477 (HO), JP21K13923 (KY), and 
IBS under the project code, IBS-R018-D1 (HO).

\newpage
\noindent
{\LARGE \bf Appendix}
\appendix

\section{SMEFT operators} \label{App:SMEFT}
\newcommand{\OpScale}{.8} 
\begin{table}[h]
\footnotesize
		\centering
		\renewcommand{\arraystretch}{1.3}
		\scalebox{\OpScale}{
			\begin{tabular}{c|  lc | lc } 
				\multicolumn{5}{c}{\bf{Class 5--7: Fermion Bilinears}} \\ [.1cm] \hline
				\multicolumn{1}{c}{} & 
				\multicolumn{4}{c}{$(\bar L R)$} \\ \hline
				\multicolumn{1}{c}{} & 
				\multicolumn{2}{c}{5: $\psi^2 H^3 +$ h.c.} &
				\multicolumn{2}{c}{6: $\psi^2 X H +$ h.c.} \\ \hline
				$(\bar \ell e)$ 
				& $Q_{eH}$ 
				& $(H^\dagger H)(\bar\ell_p e_r H)$ 
				& $Q_{eW}$ 
				& $(\bar\ell_p \sigma^{\mu\nu}e_r)\tau^I H W_{\mu\nu}^I$ \\ 
				
				&  
				&  
				& $Q_{eB}$ 
				& $(\bar\ell_p \sigma^{\mu\nu}e_r) H B_{\mu\nu}$ \\ 
				\hline
				$(\bar q u)$ 
				& $Q_{uH}$ 
				& $(H^\dagger H)(\bar q_p u_r \tilde{H})$ 
				& $Q_{uG}$ 
				& $(\bar q_p \sigma^{\mu\nu}T^A u_r)\tilde{H}G_{\mu\nu}^A$ \\ 
				
				&  
				&  
				& $Q_{uW}$ 
				& $(\bar q_p \sigma^{\mu\nu}u_r)\tau^I \tilde{H}W_{\mu\nu}^I$ \\ 
				
				&  
				&  
				& $Q_{uB}$ 
				& $(\bar q_p \sigma^{\mu\nu}u_r)\tilde{H}B_{\mu\nu}$ \\ 
				\hline
				$(\bar q d)$ 
				& $Q_{dH}$ 
				& $(H^\dagger H)(\bar q_p d_r H)$
				& $Q_{dG}$ 
				& $(\bar q_p \sigma^{\mu\nu}T^A d_r)H G_{\mu\nu}^A$ \\ 
				
				&  
				&  
				& $Q_{dW}$ 
				& $(\bar q_p \sigma^{\mu\nu}d_r)\tau^I H W_{\mu\nu}^I$ \\
				
				&  
				&  
				& $Q_{dB}$ 
				& $(\bar q_p \sigma^{\mu\nu}d_r) H B_{\mu\nu}$ \\
				\hline 
			\end{tabular}
		} \\
		\vspace{0.2cm}
		\scalebox{\OpScale}{
			\begin{tabular}{c| cc| cc| cc}
				\hline \multicolumn{7}{c}{7: $\psi^2 H^2 D$} \\ \hline
				\multicolumn{1}{c|}{} & 
				\multicolumn{2}{c|}{$(\bar L L)$} & 
				\multicolumn{2}{c|}{$(\bar R R)$} & 
				\multicolumn{2}{c}{$(\bar R R^\prime)$} \\ \hline
				lepton &
				$Q_{H\ell}^{(1)}$ & $(H^\dagger i \overleftrightarrow{D}_\mu H)(\bar\ell_p \gamma^\mu \ell_r)$ & 
				$Q_{H e}$ & $(H^\dagger i \overleftrightarrow{D}_\mu H)(\bar e_p \gamma^\mu e_r)$ & 
				& \\
				&
				$Q_{H\ell}^{(3)}$ & $(H^\dagger i \overleftrightarrow{D}_\mu^I H)(\bar\ell_p \tau^I\gamma^\mu \ell_r)$ & 
				& & 
				& \\
				\hline
				quark &
				$Q_{Hq}^{(1)}$ & $(H^\dagger i \overleftrightarrow{D}_\mu H)(\bar q_p \gamma^\mu q_r)$ & 
				$Q_{H u}$ & $(H^\dagger i \overleftrightarrow{D}_\mu H)(\bar u_p \gamma^\mu u_r)$ & 
				$Q_{Hud}$ + h.c. & $i(\tilde{H}^\dagger D_\mu H)(\bar u_p \gamma^\mu d_r)$\\
				&
				$Q_{Hq}^{(3)}$ & $(H^\dagger i \overleftrightarrow{D}_\mu^I H)(\bar q_p \tau^I\gamma^\mu q_r)$ & 
				$Q_{H d}$ & $(H^\dagger i \overleftrightarrow{D}_\mu H)(\bar d_p \gamma^\mu d_r)$
				& 
				& \\ \hline
			\end{tabular}
		}
		\vspace{0.4cm}
		\newline \centering
		\scalebox{\OpScale}{
			\begin{tabular}{c| lc | lc | lc }
				\multicolumn{7}{c}{\bf{Class 8: Fermion Quadrilinears}} \\[.1cm] \hline
				\multicolumn{1}{c}{} & 
				\multicolumn{2}{c|}{$(\bar L L)(\bar L L)$} & 
				\multicolumn{2}{c|}{$(\bar R R)(\bar R R)$} & 
				\multicolumn{2}{c}{$(\bar L L)(\bar R R)$} \\ \hline
				semileptonic &
				$Q_{\ell q}^{(1)}$  & $(\bar\ell_p \gamma_\mu \ell_r)(\bar q_s \gamma^\mu q_t)$ & 
				$Q_{eu}$  & $(\bar e_p \gamma_\mu e_r)(\bar u_s \gamma^\mu u_t)$ & 
				$Q_{\ell u}$  & $(\bar\ell_p \gamma_\mu \ell_r)(\bar u_s \gamma^\mu u_t)$ \\
				&
				$Q_{\ell q}^{(3)}$  & $(\bar\ell_p \gamma_\mu \tau^I \ell_r)(\bar q_s \gamma^\mu \tau^I q_t)$ & 
				$Q_{ed}$  & $(\bar e_p \gamma_\mu e_r)(\bar d_s \gamma^\mu d_t)$ & 
				$Q_{\ell d}$  & $(\bar\ell_p \gamma_\mu \ell_r)(\bar d_s \gamma^\mu d_t)$ \\
				&
				&& 
				& & 
				$Q_{q e}$  & $(\bar q_p \gamma_\mu q_r)(\bar e_s \gamma^\mu e_t)$ \\ 
				\hline
				4-quark &
				$Q_{qq}^{(1)}$  & $(\bar q_p \gamma_\mu q_r)(\bar q_s \gamma^\mu q_t)$ & 
				$Q_{uu}$  & $(\bar u_p \gamma_\mu u_r)(\bar u_s \gamma^\mu u_t)$ & 
				$Q_{qu}^{(1)}$  & $(\bar q_p \gamma_\mu q_r)(\bar u_s \gamma^\mu u_t)$  \\
				&
				$Q_{qq}^{(3)}$  & $(\bar q_p \gamma_\mu \tau^I q_r)(\bar q_s \gamma^\mu \tau^I q_t)$ 
				& 
				$Q_{dd}$  & $(\bar d_p \gamma_\mu d_r)(\bar d_s \gamma^\mu d_t)$ & 
				$Q_{qu}^{(8)}$  & $(\bar q_p \gamma_\mu T^A q_r)(\bar u_s \gamma^\mu T^A u_t)$ \\
				& 
				& 
				& 
				$Q_{ud}^{(1)}$  & $(\bar u_p \gamma_\mu u_r)(\bar d_s \gamma^\mu d_t)$ & 
				$Q_{qd}^{(1)}$  & $(\bar q_p \gamma_\mu q_r)(\bar d_s \gamma^\mu d_t)$ \\
				& 
				& 
				& 
				$Q_{ud}^{(8)}$  & $(\bar u_p \gamma_\mu T^A u_r)(\bar d_s \gamma^\mu T^A d_t)$ & 
				$Q_{qd}^{(8)}$  & $(\bar q_p \gamma_\mu T^A q_r)(\bar d_s \gamma^\mu T^A d_t)$ \\
				\hline
				4-lepton& 
				$Q_{\ell\ell}$  & $(\bar\ell_p \gamma_\mu \ell_r)(\bar\ell_s \gamma^\mu \ell_t)$ & 
				$Q_{ee}$  & $(\bar e_p \gamma_\mu e_r)(\bar e_s \gamma^\mu e_t)$ & 
				$Q_{\ell e}$  & $(\bar\ell_p \gamma_\mu \ell_r)(\bar e_s \gamma^\mu e_t)$ \\ \hline
			\end{tabular}
		}
		\vspace{0.2cm}
		\newline\centering
		\scalebox{\OpScale}{
			\begin{tabular}{c|  lc | lc } \hline
				\multicolumn{1}{c}{} & 
				\multicolumn{2}{c|}{$(\bar L R)(\bar R L)$ + h.c.} & 
				\multicolumn{2}{c}{$(\bar L R)(\bar L R)$ + h.c.} \\ \hline
				semi-leptonic&
				$Q_{\ell e d q }$  & $(\bar \ell _p^j e_r)(\bar d_s q_{tj})$ & 
				$Q_{\ell equ}^{(1)}$  & $(\bar \ell_p^j e_r)\epsilon_{jk}(\bar q_s^k u_t)$ \\
				\hline
				4-quark &
				& & 
				$Q_{quqd}^{(1)}$  & $(\bar q_p^j u_r)\epsilon_{jk}(\bar q_s^k d_t)$ \\ 
				&
				& & 
				$Q_{quqd}^{(8)}$  & $(\bar q_p^j T^A u_r)\epsilon_{jk}(\bar q_s^k T^A d_t)$ \\ 
				&
				& & 
				$Q_{\ell equ}^{(3)}$  & $(\bar \ell_p^j \sigma_{\mu\nu} e_r)\epsilon_{jk}(\bar q_s^k \sigma^{\mu\nu} u_t)$ \\ \hline 
			\end{tabular}
		}
		\caption{List of all fermionic SMEFT operators in the Warsaw basis~\cite{Grzadkowski:2010es}. 
			The division in classes is adopted from Ref. \cite{Alonso:2013hga}. 
			The $p,r,s,t$ are flavor index, and $j,k$ stand for SU(2) index.
			The operator classes 1--4 without fermion fields are irrelevant in this paper, and not listed here.
			\label{tab:SMEFTOpe}
		}
\normalsize
\end{table}

\newpage
\section{A model of  mass matrices in the $A_4$ modular symmetry}
\label{massmatrixmodel}
We present a viable model for quarks \cite{Okada:2020ukr}.
Suppose that 
three left-handed quark doublets are of a triplet of the $A_4$ group.
The three right-handed quarks are three different singlets of $A_4$.
On the other hand, the Higgs doublets are supposed to be singlets of $A_4$.
The generic assignments of representations and modular weights to 
the minimal supersymmetric standard model (MSSM) fields
are presented in Table \ref{quark-model}, where weight 2 and 6 modular forms are presented. 

\begin{table}[H]
	\centering
	\begin{tabular}{|c||c|c|c|c|ccc|} \hline
		&$Q_L$&$(d^c_R,\,s^c_R,\,b^c_R)$& $(u^c_R,\,c^c_R,\,t^c_R)$&$H_q$&
		$Y_{\bf 3}^{(2)}$
		&$ Y_{\bf 3}^{(6)}$
		&$Y_{\bf 3'}^{(6)}$ \\  \hline\hline 
		\rule[14pt]{0pt}{0pt}
		$SU(2)$&$\bf 2$&$\bf 1$&$\bf 1$&$\bf 2$& \multicolumn{3}{c|}{$\bf 1$} \\
		$A_4$&$\bf 3$& \bf (1,\ 1$''$,\ 1$'$)&\bf (1,\ 1$''$,\ 1$'$)&$\bf 1$
		& \multicolumn{3}{c|}{$\bf 3$} \\
		$k$ & $2$ &$(0,\ 0,\ 0)$& $(4,4,4)$ &0& $2$
		& $6$ & $6$ \\
		\hline
	\end{tabular}
	\caption{An example of assignments of  weights
		$k$ for quarks and  modular forms.
	}
	\label{quark-model}
\end{table}
Then, up-type $M_u$ and down-type $M_d$ quark mass matrices are given, respectively as:
\footnotesize
\begin{align}
M_u&=v_u\left [
\begin{pmatrix}
\alpha_{u(m)} & 0 & 0 \\
0 &\beta_{u(m)} & 0\\
0 & 0 &\gamma_{u(m)}
\end{pmatrix} 
\begin{pmatrix}
Y_1^{(6)} & Y_3^{(6)}& Y_2^{(6)} \\
Y_2^{(6)} & Y_1^{(6)} &  Y_3^{(6)} \\
Y_3^{(6)} &  Y_2^{(6)}&  Y_1^{(6)}
\end{pmatrix}
+ 
\begin{pmatrix}
\alpha'_{u(m)} & 0 & 0 \\
0 &\beta'_{u(m)} & 0\\
0 & 0 &\gamma'_{u(m)}
\end{pmatrix}
\begin{pmatrix}
Y_1^{'(6)} & Y_3^{'(6)}& Y_2^{'(6)} \\
Y_2^{'(6)} & Y_1^{'(6)} &  Y_3^{'(6)} \\
Y_3^{'(6)} &  Y_2^{'(6)}&  Y_1^{'(6)}
\end{pmatrix}
\right ]_{RL} \nonumber\\
&=v_u
\begin{pmatrix}
\alpha_{u(m)} & 0 & 0 \\
0 &\beta_{u(m)} & 0\\
0 & 0 &\gamma_{u(m)}
\end{pmatrix}
\begin{pmatrix}
\tilde Y_1^{(6)} & \tilde Y_3^{(6)}& \tilde Y_2^{(6)} \\
\tilde Y_2^{(6)} & \tilde Y_1^{(6)} &  \tilde Y_3^{(6)} \\
\tilde Y_3^{(6)} &  \tilde Y_2^{(6)}&  \tilde Y_1^{(6)}
\end{pmatrix}_{RL}\,,  
\nonumber \\
M_d&= v_d
\begin{pmatrix}
\alpha_{d(m)} & 0 & 0 \\
0 &\beta_{d(m)} & 0\\
0 & 0 &\gamma_{d(m)}
\end{pmatrix}
\begin{pmatrix}
Y_1 & Y_3& Y_2\\
Y_2 & Y_1 &  Y_3 \\
Y_3 &  Y_2&  Y_1
\end{pmatrix}_{RL}\,,  
\label{quark-mass-matrix}
\end{align} 
\normalsize
where $\tilde Y_i^{(6)}\equiv Y_i^{(6)}+g_i Y_i^{'(6)}$ with 
$g_{u1}=\alpha'_{u(m)}/\alpha_{u(m)}$, $g_{u2}=\beta'_{u(m)}/\beta_{u(m)}$, $g_{u3}=\gamma'_{u(m)}/\gamma_{u(m)}$ and $g_q\equiv \alpha_{q(m)}'/\alpha_{q(m)}$.
The VEV of the Higgs field $H_q$ is denoted by  $ v_q$.
Parameters $\alpha_q$,  $\beta_q$,  $\gamma_q$  can be taken to be  real,
on the other hand, $g_{u1}$, $g_{u2}$, $g_{u3}$ and  $g_{u}$ are  complex parameters.


For the lepton sector, we also present 
mass matrices of a successful model \cite{Okada:2020brs}, where
the neutrino mass matrix is given 
in terms of weight 4  modular forms by using  Weinberg operator.
The  assignments of representations and modular weights to the lepton fields
are presented in Table \ref{tb:lepton}. 
\begin{table}[h]
	\centering
	\begin{tabular}{|c||c|c|c|cc|} \hline
		\rule[14pt]{0pt}{1pt}
		&$L_L$&$(e^c_R,\mu^c_R,\tau^c_R)$&$H_q$
		&$Y_{\bf r}^{\rm (2)}$ & $Y_{\bf r}^{\rm (4)}$ \\  \hline\hline 
		\rule[14pt]{0pt}{1pt}
		$SU(2)$&$\bf 2$&$\bf 1$&$\bf 2$ & \multicolumn{2}{c|}{$\bf 1$}\\
		\rule[14pt]{0pt}{1pt}
		$A_4$&$\bf 3$& \bf (1,\ 1$''$,\ 1$'$)&$\bf 1$ & $\bf 3$ &$\bf \{3, 1, 1'\} $ \\
		\rule[14pt]{0pt}{1pt}
		$k$& $2$ &$(0,\ 0,\ 0)$ &0 & $2$ & $4$ \\ \hline
	\end{tabular}	
	\caption{ Representations and  weights
		$k$ for MSSM fields and  modular forms of weight $2$ and $4$.
	}
	\label{tb:lepton}
\end{table}

The charged lepton mass matrix and neutrino ones are given as:
\begin{align}
&M_E=v_d
\begin{pmatrix}
\alpha_{e(m)} & 0 & 0 \\
0 &\beta_{e(m)} & 0\\
0 & 0 &\gamma_{e(m)}
\end{pmatrix} 
\begin{pmatrix}
Y_1 & Y_3 & Y_2 \\
Y_2 & Y_1 &  Y_3 \\
Y_3&  Y_2&  Y_1
\end{pmatrix}_{RL}, \nonumber\\
&M_\nu=\frac{v_u^2}{\Lambda} \left [
\begin{pmatrix}
2Y_1^{(4)} & -Y_3^{(4)} & -Y_2^{(4)}\\
-Y_3^{(4)} & 2Y_2^{(4)} & -Y_1^{(4)} \\
-Y_2^{(4)} & -Y_1^{(4)} & 2Y_3^{(4)}
\end{pmatrix}
+g^{\nu}_1 Y_{\bf 1}^{\rm (4)} 
\begin{pmatrix}
1 & 0 &0\\ 0 & 0 & 1 \\ 0 & 1 & 0
\end{pmatrix}
+g^{\nu}_2 Y_{\bf 1'}^{\rm (4)}
\begin{pmatrix}
0 & 0 &1\\ 0 & 1 & 0 \\ 1 & 0 & 0
\end{pmatrix}
\right ] \, ,
\label{ME222}
\end{align}
respectively,
where  $\alpha_{e(m)}$, $\beta_{e(m)}$ and $\gamma_{e(m)}$ are real parameters
and  $g_1^\nu$,  $g_2^\nu$ are supposed to be real.

\section{Unitary transformation of  $S$ and  $ST$ and mass matrix}
\label{appe-base}
The mass matrix is transformed by the unitary transformation,
which transforms the generator $S$ and  $ST$.
We discuss details at the fixed points $\tau=i$ and $\tau=\omega$.

\subsection{Diagonal base of $S$ for $A_4$ triplet}
\label{appe-S}
The generators of $A_4$ group for the triplet are:
\begin{align}
\begin{aligned}
S=\frac{1}{3}
\begin{pmatrix}
-1 & 2 & 2 \\
2 &-1 & 2 \\
2 & 2 &-1
\end{pmatrix},
\end{aligned}
\qquad 
\begin{aligned}
T=
\begin{pmatrix}
1 & 0& 0 \\
0 &\omega& 0 \\
0 & 0 & \omega^2
\end{pmatrix}, 
\end{aligned}
\label{STbase}
\end{align}
where $\omega=e^{i\frac{2}{3}\pi}$ for a triplet.
The eigenvalues of $S$ is $(1\,,-1\,,-1)$. 
This is the diagonal base of $T$.

In order to present the mass matrices in the diagonal base of $S$,
we move to the diagonal base of $S$ as follows:
\begin{align}
\begin{aligned}
U_{S1}\, { S}\, U_{S1}^\dagger=
\begin{pmatrix} 
-1 & 0 & 0 \\
0 & 1 & 0 \\
0 &0 &-1
\end{pmatrix},  \quad 
U_{S2}\, { S}\, U_{S2}^\dagger=
\begin{pmatrix} 
1 & 0 & 0 \\
0 & -1 & 0 \\
0 &0 &-1
\end{pmatrix},  \quad
U_{S3}\, { S} \, U_{S3}^\dagger=
\begin{pmatrix} 
-1 & 0 & 0 \\
0 & -1 & 0 \\
0 &0 &1
\end{pmatrix}, 
\end{aligned}
\end{align}
where
\begin{align}
\begin{aligned}
U_{Si}\equiv P_i\begin{pmatrix} 
\ \frac{2}{\sqrt{6}} & -\frac{1}{\sqrt{6}} & -\frac{1}{\sqrt{6}} \\
\frac{1}{\sqrt{3}} &\frac{1}{\sqrt{3}} & \frac{1}{\sqrt{3}} \\
0 &-\frac{1}{\sqrt{2}} &\ \frac{1}{\sqrt{2}}
\end{pmatrix},
\quad P_1=
\begin{pmatrix} 
1 & 0 & 0 \\
0 & 1 & 0 \\
0 &0 &1
\end{pmatrix}, \ \ 
P_2=
\begin{pmatrix} 
0 & 1 & 0 \\
1 & 0 & 0 \\
0 &0 &1
\end{pmatrix}, \ \
P_3=
\begin{pmatrix} 
1 & 0 & 0 \\
0 & 0 & 1 \\
0 &1 &0
\end{pmatrix} \,.
\end{aligned}
\label{Sdiagonal}
\end{align}
Then, the  generator $T$ is not anymore diagonal.
However, the eigenvalue $-1$ of $S$ is degenerated, 
there is a freedom of the rotation between   corresponding 
rows and between  columns.

As seen in subsection \ref{appe-massmatrix}, the Dirac mass matrix $M_{RL}$ is transformed as:
\begin{align}
\hat M_{RL}=  M_{RL} U_{Si}^\dagger\,, 
\label{diagonalbaseS}
\end{align}
If there is a residual symmetry of $A_4$ in the Dirac mass matrix, 
$\mathbb{Z}_2^{S}=\{  I, S \}$,
which is realized at the fixed point $\tau=i$
as presented in Eq.\,\ref{fixed-points},
the generator $S$ commutes with $\hat M_{RL}^\dagger\hat  M_{RL}$,
\begin{align}
\begin{aligned}
\left [\hat M_{RL}^\dagger \hat M_{RL}\, ,\,  S\right ]=0 \, .
\end{aligned}
\end{align}
Therefore, the mass matrix could  be diagonal
in the diagonal base of $S$.

\subsection{Diagonal bases of $ST$ and $T$ for $A_4$ triplet }
\label{appe-ST}

We can move $ST$ to the diagonal base  by the 
six-type unitary transformations $V_{ST}$
as follows:
\begin{align}
\begin{aligned}
U_{STi}\, {ST} \, U_{STi}^\dagger=P_i
\begin{pmatrix} 
\omega^2 & 0 & 0 \\
0 & \omega & 0 \\
0 &0 &1
\end{pmatrix}P_i^T, 
\end{aligned} 
\end{align}
where
\begin{align}
\begin{aligned}
U_{STi}=\frac{1}{3}P_i
\begin{pmatrix} 
-2 \omega^2  & -2 \omega & 1 \\
- \omega^2  & \ 2 \omega & 2 \\
\ 2 \omega^2 &- \omega &2
\end{pmatrix}, \ \ 
P_4=
\begin{pmatrix} 
0  & 0 & 1 \\
0  &1 & 0 \\
1  & 0 & 0
\end{pmatrix}, \ \
P_5=
\begin{pmatrix} 
0  & 0 & 1 \\
1  &0 & 0 \\
0  & 1 & 0
\end{pmatrix},\ \
P_6=
\begin{pmatrix} 
0  & 1 & 0 \\
0  &0 & 1 \\
1  & 0 & 0
\end{pmatrix},
\end{aligned}
\label{STdiagonal}
\end{align}
and $P_i(i=1,2,3)$ are given in Eq.\,(\ref{Sdiagonal}).
The order of eigenvalues of $ST$ depends on $P_i$. We have 
eigenvalues 
$\{\omega^2,\, \omega,\, 1 \}$ for $P_1$, $\{\omega,\, \omega^2,\, 1 \}$ for $P_2$,
$\{\omega^2,\, 1, \, \omega\}$ for $P_3$,  
$\{1,\, \omega,\, \omega^2\}$ for $P_4$,
$\{1,\, \omega^2,\, \omega\}$ for $P_5$
and  $\{\omega,\, 1, \, \omega^2\}$ for $P_6$, respectively.
Thus, there are independent six unitary transformations
to move  $ST$ to the diagonal base.

As seen in subsection \ref{appe-massmatrix}, the Dirac mass matrix $M_{RL}$ is transformed as:
\begin{align}
\hat M_{RL}=  M_{RL} U_{STi}^\dagger\,. 
\label{diagonalbaseST}
\end{align}
If there exists the residual symmetries of the $A_4$
group $\mathbb{Z}_3^{ST}=\{  I, ST, (ST)^2 \}$ 
which is realized at the fixed point $\tau=\omega$ 
as presented  in  Eq.\,\eqref{fixed-points},
we have
\begin{align}
\begin{aligned}
\left [\hat M_{RL}^\dagger \hat M_{RL}\, ,\,  ST\right ]=0 \, ,
\end{aligned}
\end{align} 
which leads to the diagonal  $\hat M_{RL}^\dagger \hat M_{RL}$
because $ST$ has three different eigenvalues.

On the other hand, the generator $T$ is already diagonal in the original base of Eq.\,(\ref{STbase}).
If there exists the residual symmetries of the $A_4$
group 
$\mathbb{Z}_3^{T}=\{  I, T, T^2 \}$, which is realized at $\tau=i\infty$,
\begin{align}
\begin{aligned}
\left [ M_{RL}^\dagger  M_{RL}\, ,\,  T\right ]=0 \, ,
\end{aligned}
\end{align} 
which also gives the diagonal  $ M_{RL}^\dagger  M_{RL}$.
\subsection{Mass matrix in the new bases of generators $S$ and $ST$}
\label{appe-massmatrix}
Define the new basis of generators, $\tilde G\ (G=S,ST)$ 
by a unitary transformation $U$ as:
\begin{align}
\begin{aligned}
\hat G= U G U^\dagger  \  ,
\end{aligned}
\end{align}
where $\hat G$, $G$ and $U$ are $3\times 3$ matrices.
Since the $A_4$ triplet transforms under the unitary  transformation as:
\begin{align}
\begin{pmatrix}
a_1\\
a_2\\
a_3
\end{pmatrix}_{\bf 3}
\to G
\begin{pmatrix}
a_1\\
a_2\\
a_3
\end{pmatrix}_{\bf 3}
= U^\dagger \hat G U
\begin{pmatrix}
a_1\\
a_2\\
a_3
\end{pmatrix}_{\bf 3} \ .
\end{align}
Thus, in the new basis,  the $A_4$ triplet transforms as:
\begin{align} 
\begin{pmatrix}
\hat a_1\\
\hat a_2\\
\hat a_3
\end{pmatrix}_{\bf 3}
\to \hat G \  
\begin{pmatrix}
\hat a_1\\
\hat a_2\\
\hat a_3
\end{pmatrix}_{\bf 3},
\end{align}
where 
\begin{align} 
\begin{pmatrix}
\hat a_1\\
\hat a_2\\
\hat a_3
\end{pmatrix}_{\bf 3}
= U
\begin{pmatrix}
a_1\\
a_2\\
a_3
\end{pmatrix}_{\bf 3}.
\end{align}

Let us rewrite the Dirac mass matrix $M_{RL}$
in the new base ($\hat G$) of
the triplet left-handed fields.
Denoting $L$ and $\hat L$ to be  triplets of the left-handed fields in the base of $G$ and $\hat G$, respectively, and $R$ to be right-handed singlets,
the Dirac mass matrix is written as:
\begin{align}
\begin{aligned}
\bar R M_{RL} L=\bar R M_{RL} U^\dagger  \hat L  \,,
\end{aligned}
\end{align}
where 
\begin{align}
L=U^\dagger \hat L\, .
\label{newSbase}
\end{align}
Then, the Dirac mass matrix $\hat M_{RL}$ in the new base is given as:
\begin{align}
\hat M_{RL}=  M_{RL}U^\dagger \, .
\label{newmassmatrix}
\end{align}

\section{Mass matrix at nearby $\tau=i$}
\label{massmatrix-i}
Both  down-type quark mass matrix of  Eq.\,(\ref{quark-mass-matrix})
and the charged lepton mass matrix of Eq.\,(\ref{ME222}) 
are given in terms of weight $2$ modular forms.

Perform  the unitary transformation
of $D_L\to U_S D_L$, where   $ U_S=U_{23}(15^\circ) U_{S2}$ 
(see Appendix \ref{appe-massmatrix}) 
\begin{align}
U_S =U_{23}(-15^\circ) U_{S2}=
\begin{pmatrix}
1  &0 & 0\\
0& \cos 15^\circ&  \sin 15^\circ \\ 
0& -\sin 15^\circ&  \cos 15^\circ \\ 
\end{pmatrix}\,U_{S2}=
\frac{1}{2\sqrt{3}}
\begin{pmatrix}
2  & 2 & 2\\
\sqrt{3}+1  &  -2 & \sqrt{3}-1\\
\sqrt{3}-1 & -2  & \sqrt{3}+1
\end{pmatrix}\,.
\label{U-S0}
\end{align}

Then, the down-type quark mass matrix at $\tau=i$
in Eq.\,(\ref{quark-mass-matrix}) is simply given as:
\begin{align} 
&M_{d}\,=\frac12 v_d
\begin{pmatrix}
0& 3(\sqrt{3}-1) \tilde\alpha_{d(m)} &-(3-\sqrt{3}) \tilde\alpha_{d(m)}\\
0 & -3(\sqrt{3}-1) \tilde\beta_{d(m)} & -(3-\sqrt{3}) \tilde\beta_{d(m)}\\
0 &0& 2(3-\sqrt{3})\tilde\gamma_{d(m)}
\end{pmatrix}_{RL}\,, \nonumber\\
&M_{d}^\dagger M_{d}=\frac12 v_d^2
\begin{pmatrix}
0  & 0 & 0\\
0  &  9(2-\sqrt{3}) (\tilde\alpha_{d(m)}^2+\tilde\beta_{d(m)}^2) &
3(3-2\sqrt{3}) (\tilde\alpha_{d(m)}^2-\tilde\beta_{d(m)}^2)\\
0 &  3(3-2\sqrt{3}) (\tilde\alpha_{d(m)}^2-\tilde\beta_{d(m)}^2) & 
3(2-\sqrt{3}) (\tilde\alpha_{d(m)}^2+\tilde\beta_{d(m)}^2+4\tilde\gamma_{d(m)}^2)
\end{pmatrix}_{LL}\,,
\label{downmassmatrix-S}
\end{align}
where $\tilde \alpha_{d(m)}=(6-3\sqrt{3})  Y_1(i)\alpha_{d(m)}$,
$\tilde \beta_{d(m)}= (6-3\sqrt{3})  Y_1(i) \beta_{d(m)}$ and 
$\tilde \gamma_{d(m)} =(6-3\sqrt{3}) Y_1(i)  \gamma_{d(m)}$
and 
$\tilde \gamma_{d(m)}$ is supposed to be  much larger than  $\tilde \alpha_{d(m)}$
and  $\tilde \beta_{d(m)}$.  
The rotation $U_{23}(-15^\circ)$ realizes that the (3,3) entry is  much large than other entries.
Since two eigenvalues of $S$ are degenerate
such as $(1,-1,-1)$, there is still a freedom of the  $2$--$3$ family rotation.
Therefore, $M_{d}^\dagger M_{d}$ could be diagonal
after the small $2$--$3$ family rotation of
${\cal O}(\tilde \alpha_{d(m)}^2\tilde/ \gamma_{d(m)}^2,\,\tilde \beta_{d(m)}^2/\tilde \gamma_{d(m)}^2)$.
The charged lepton mass matrix is the same one 
in Eq.\,(\ref{downmassmatrix-S}) by replacing
the subscript $d$ with $e$.


Since the quark and lepton mass matrices cannot reproduce
the observed CKM and PMNS matrices at fixed points as discussed in Ref.\cite{Okada:2020ukr}.
Therefore, the deviations from the fixed points are required
to realize observed masses and mixing angles.

Let us consider the small deviation from $\tau=i$.
By using the approximate modular forms of weight 2
at  $\tau=i+\epsilon$, we present
$M_d^\dagger M_d$ 
including of  order $\epsilon$.
By the   unitary transformation $D_L\to U_{12}  U_{S} D_L$, where $U_{12}$ is the first-second family exchange, 
\begin{align}
U_{12}=
\begin{pmatrix}
0 & 1 & 0\\  1& 0 & 0 \\ 0 & 0 &1
\end{pmatrix}\,,
\label{U12}
\end{align}
the mass matrix of Eq.\,(\ref{downmassmatrix-S}) is modified as:    
\scriptsize
\begin{align} 
M_{d}^\dagger M_{d}\simeq v_d^2
\begin{pmatrix}
\frac92 (2-\sqrt{3}) (\tilde\alpha_{d(m)}^2+\tilde\beta_{d(m)}^2) 
& 3(\sqrt{3}-2) (\tilde\alpha_{d(m)}^2-\tilde\beta_{d(m)}^2)\epsilon_1 
& \frac32 (3-2\sqrt{3}) (\tilde\alpha_{d(m)}^2-\tilde\beta_{d(m)}^2) \\
3(\sqrt{3}-2) (\tilde\alpha_{d(m)}^2-\tilde\beta_{d(m)}^2)\epsilon_1^*   &  2(2-\sqrt{3}) (\tilde\gamma_{d(m)}^2+\tilde\alpha_{d(m)}^2+\tilde\beta_{d(m)}^2)|\epsilon_1|^2 &
(3-2\sqrt{3}) (2\tilde\gamma_{d(m)}^2-\tilde\alpha_{d(m)}^2-\tilde\beta_{d(m)}^2)\epsilon_1^*\\
\frac32 (3-2\sqrt{3}) (\tilde\alpha_{d(m)}^2-\tilde\beta_{d(m)}^2) &  (3-2\sqrt{3}) (2\tilde\gamma_{d(m)}^2-\tilde\alpha_{d(m)}^2-\tilde\beta_{d(m)}^2)\epsilon_1 & 
\frac32(2-\sqrt{3}) (4\tilde\gamma_{d(m)}^2+\tilde\alpha_{d(m)}^2+\tilde\beta_{d(m)}^2)
\end{pmatrix}\,,
\label{downmassmatrix-m}
\end{align} 
\normalsize
where $\gamma_{d(m)}^2\gg \alpha_{d(m)}^2, \beta_{d(m)}^2$ is taken 
and  $\epsilon_2=2\epsilon_1$ in Eq.\,(\ref{epS12}) is put.
The extra transformation $U_{12}$ is performed 
to keep the hierarchy among matrix elements  $(3,3)\gg (2,3) \gg (1,3)$
because the CKM mixing angles are hierarchical.

Then, we can estimate the mixing matrix $U_{dm}$ of order $ {\cal O}(\epsilon_1)$
to diagonarize $M_{d}^\dagger M_{d}$:  
\begin{align}
U_{md}^\dagger M_{d}^\dagger M_{d}U_{md}={\rm diag}(m_d^2, m_s^2, m_b^2)\,,
\qquad\quad 
U_{md}\simeq 
\begin{pmatrix}
1 & s_{12}^{d} e^{i\eta} & 0\\
-s_{12}^{d}e^{-i\eta} & 1 & s_{23}^{d} \\
s_{12}^d s_{23}^de^{-i\eta} & -s_{23}^{d} & 1
\end{pmatrix}
\label{}\,,
\end{align}
where $\eta=2 {\rm arg}\,[\epsilon_1]$.
The mixing angle  $s_{23}^{d}$ is of order $ {\cal O}(\epsilon_1)$.
On the other hand, $s_{12}^{d}$ depends on 
$\tilde\alpha_{d(m)}^2/\tilde\gamma_{d(m)}^2$, $\tilde\beta_{d(m)}^2/\tilde\gamma_{d(m)}^2$
and $\epsilon_1$. In the numerical fit,
$s_{12}^{d}\sim 0.1$ has been obtained \cite{Okada:2019uoy}.

On the other hand, the transformation of the charged leptons
is still  $E_L\to   U_{S} E_L$, 
where $U_{12}$ of Eq.\,\eqref{U12} does not appear
because the large mixing angle between the second and third families is reproduced in the neutrino mass matrix under this transformation,
which is preferable for the lepton mixing matrix.
Then, at  $\tau=i+\epsilon$, we have 
\scriptsize
\begin{align} 
M_{E}^\dagger M_{E}\simeq v_d^2
\begin{pmatrix}
2(2-\sqrt{3}) (\tilde\gamma_{e(m)}^2+\tilde\alpha_{e(m)}^2+\tilde\beta_{e(m)}^2)|\epsilon_1|^2
& 3(\sqrt{3}-2) (\tilde\alpha_{e(m)}^2-\tilde\beta_{e(m)}^2)\epsilon_1^* 
& 
(3-2\sqrt{3}) (2\tilde\gamma_{e(m)}^2-\tilde\alpha_{e(m)}^2-\tilde\beta_{e(m)}^2)\epsilon_1^* \\
3(\sqrt{3}-2) (\tilde\alpha_{e(m)}^2-\tilde\beta_{e(m)}^2)\epsilon_1   & 
\frac92 (2-\sqrt{3}) (\tilde\alpha_{e(m)}^2+\tilde\beta_{e(m)}^2)  &
\frac32 (3-2\sqrt{3}) (\tilde\alpha_{e(m)}^2-\tilde\beta_{e(m)}^2) \\
(3-2\sqrt{3}) (2\tilde\gamma_{e(m)}^2-\tilde\alpha_{e(m)}^2-\tilde\beta_{e(m)}^2)\epsilon_1 & 
\frac32 (3-2\sqrt{3}) (\tilde\alpha_{e(m)}^2-\tilde\beta_{e(m)}^2) & 
\frac32(2-\sqrt{3}) (4\tilde\gamma_{e(m)}^2+\tilde\alpha_{e(m)}^2+\tilde\beta_{e(m)}^2)
\end{pmatrix}\,,
\label{Emassmatrix-m}
\end{align}
\normalsize
where   $\gamma_{e(m)}^2\gg \alpha_{e(m)}^2, \beta_{e(m)}^2$ is taken 
and  $\epsilon_2=2\epsilon_1$ in Eq.\,(\ref{epS12}) is put.
It is remarked that the phase of $\epsilon_1$ can be removed.
That is a real matrix.
Since $\alpha_{e(m)}^2/\gamma_{e(m)}^2$ and $\beta_{e(m)}^2/\gamma_{e(m)}^2$ are  smaller than
$10^{-3}$,
it is noticed that this matrix is a rank one matrix in the limit of  $\alpha_{e(m)}^2=\beta_{e(m)}^2=0$. 
Taking relevant value of  $\alpha_{e(m)}^2/\gamma_{e(m)}^2$, $\beta_{e(m)}^2/\gamma_{e(m)}^2$
and $|\epsilon_1|$, desired lepton masses could be obtained.

The mixing matrix $U_{me}$ 
to diagonarize $M_{E}^\dagger M_{E}$:  
\begin{align}
U_{me}^\dagger M_{E}^\dagger M_{E}U_{me}={\rm diag}(m_e^2, m_\mu^2, m_\tau^2)\,,
\qquad\quad 
U_{me}\simeq 
\begin{pmatrix}
1 & s_{12}^{e}  &s_{13}^{e}\\
-s_{12}^{e} & 1 & 0\\
-s_{13}^d  & 0 & 1
\end{pmatrix}\,,
\end{align}
where   $s_{13}^{e}$ is of order $ {\cal O}(|\epsilon_1|)$.
On the other hand, $s_{12}^{e}$ depends on 
$\tilde\alpha_{e(m)}^2/\tilde\gamma_{e(m)}^2$, $\tilde\beta_{e(m)}^2/\tilde\gamma_{e(m)}^2$
and $|\epsilon_1|$. 
In the numerical fit,
$s_{12}^{e}\sim 0.1$ has been obtained \cite{Okada:2020brs}.

\section{Decomposition of products of triplets}
\label{decomposition}
We take the generators of $A_4$ group for the triplet as follows:
\begin{align}
\begin{aligned}
S=\frac{1}{3}
\begin{pmatrix}
-1 & 2 & 2 \\
2 &-1 & 2 \\
2 & 2 &-1
\end{pmatrix},
\end{aligned}
\qquad 
\begin{aligned}
T=
\begin{pmatrix}
1 & 0& 0 \\
0 &\omega& 0 \\
0 & 0 & \omega^2
\end{pmatrix}, 
\end{aligned}
\end{align}
where $\omega=e^{i\frac{2}{3}\pi}$ for a triplet.
In this base,
the multiplication rule is
\begin{align}
\begin{pmatrix}
a_1\\
a_2\\
a_3
\end{pmatrix}_{\bf 3}
\otimes 
\begin{pmatrix}
b_1\\
b_2\\
b_3
\end{pmatrix}_{\bf 3}
&=\left (a_1b_1+a_2b_3+a_3b_2\right )_{\bf 1} 
\oplus \left (a_3b_3+a_1b_2+a_2b_1\right )_{{\bf 1}'} \nonumber \\
& \oplus \left (a_2b_2+a_1b_3+a_3b_1\right )_{{\bf 1}''} \nonumber \\
&\oplus \frac13
\begin{pmatrix}
2a_1b_1-a_2b_3-a_3b_2 \\
2a_3b_3-a_1b_2-a_2b_1 \\
2a_2b_2-a_1b_3-a_3b_1
\end{pmatrix}_{{\bf 3}}
\oplus \frac12
\begin{pmatrix}
a_2b_3-a_3b_2 \\
a_1b_2-a_2b_1 \\
a_3b_1-a_1b_3
\end{pmatrix}_{{\bf 3}\  } \ , \nonumber \\
\nonumber \\
{\bf 1} \otimes {\bf 1} = {\bf 1} \ , \qquad &
{\bf 1'} \otimes {\bf 1'} = {\bf 1''} \ , \qquad
{\bf 1''} \otimes {\bf 1''} = {\bf 1'} \ , \qquad
{\bf 1'} \otimes {\bf 1''} = {\bf 1} \  ,
\end{align}
where
\begin{align}
T({\bf 1')}=\omega\,,\qquad T({\bf 1''})=\omega^2. 
\end{align}
More details are shown in the review~\cite{Ishimori:2010au,Ishimori:2012zz}.

By using these multiplication rules, we have 
\begin{align}
(YD_L)_{\bf 1} &= Y_1 d_{L1} + Y_2d_{3L} + Y_3d_{2L}\,, \nonumber\\
(YD_L)_{\bf 1'} &= Y_3 d_{3L}  + Y_1d_{2L} + Y_2d^L_1\,, \nonumber \\
(YD_L)_{\bf 1''} &= Y_2 d_{2L} + Y_1d_{3L} + Y_3d_{1L}\,,  
\label{product1}
\end{align}
\begin{eqnarray}
(YD_L)_{\bf 3s}=\left(
\begin{array}{c}
(YD_L)_{3s1} \\ (YD_L)_{3s2} \\
(YD_L)_{3s3}
\end{array}\right)=\frac13 \left(
\begin{array}{c}
2Y_1d_{1L}- Y_2d_{3L} - Y_3 d_{2L} \\
2Y_3 d_{3L} -Y_1d_{2L} -Y_2d_{1L} \\ 
2Y_2 d_{2L} - Y_1 d_{3L} - Y_3 d_{1L}
\end{array}
\right)_{\bf 3s}\,,
\label{product3s}
\end{eqnarray}
\begin{eqnarray}
(YD_L)_{\bf 3a}=\left(
\begin{array}{c}
(YD_L)_{3a1} \\
(YD_L)_{3a2} \\
(YD_L)_{3a3}
\end{array}\right)=\frac12 \left(
\begin{array}{c}
Y_2d_{3L} - Y_3 d_{2L} \\
Y_1d_{2L} -Y_2d_{1L} \\ 
Y_3 d_{1L} - Y_1 d_{3L}
\end{array}
\right)_{\bf 3a}\, ,
\label{product3a}
\end{eqnarray}
where the $A_4$ triplets
$Y_{\bf 3}$ and $D_{L\bf 3}$ are abbreviated as
$Y$ and $D_{L}$, respectively.

The similar formulae are obtained 
for$Y_k^* \bar d^L_j$,
but it is noticed  that
the exchange between $1'$ and  $1''$, and  the second entry and third one
in the triplet representation, respectively,
through their complex conjugates.
That is:
\begin{align}
(Y^* \bar D_L)_{\bf 1} &=  Y_1^*\bar d^L_1 + Y_2^* \bar d^L_3 + Y_3^* \bar d^L_2\,, \nonumber \\
(Y^* \bar D_L)_{\bf 1''} &=  Y_3^* \bar d^L_3 +  Y_1^* \bar d^L_2 + Y^*_2 \bar d^L_1 \,, \nonumber\\
(Y^* \bar D_L)_{\bf 1'} &= Y_2^* \bar d^L_2 +  Y_1^* \bar d^L_3 +  Y_3^* \bar d^L_1 \,,
\label{product1bar}
\end{align}
\begin{eqnarray}
(Y^* \bar D_L)_{\bf 3s}=\left(
\begin{array}{c}
(Y^* \bar D_L)_{3s1} \\ (Y^* \bar D_L)_{3s2} \\ 
(Y^* \bar D_L)_{3s3}
\end{array}\right)=\frac13 \left(
\begin{array}{c}
2Y_1^* \bar d_{1L} -  Y_2^* \bar d_{3L}  -  Y_3^* \bar d_{2L}  \\ 
2 Y_2^* \bar d_{2L} -  Y_1^* \bar d_{3L}  -  Y_3^* \bar d_{1L} \\
2 Y_3^* \bar d_{3L}  - Y_1^* \bar d_{2L}  - Y_2^* \bar d_{1L}  
\end{array}
\right)_{\bf 3s}\,,
\label{product3sbar}
\end{eqnarray}
\begin{eqnarray}
(Y^* \bar D_L)_{\bf 3a}=\left(
\begin{array}{c}
(Y^* \bar D_L)_{3a1} \\ (Y^* \bar d_L)_{3a2} \\
(Y^* \bar D_L)_{3a3}
\end{array}\right)=\frac12 \left(
\begin{array}{c}
Y_2^* \bar d_{3L}  -  Y_3^* \bar d_{2L}  \\ 
Y_3^* \bar d_{1L}  -  Y_1^* \bar d_{3L}  \\
Y_1^*\bar d_{2L}  - Y_2^* \bar d_{1L}  
\end{array}
\right)_{\bf 3a}\,.
\label{produc3abar}
\end{eqnarray}
By using these decompositions,
we can calculate the product of following triplets:
\footnotesize
\begin{eqnarray}
&&  [( Y^* \bar D_L)_{\bf 3s} \otimes  (Y D_L)_{\bf 3s}]_{\bf 3s}=
\frac13 \left(
\begin{array}{c}
2(Y^*\bar D_L)_{3s1}(YD_L)_{3s1}- (Y^*\bar D_L)_{3s2}(YD_L)_{3s3} -
(Y^*\bar D_L)_{3s3}(YD_L)_{3s2}  \\
2(Y^*\bar D_L)_{3s3}(YD_L)_{3s3}- (Y^*\bar D_L)_{3s1}(YD_L)_{3s2} -
(Y^*\bar D_L)_{3s2}(YD_L)_{3s1} \\ 
2(Y^*\bar D_L)_{2s1}(YD_L)_{2s1}- (Y^*\bar D_L)_{3s1}(YD_L)_{3s3} -
(Y^*\bar D_L)_{3s3}(YD_L)_{3s1}
\end{array}
\right), 
\nonumber\\
&&  [( Y^* \bar D_L)_{\bf 3s} \otimes  (Y D_L)_{\bf 3s}]_{\bf 3a}=
\frac12 \left(
\begin{array}{c}
(Y^*\bar D_L)_{3s2}(YD_L)_{3s3}- (Y^*\bar D_L)_{3s3}(YD_L)_{3s2} \\
(Y^*\bar D_L)_{3s1}(YD_L)_{3s2}- (Y^*\bar D_L)_{3s2}(YD_L)_{3s1}\\ 
(Y^*\bar D_L)_{3s3}(YD_L)_{3s1}- (Y^*\bar D_L)_{3s1}(YD_L)_{3s3}
\end{array}
\right). 
\nonumber 
\\\end{eqnarray}
\normalsize
Other six decompositions
$[( Y^* \bar D_L)_{\bf 3s} \otimes  (Y D_L)_{\bf 3a}]_{\bf 3s}$,
$[( Y^* \bar D_L)_{\bf 3a} \otimes  (Y D_L)_{\bf 3s}]_{\bf 3s}$,
$[( Y^* \bar D_L)_{\bf 3a} \otimes  (Y D_L)_{\bf 3a}]_{\bf 3s}$,
$[( Y^* \bar D_L)_{\bf 3s} \otimes  (Y D_L)_{\bf 3a}]_{\bf 3a}$,
$[( Y^* \bar D_L)_{\bf 3a} \otimes  (Y D_L)_{\bf 3s}]_{\bf 3a}$
and 
$[( Y^* \bar D_L)_{\bf 3a} \otimes  (Y D_L)_{\bf 3a}]_{\bf 3a}$
are written similarly.


\section{$U(2)$ flavor symmetry}
\label{U2}
	The $U(2)^5$ flavor symmetry is the subgroup of $U(3)^5$ and distinguishes the first two families of fermions 
	from the third one~\cite{Barbieri:2011ci,Barbieri:2012uh,Blankenburg:2012nx}.
	The flavor symmetry is decomposed as 
	\begin{align}
	U(2)^5 = U(2)_L \otimes U(2)_Q \otimes U(2)_E \otimes U(2)_U \otimes U(2)_D~.
	\end{align}
	Under this symmetry,  the first two families transform as doublets of the $U(2)$ subgroups, 
	whereas third family ones as a singlet. 
	Then, the third-family Yukawa couplings are allowed by the symmetry, and  
	 it provides a natural explanation of why third-family Yukawa couplings are large.
	A minimal set of $U(2)^5$ breaking terms (spurions) which reproduce the observed SM flavor parameters,
	 without tuning and with minimal size for the breaking terms, is given by~\cite{Barbieri:2011ci}
	\begin{eqnarray}
	&	V_\ell \sim\left(2,1,1,1,1\right)~,	\qquad   V_q \sim\left(1,2,1,1,1\right)~,  & \nonumber \\
	&  \Delta_e \sim\left(2,1,\bar{2},1,1\right)~, \qquad  \Delta_u \sim\left(1,2,1,\bar{2},1\right)~, \qquad  \Delta_d \sim\left(1,2,1,1,\bar{2}\right)~, &
	\end{eqnarray}
	where $V_{q,\ell}$ are complex two-vectors and $\Delta_{e,u,d}$ are complex $2\times 2$~matrices. 
	In terms of these spurions, the $3 \times 3$ Yukawa matrices can be decomposed as
	\begin{align}
	Y_e = y_\tau\left(\begin{matrix}
	\Delta_e	 & x_\tau V_\ell \\
	0			 & 1
	\end{matrix}\right), && Y_u = y_t\left(\begin{matrix}
	\Delta_u	 & x_t V_q \\
	0			 & 1
	\end{matrix}\right), && Y_d = y_b\left(\begin{matrix}
	\Delta_d	 & x_b V_q \\
	0			 & 1
	\end{matrix}\right),
	\label{eq:YU2_5}
	\end{align}
	where $y_{\tau,t,b}$ and $x_{\tau,t,b}$ are free complex parameters, expected to be of order~$\mathcal{O}(1)$. 	
	Using the residual $U(2)^5$ invariance, we can transform the spurions to the following 
	explicit form (see Ref. \cite{Faroughy:2020ina})
	\begin{equation}
	V_{q(\ell)} =    e^{i \bar \phi_{q(\ell)}} \begin{pmatrix}0 \\   \epsilon_{q(l)}  \end{pmatrix}~,  \quad  
	\Delta_e = O_e^\intercal\, \begin{pmatrix} \delta'_e  & 0 \\  0 & \delta_e \end{pmatrix} ~, \quad
	\Delta_u= U_u^\dagger   \begin{pmatrix} \delta'_u  & 0 \\  0 & \delta_u \end{pmatrix}~,  \quad  
	\Delta_d= U_d^\dagger  \begin{pmatrix} \delta'_d  & 0 \\  0 & \delta_d \end{pmatrix} ~, 
	\end{equation}
	where $O$ and $U$ represent $2\times2$ orthogonal  and complex unitary matrices, respectively
	\begin{align}\label{eq:Uq}
	O_{e}=
	\begin{pmatrix}
	c_{e} & s_{e}  \\
	-s_{e}   & c_{e}
	\end{pmatrix}\,,
	\qquad 
	U_{q}=
	\begin{pmatrix}
	c_q & s_q\,e^{i\alpha_q}\\
	-s_q\,e^{-i\alpha_q} & c_q
	\end{pmatrix}\,,
	\end{align}
	with $s_i \equiv \sin\theta_i$ and $c_i\equiv\cos\theta_i$. 
	The $\epsilon_i$ and  $\delta^{(\prime)}_i$ are small positive real parameters 
	controlling the overall size of the spurions.
	From the observed hierarchies of the Yukawa couplings, it is expected that they have following order relation 
	\begin{equation}
	1 \gg \epsilon_i   \gg  \delta_i      \gg \delta'_i     > 0
	\end{equation}
	where 
	\small
	\begin{eqnarray}
	\epsilon_i 
	= \mathcal O( y_t |V_{ts}| ) = \mathcal O(10^{-1})~, 
	\delta_i  =  \mathcal O\left( \frac{y_c}{y_t},   \frac{y_s}{y_b},   \frac{y_\mu}{y_\tau}  \right) = \mathcal O(10^{-2})~, 
	\delta'_i  =  \mathcal O\left( \frac{y_u}{y_t},   \frac{y_d}{y_b},   \frac{y_e}{y_\tau}  \right)   = \mathcal O(10^{-3}).
	\label{eq:U2SpSize}
	\end{eqnarray}
	\normalsize
	
	In terms of the spurion expressions discussed above,  
	we can classify the number of independent operators in the SMEFT with a $U(2)^5$ flavor symmetry.  
	Following the complete classification in Ref. \cite{Faroughy:2020ina}, 
	we focus on the chirality-flipped $d_i \to d_j$ and $\ell_i \to \ell_j$ transitions in this paper.
	Here, we classify the operators up to $\mathcal O(V^3, \Delta V) \sim \mathcal O(10^{-3})$,  
	given the size of the spurions in Eq.\,(\ref{eq:U2SpSize}). 
	The $b_i \to b_j$ transition allowed by different $U(2)$ breaking terms 
	are summarized in Table \ref{tab:U2RLbs},  where we follow the result of Table 4 in Ref.\cite{Faroughy:2020ina}. 
	The case of $\ell_i \to \ell_j$ transition is given in the same way, and is summarised in Table \ref{tab:leptonLRatU2}.  
	We denote with latin (greek) letters the real (complex) couplings appearing in hermitian (non-hermitian) structures. 
	Terms with the same number of spurions are denoted by the same latin or greek letter with different subscript.
 

\end{document}